\documentclass[%
,aps%
 ,onecolumn%
 ,secnumarabic%
,amssymb, amsmath,nobibnotes, aps, prl, floatfix, superscriptaddress, nofootinbib, showkeys, 10pt]{revtex4-2}
\usepackage{graphicx}
\usepackage{bm}%
\usepackage[colorlinks=true,linkcolor=blue]{hyperref}%
\usepackage{orcidlink}
\expandafter\ifx\csname package@font\endcsname\relax\else
 \expandafter\expandafter
 \expandafter\usepackage
 \expandafter\expandafter
 \expandafter{\csname package@font\endcsname}%
\fi

\begin{document}


\title{{$\Phi$} index: A standardized scale-independent and field-normalized citation indicator}




\author{Manolis Antonoyiannakis\,\orcidlink{0000-0001-6174-0668}}
\email{manolis@bibliostatistics.org}
\affiliation{American Physical Society, 1 Physics Ellipse, College Park, 20740 Maryland, USA}


\date{\today}

\keywords{Science of science  $|$ bibliostatistics $|$ citation distributions $|$ Journal Impact Factor $|$ standardized indicators $|$ Monte Carlo simulations $|$ $\Phi$ index}

\begin{abstract}


The Impact Factor (IF), despite its widespread use, suffers from 
well-known biases that remain incompletely addressed in practice---most 
notably its sensitivity to journal size and its lack of field 
normalization. Because of size sensitivity, a randomly formed journal 
of $n$ papers can attain a range of IF values that decreases sharply 
with size, as $\sim 1/\sqrt{n}$. The Central Limit Theorem, which 
underlies this effect, also allows us to correct for it by standardizing 
citation averages for scale \textit{and} field in a manner analogous to 
calculating the $z$-score in statistics. We thus introduce the $\Phi$ 
(Phi) index, defined as $\Phi = (f - \mu)\sqrt{n}/\sigma$, where $f$ 
is a journal's average citation count (akin to the IF), $n$ its publication count, and 
$\mu, \sigma$ the mean and standard deviation of citations in its field. 
Applying the $\Phi$ index to 12,173 journals in Clarivate's Journal 
Citation Reports, we obtain rankings that correct for size bias and 
elevate journals from underrepresented fields such as mathematics, law, 
and history. We validate the $\Phi$ index via a Monte Carlo random 
sample test, which we propose as a standard diagnostic for any citation 
indicator. The methodology extends readily to departments, universities, 
and countries.

\end{abstract}








\maketitle



\section{Why a scale-independent citation indicator?}

\subsection{Impact Factors are scale-dependent}

\begin{figure}
\centering
\includegraphics[width=0.4\linewidth]{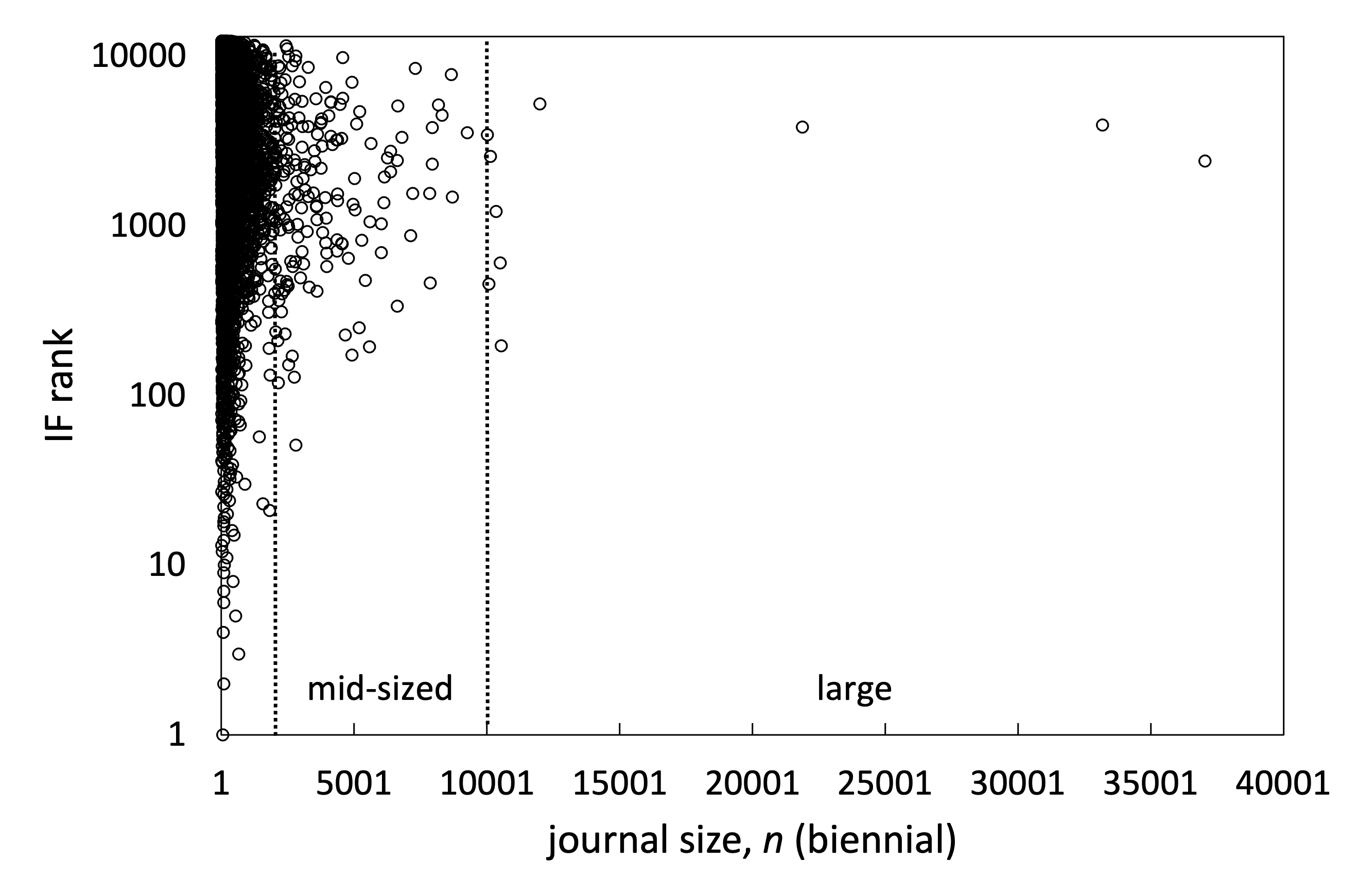}
\includegraphics[width=0.4\linewidth]{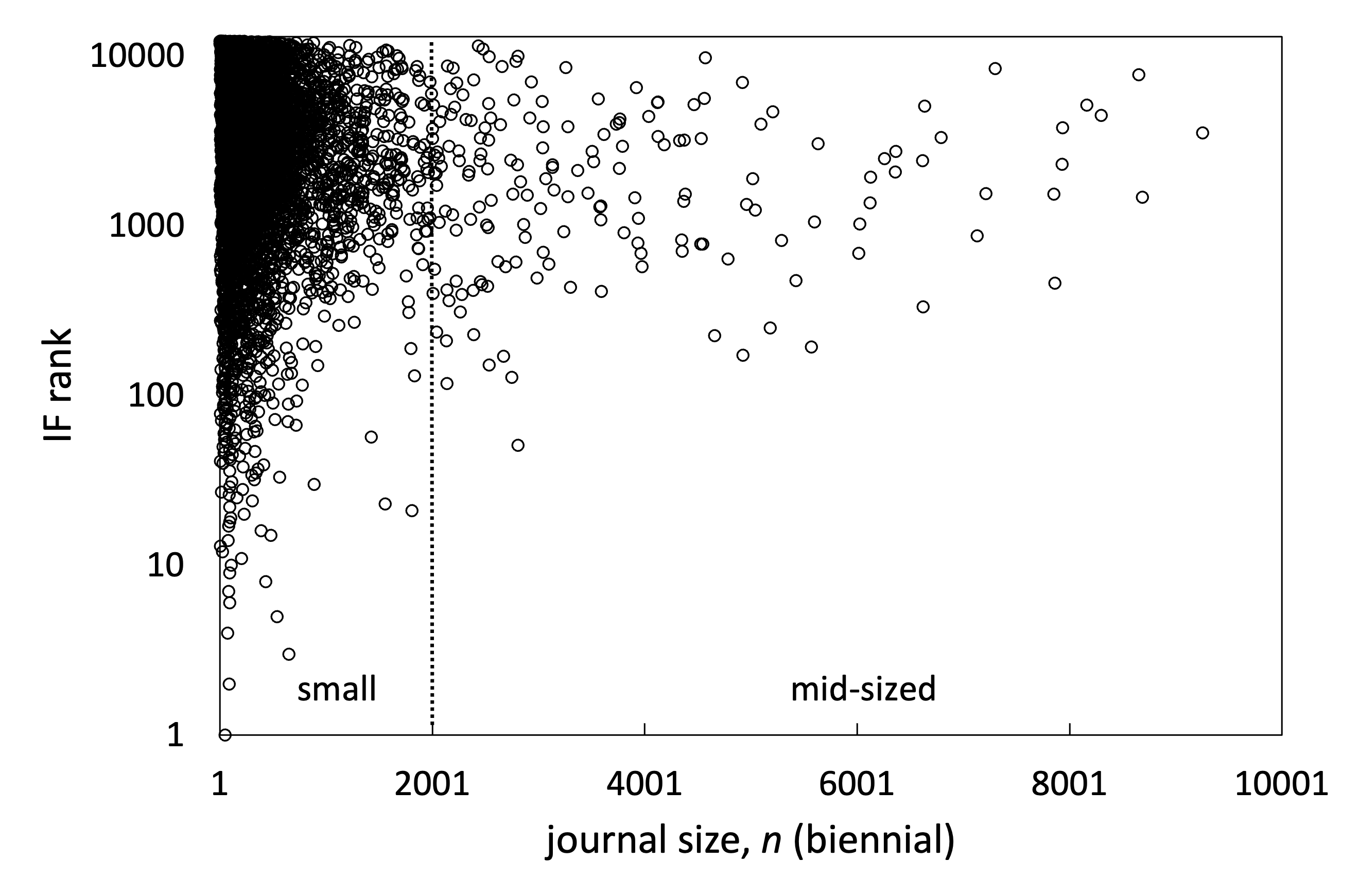}
\caption{How journal IF rankings are related to journal size. Small journals span all ranks. Mid-sized journals span more middle ranks, while large journals take even more mediocre ranks. Left panel: All journals. Right panel: Journals sized $n \le 10,000$.}
\label{fig:size_rank_scatter} 
\end{figure}

Every year at the end of June, when the Journal Impact Factors are announced by Clarivate Analytics, a familiar pattern occurs (Antonoyiannakis, 2018): (a) The high-Impact-Factor ranks are always populated by {\it small} journals that typically publish fewer than 1000 papers per discipline per year. For example, all the top 50 IF ranks---and all but 13 of the top 51--250 ranks---of the 2020 Journal Citation Reports (JCR) are filled by small journals. Of the 13 slots in the ranks 51--250 that are not occupied by small journals, we find 12 {\it mid-sized} journals (1001--5000 papers/year), and only one {\it large} journal (more than 5000 papers/year), in the 196$^{th}$ rank. (b) Small journals also fill the low-Impact-Factor ranks---for example, all but one of the bottom 1000 ranks in the 2020 JCR, the one exception being a mid-sized journal. (c) Large journals have modest Impact Factors that are neither very small nor very large, and tend to take unremarkable positions in the middle ranks---for example, there are 10 large journals with IF ranging from 2.5--14 and rank positions from 196 to 5189. Thus, large journals are precluded from reaching high IF ranks but are also protected from falling to the lowest of ranks as well. (d) Single papers affect the Impact Factors of hundreds---sometimes thousands---of {\it small} journals. For example, in 2017, the IF of 818 journals increased by more than 25\% by their {\it single} top-cited paper, while one in 10 journals had their IF boosted by more than 50\% by their top {\it three} cited papers (Antonoyiannakis, 2020). Figure {\ref{fig:size_rank_scatter}} shows a scatter plot of journal rank vs. journal size. The vertical dotted lines at {\it biennial} size $n=2000$ and $n=10,000$ denote our classification of journals into the three size categories (small, mid-sized, large). A glance at this figure confirms points (a), (b), and (c) above. 


Clearly, therefore, journal size affects IF rankings. This happens because despite the averaging process entailed in the IF calculation, IFs are not {\it size-independent}. They are {\it size-normalized} but size-dependent, because they are drawn from populations with high variance in citations, a behavior explained by the Central Limit Theorem (Antonoyiannakis, 2018). Remarkably, and despite this recurrent size-dependent behavior of Impact Factors every year,  ``the overwhelming majority of researchers, bibliometricians, administrators, publishers, and editors continue to use them without realizing or acknowledging the problem.''  (Antonoyiannakis, 2020). 

To the extent that IF rankings are important to the research community, high-IF journals have a perverse incentive to stay small to remain competitive in IF rankings (Antonoyiannakis, 2019; Antonoyiannakis \& Mitra, 2009). The editorial choice of the vast majority of high-IF journals to publish in the low hundreds of research papers per year in any field may thus be influenced more by the awareness of IF sensitivity to journal scale than by a genuine desire to stay small (Campbell, 2008).  


What can be done to remedy this situation and level the playing field? One approach would be to abandon citation averages (and thus Impact Factors) altogether in assessing research journals. Indeed, several {\it other} flaws of IF rankings have long been identified, such as the skewness of citation distributions, the field dependence of citation averages, the numerator/denominator asymmetry, journal self-citations, the length of citation window, IF inflation, IF gaming and engineering, or the document-type citation disparity (Amin \& Mabe, 2004; Moed, 2005; Larivi{\`e}re \& Sugimoto, 2019; Pulverer, 2013; Garfield, 1996; Gingras, 2016; Wouters {\it et al.}, 2019; Siler \& Larivi{\` e}re, 2022). Growing concern for these shortcomings and their repercussions in IF rankings has led to public calls for abandoning, de-emphasizing, or at least complementing IFs with more reliable metrics (San Francisco Declaration on Research Assessment, 2012; Gaind, 2018) and with detailed, statistically sensible visualizations (Adams {\it et al.}, 2019). 

In our opinion, Impact Factors are currently too deeply ingrained in the minds of scientists and administrators to be completely eliminated. Besides, the concept of an average is appealing in its simplicity. A more sensible approach may be therefore to provide a technical solution that corrects the scale-dependence of citation averages---in the same fashion that the field dependence of IFs has been addressed by indicators like the Article Influence Score (Clarivate, 2017), the Source Normalized Impact per Paper (SNIP) (Moed, 2010; Waltman, 2013), or the Journal Citation Indicator (Clarivate, 2021; Szomszor, 2021)---and argue for its acceptance by the community. This is the main aim of our paper.

Fortunately, such a solution is simple and can be found with the help of the Central Limit Theorem. Our methodology also enables us to correct for the field dependence of citation averages in an elegant and statistically sound manner, distinct from other approaches.  

In this paper, we introduce the concept of standardized citation averages. First, we perform a numerical experiment, where we show how a randomly formed journal---by consecutive unbiased samplings of actual papers in a given field---produces a citation average whose range of values clearly depends on its size, hence the need for a correction. We call this experiment the ``random sample test.'' Then, we express the Central Limit Theorem algebraically, thus bringing out the standardization of the citation average to correct for the scale dependence, which leads us to introduce the $\Phi$ index. We present an elegant and intuitive geometric interpretation of the $\Phi$ index, followed by a more formal interpretation as a $z$-score. We show how it accounts for both scale- and field-dependence of citation averages, thus passing the random sample test. We then present $\Phi$ index data for journals, using data from the 2020 JCR. 

Our work is aligned with the recommendations of scrutinizing and updating metrics (Hicks {\it et al.}, 2015), adding to the variety of journal-based metrics and accounting for article-type citation disparity and field-dependent citation practices (San Francisco Declaration on Research Assessment, 2012; Moed {\it et al.}, 2012).

\subsection{A randomly formed journal reveals the problem}

Consider a population of all papers published in a two-year window for which we collect their citations during the following (third) year. (We use these time windows for publications and citations merely for relevance to Impact Factors, but our analysis applies to any time windows.) Let us call the population's global citation average $\mu$ and its citation standard deviation $\sigma$.
For example, the 3,537,118 papers (articles and reviews) published in 2018--2019 have $\mu$=4.11 and $\sigma$=12.5 (2020 JCR).

\begin{figure}
\centering
\includegraphics[width=0.45\linewidth]{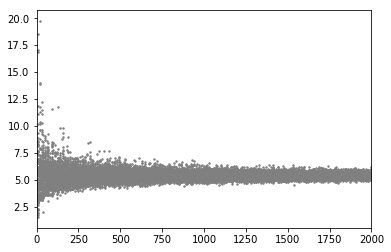}
\caption{Dependence of citation average $f_n$ on journal size $n$ for a random journal. Each gray dot represents $f_n$ for a random selection of $n$ papers within the field of cell biology. The range of $f_n$ values is centered on the global citation average $\mu$, while the upper and lower envelopes are proportional to $k \sigma/\sqrt{n}$ as shown in Eqs. (\ref{eq:3}, \ref{eq:4}).}
\label{fig:f-mu_1/sqrt(n)}
\end{figure}

Now, imagine we draw thousands of random samples of size $n$ from this population and calculate their citation average, $f_n$, which is essentially the Impact Factor of an $n$-sized ``random journal.'' We could perform this experiment for the entire set of 3,537,118 papers (in all fields) or for the papers in any field of choice, without loss of generality. For demonstration purposes, we choose here the subject (field) of cell biology, as defined by Clarivate Analytics' Journal Citation Reports (we use the terms ``subject'' and ``field'' as synonyms throughout the paper). We thus form a scatter plot of the values ($n, f_n$) as shown in Figure \ref{fig:f-mu_1/sqrt(n)}. As we can see, the range of $f_n$ values is much wider for smaller journals. Smaller journals thus have a better chance of scoring a high IF (but also low IF), while as the journal size increases, variations from the global mean decrease. The playing field is not level. (We will soon see that this behavior is typical across fields.) 

Let us understand this effect quantitatively. Under random sampling conditions, the sampling distribution of the sample means is normally distributed, so that
\begin{equation}
\mu - k \sigma_n \le f_n \le \mu + k \sigma_n \;, \qquad (k=1,2,3,...), \label{eq:1}
\end{equation}
where $k$ is an integer that increases with statistical accuracy, e.g., when $k=1$ then Eq. (\ref{eq:1}) is expected to capture 68\% of observations, $k=2$ for 95\%, $k=3$ for 99.7\% of observations, etc.
Crucially, the Central Limit Theorem (De Veaux, Velleman, \& Bock, 2014) allows us to express the sample standard deviation $\sigma_n$ in terms of the population standard deviation $\sigma$ as follows,
\begin{equation}
\sigma_n = \frac{\sigma}{\sqrt{n}}. \label{eq:2}
\end{equation}
Equations (\ref{eq:1}) and (\ref{eq:2}) imply that 
\begin{equation}
\mu -   \frac{k \sigma}{\sqrt{n}} \le f_n \le \mu + \frac{k \sigma}{\sqrt{n}}\;, \qquad \text{where} \;\;
    \begin{cases}
      k=1, & \text{for 68\% of random journals}\\
      k=2, & \text{for 95\% of random journals}\\
      k=3, & \text{for 99.7\% of random journals}
    \end{cases}       
\label{eq:3}
\end{equation}
or
\begin{equation}
-   \frac{k \sigma}{\sqrt{n}} \le f_n - \mu  \le + \frac{k \sigma}{\sqrt{n}}\;,
\label{eq:4}
\end{equation}
which can be rewritten as 
\begin{equation}
\text{max}(\Delta f_n) \cdot \sqrt{n} = k \sigma \;, \qquad
\text{``citation average uncertainty relation''}
\label{eq:5}
\end{equation}
where $\text{max}(\Delta f_n) \equiv \text{max}(|f_n - \mu|)$ is the maximum distance (or {\it range}) of $f_n$ values from the global mean $\mu$.
Equation (\ref{eq:5}) says that for randomly formed journals, the range of $f_n$ values from the global mean $\mu$ is inversely proportional to the square root of the journal size. Herein lies the scale dependence of citation averages.  

\begin{figure}
\centering
\includegraphics[width=0.45\linewidth]{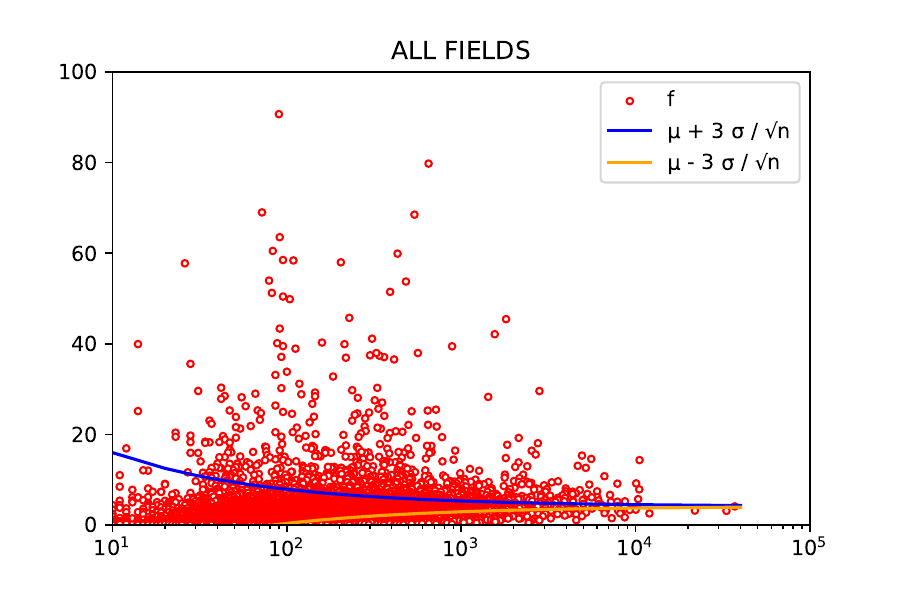}
\caption{
Dependence of citation average $f$ (red circles) on journal size $n$ for all 12173 journals in the 2020 JCR. Shown in blue and orange are the upper and lower limits, respectively, predicted by the Central Limit Theorem for random journals. There is a lot of overlap of data points in the low $f$ values: 78.6\% of all journals lie within these bounds.}
\label{fig:CLT_bounds_all} 
\end{figure}

While Eq. (\ref{eq:5}) applies to {\it randomly} formed journals, it turns out that more than 3/4 of {\it actual} journals have IFs within the range predicted by the Central Limit Theorem (with $k=3$)---for example, 78.6\% (or 9571 out of 12173) journals in the 2020 JCR.  So the Central Limit Theorem is directly relevant to journals in practice. It is not just of academic interest, but also allows us to predict and understand citation averages of real journals. See Figure \ref{fig:CLT_bounds_all}.

\begin{figure}
\centering

\includegraphics[width=0.3\linewidth]{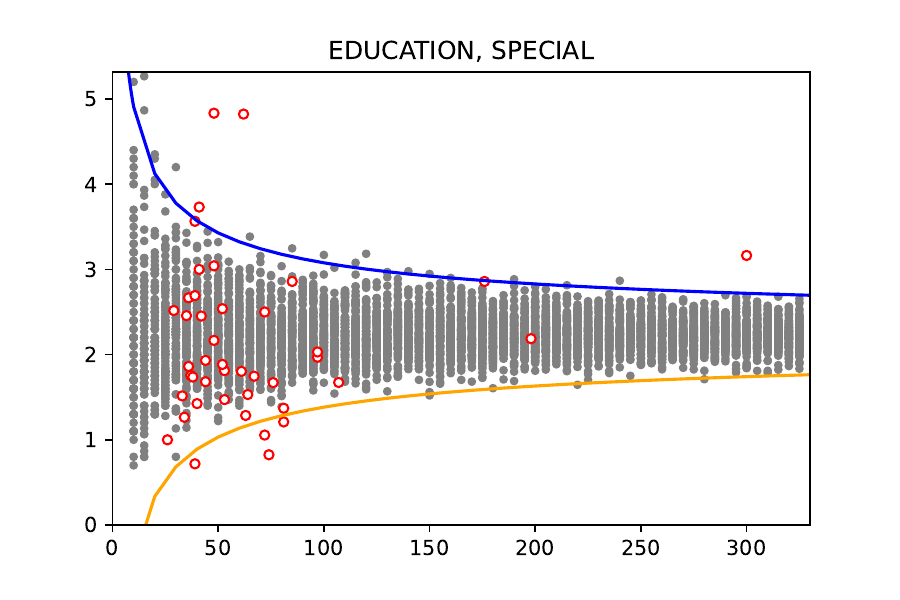} 
\includegraphics[width=0.3\linewidth]{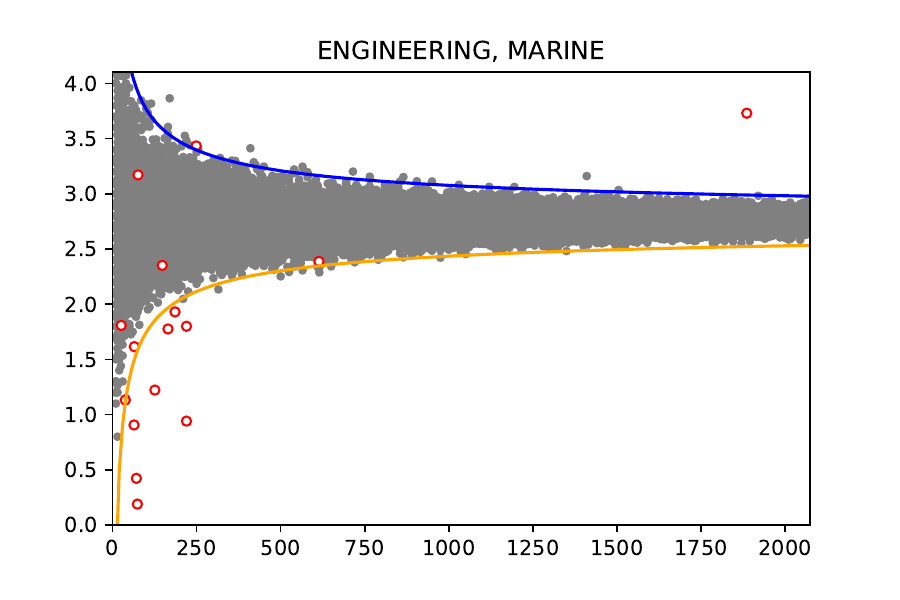}
\includegraphics[width=0.3\linewidth]{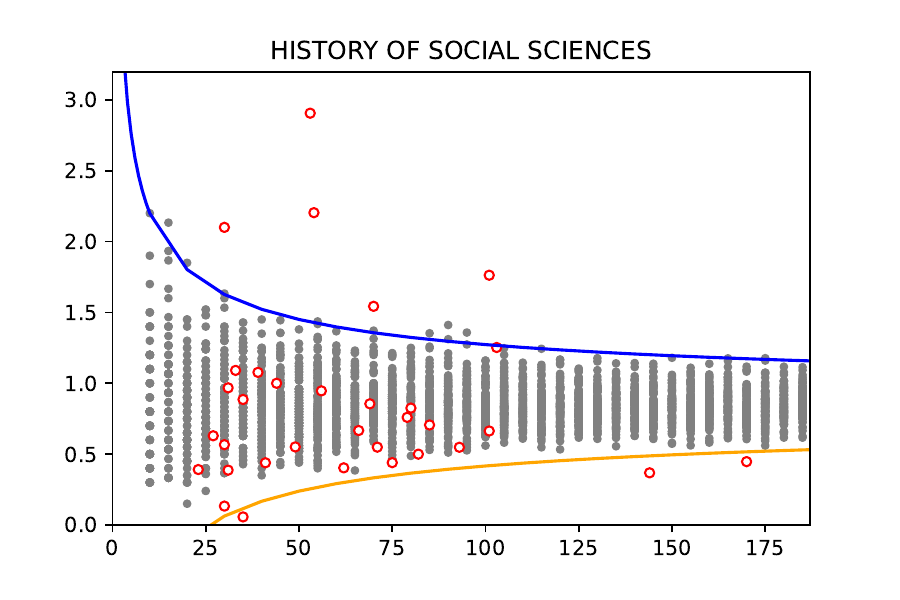}
\includegraphics[width=0.3\linewidth]{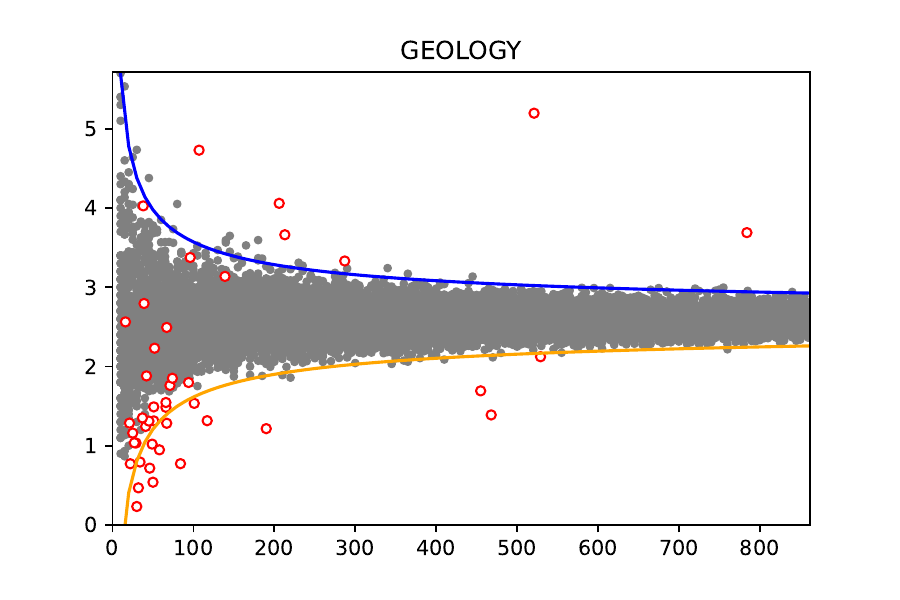}
\includegraphics[width=0.3\linewidth]{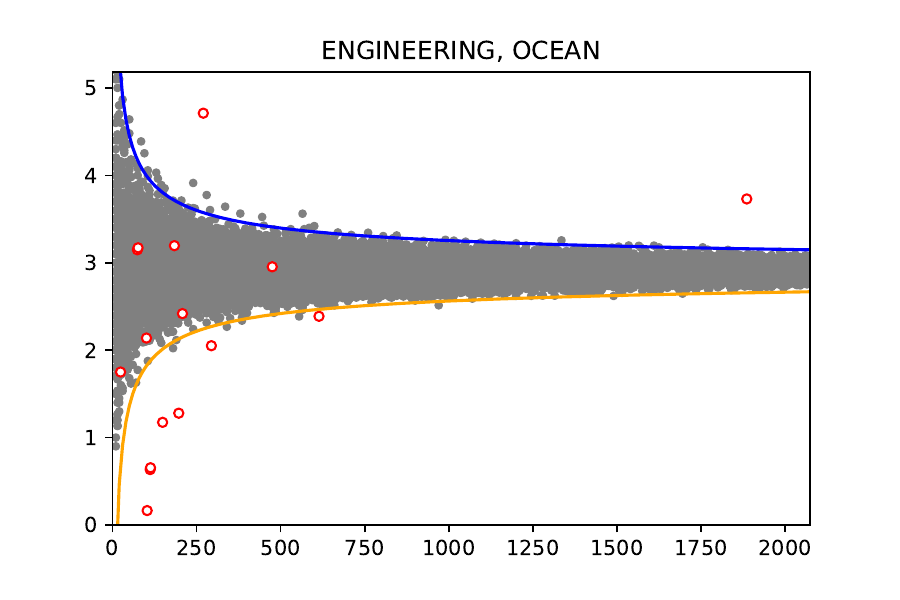}
\includegraphics[width=0.3\linewidth]{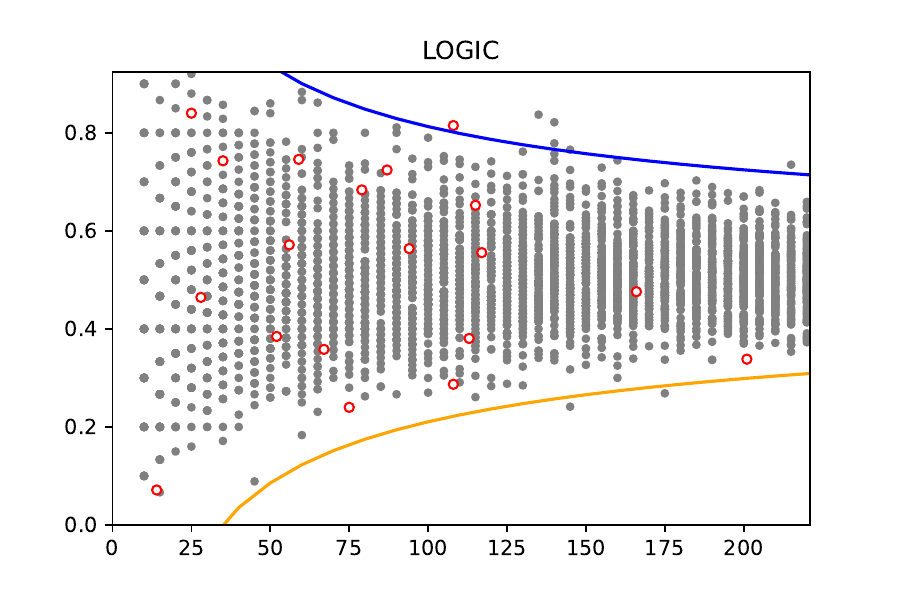}
\includegraphics[width=0.3\linewidth]{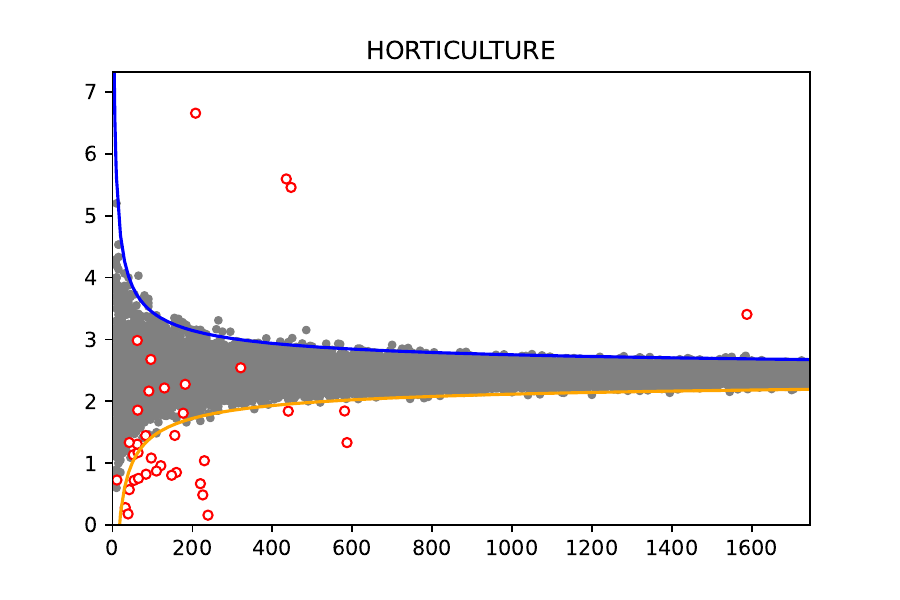}
\includegraphics[width=0.3\linewidth]{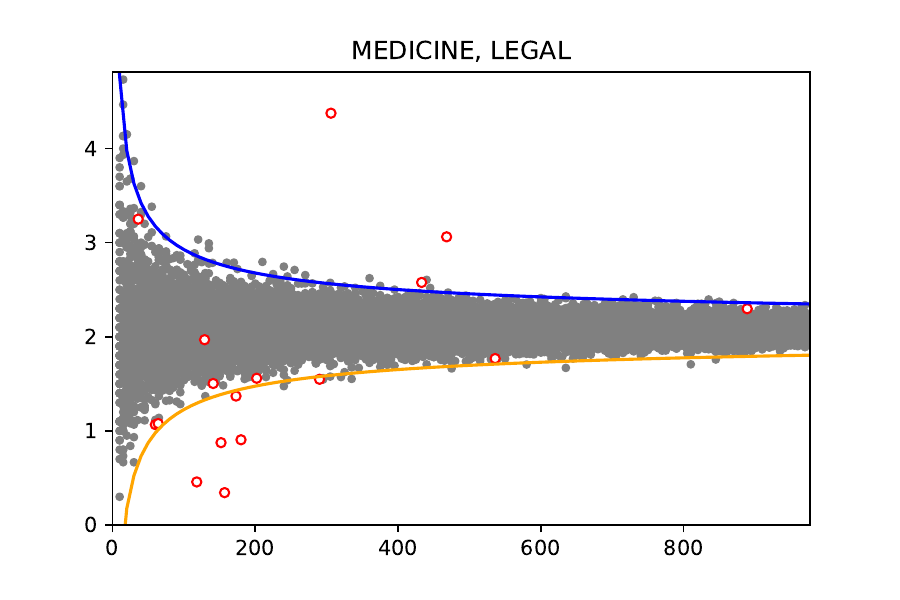}
\includegraphics[width=0.3\linewidth]{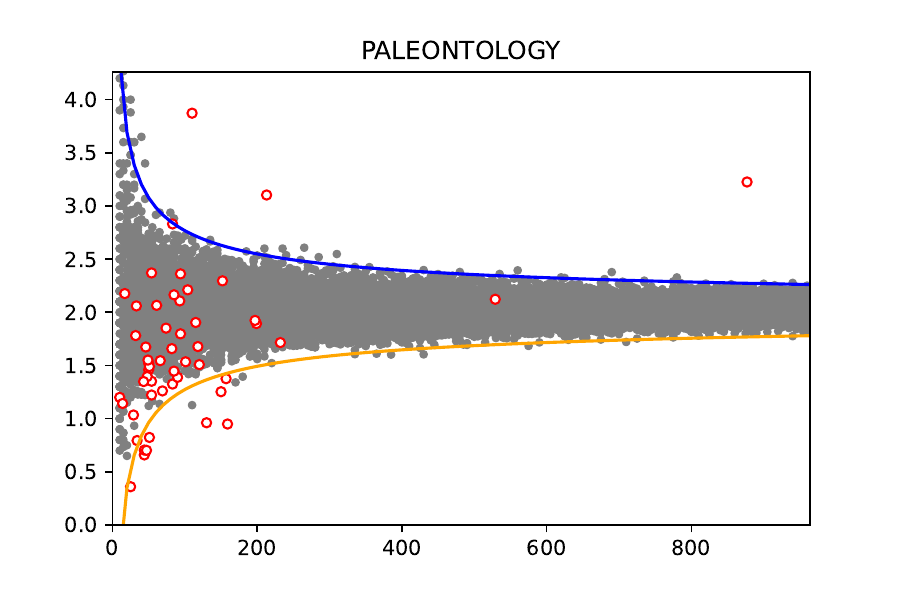}
\includegraphics[width=0.3\linewidth]{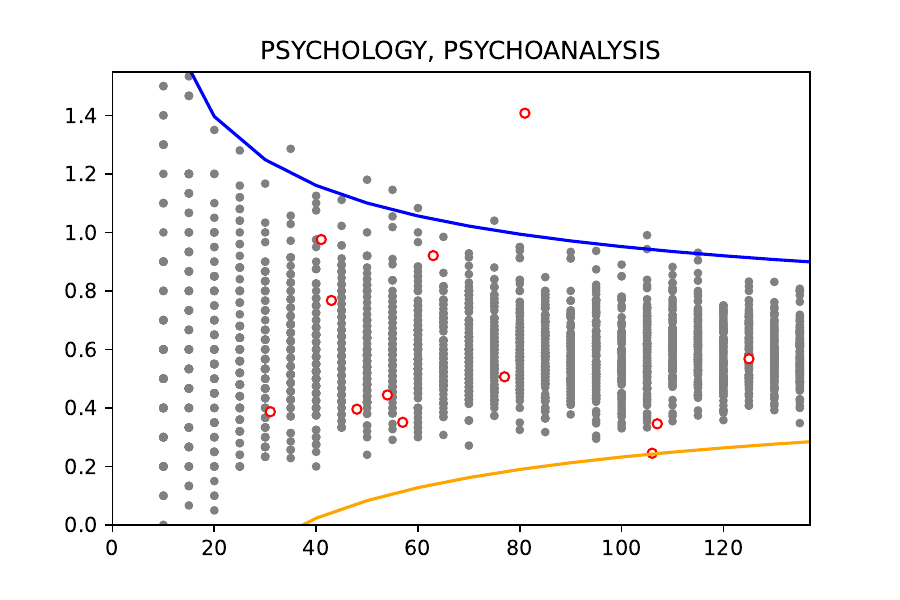}
\includegraphics[width=0.3\linewidth]{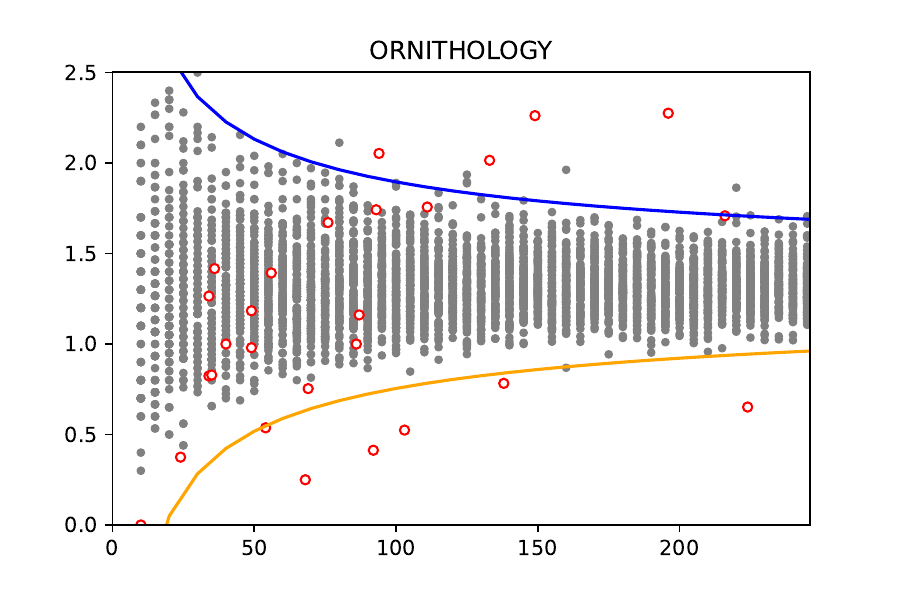}
\includegraphics[width=0.3\linewidth]{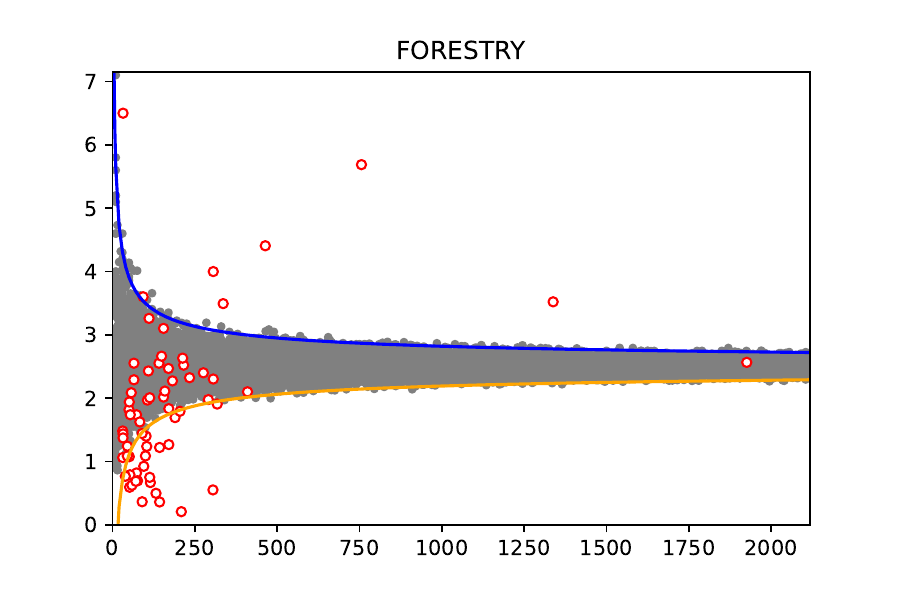}
\includegraphics[width=0.3\linewidth]{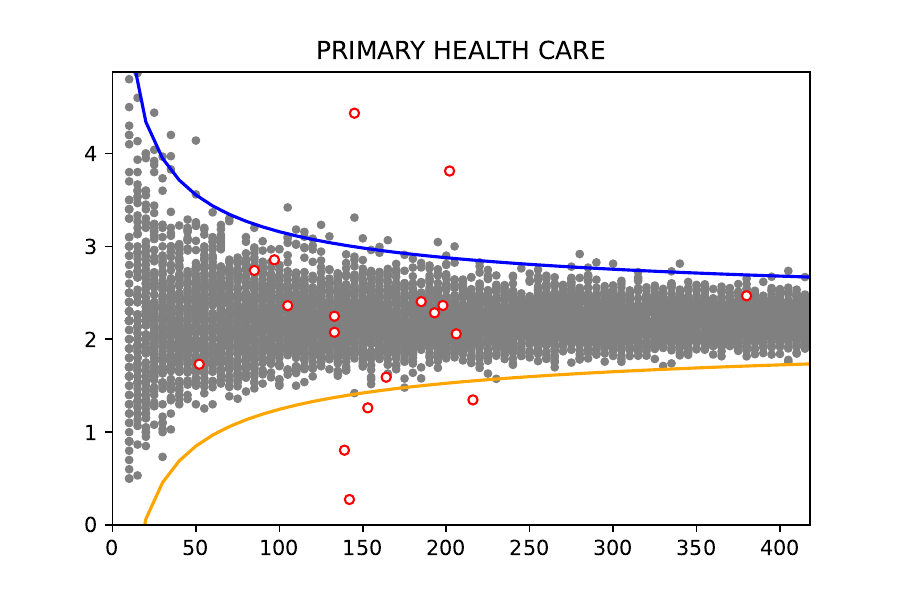}
\includegraphics[width=0.3\linewidth]{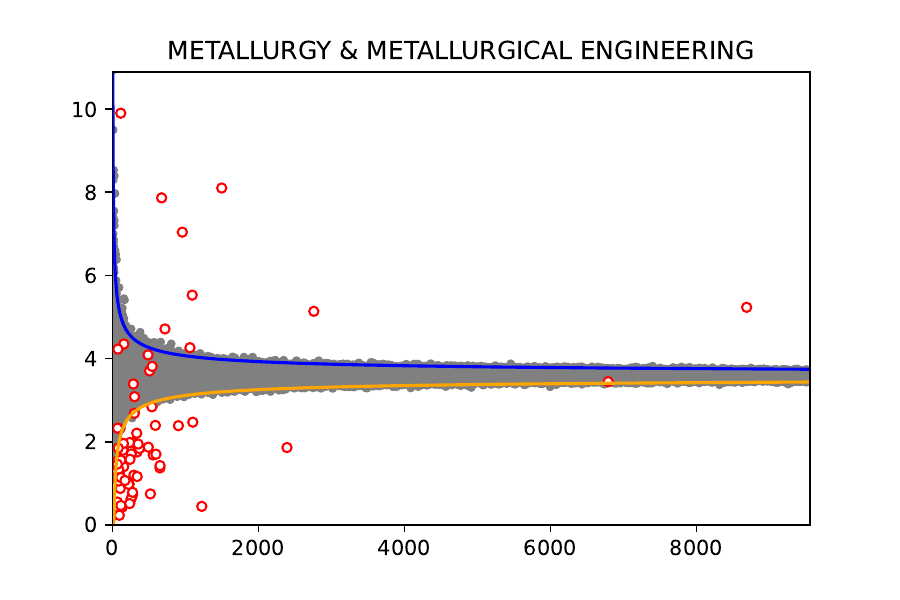}
\includegraphics[width=0.3\linewidth]{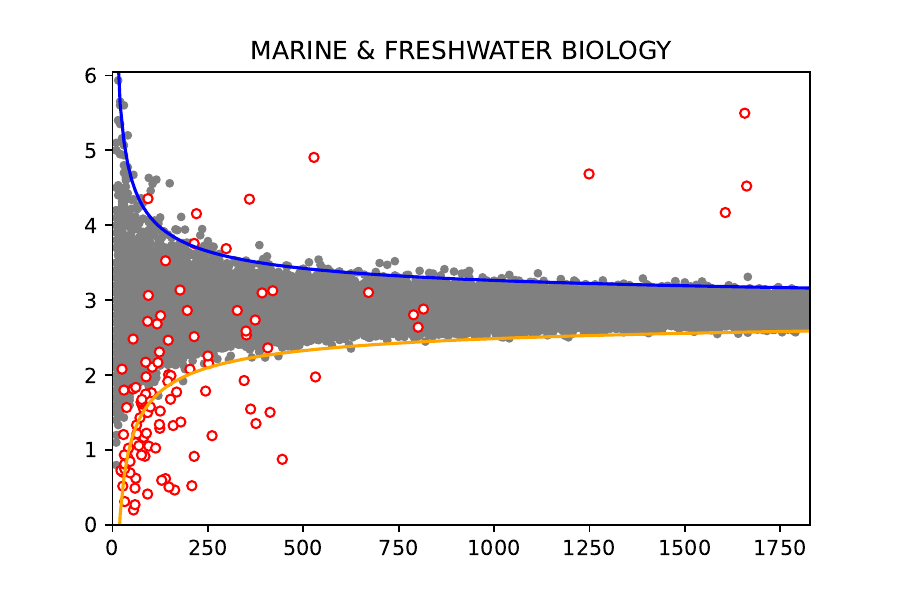}
\includegraphics[width=0.3\linewidth]{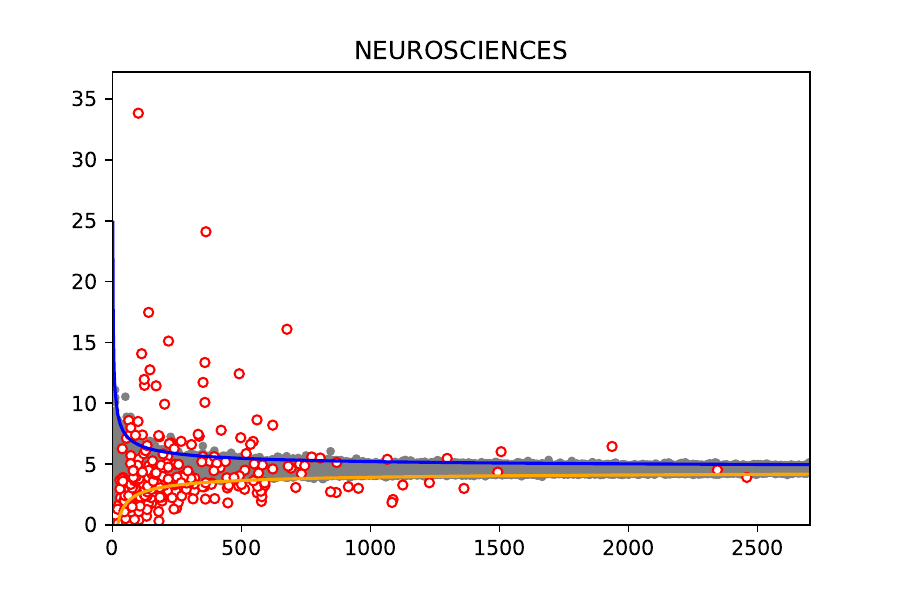}
\includegraphics[width=0.3\linewidth]{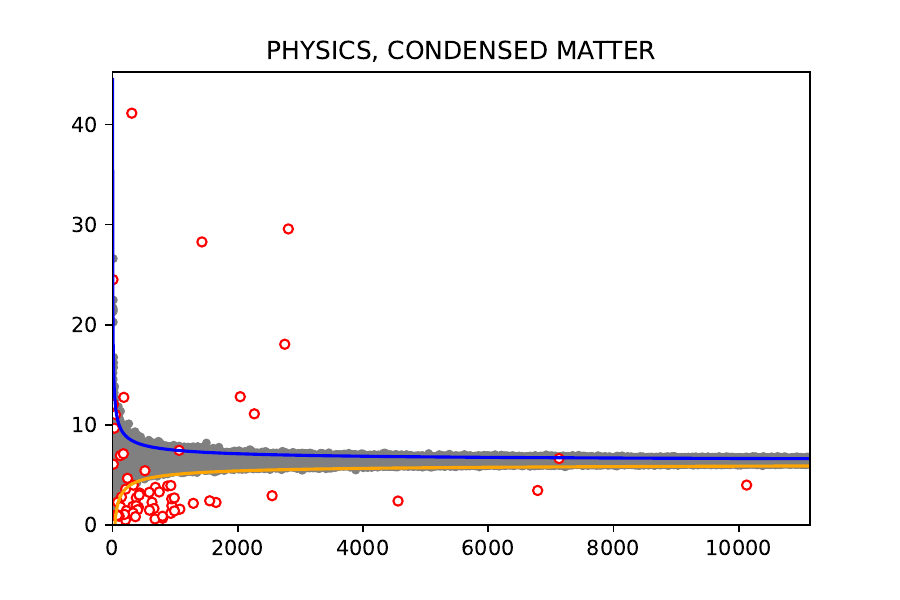}
\includegraphics[width=0.3\linewidth]{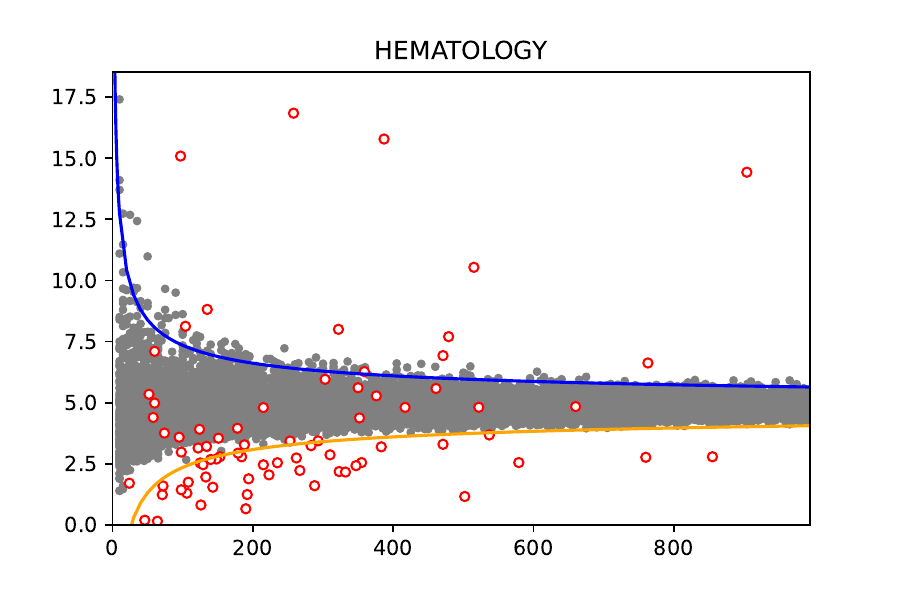}

\caption{Dependence of citation average $f_n$ (gray dots) on journal size $n$ for randomly formed journals in 18 field categories in the 2020 JCR. The fields are selected for skewness values ranging from ``low'' (the lowest among the 229 fields) to high. 
Shown in blue and orange lines are the upper and lower limits, respectively, predicted by the Central Limit Theorem for random journals. Also shown (in red circles) are the citation averages, $f$, for actual journals in each category. The $f_n$ values tend to fall within the theorem's predicted limits for all fields shown. The skewness increases from left to right and from row to row, from EDUCATION, SPECIAL (skewness = 2.7) to HEMATOLOGY (skewness = 9).
}
\label{fig:CLT_bounds_1}
\end{figure}

\begin{figure}
\centering

\includegraphics[width=0.3\linewidth]{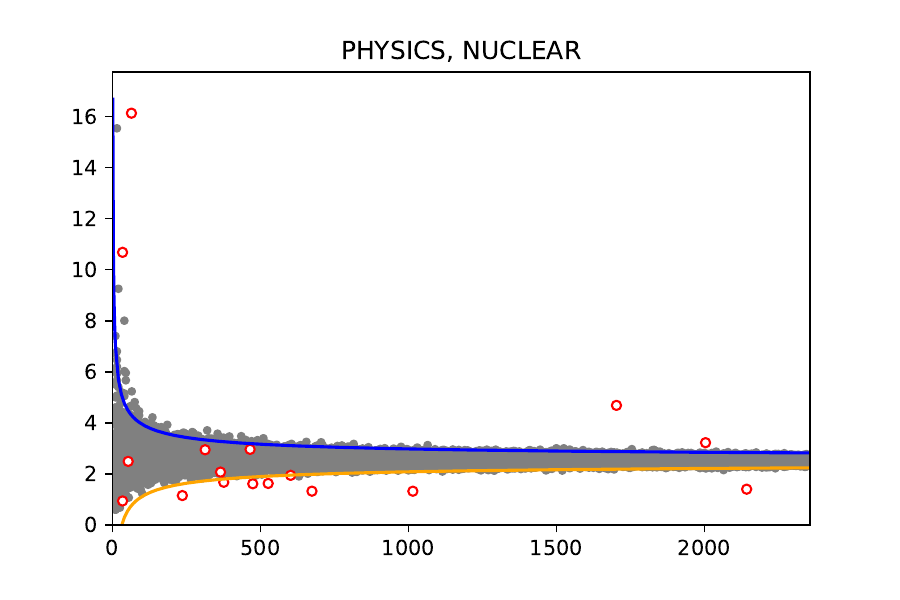}
\includegraphics[width=0.3\linewidth]{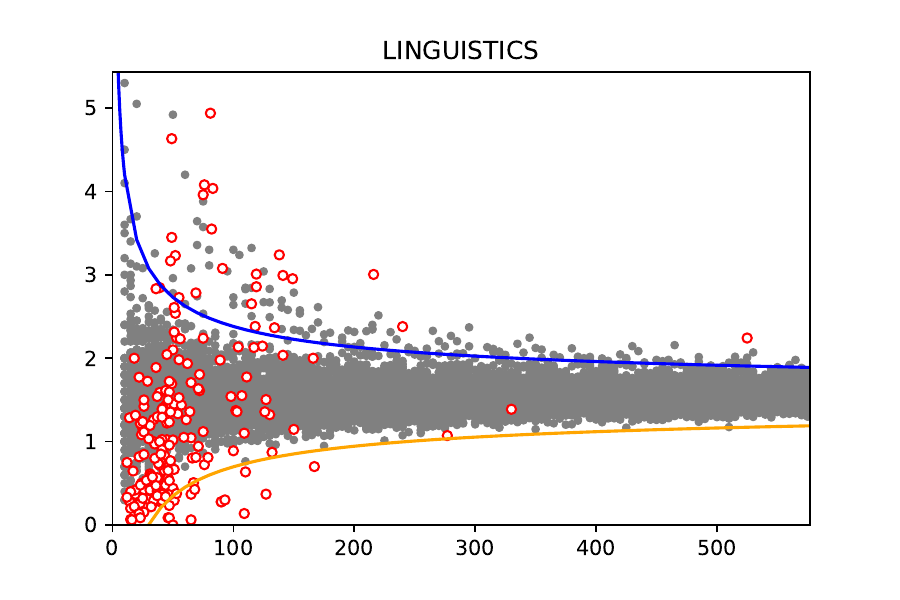}
\includegraphics[width=0.3\linewidth]{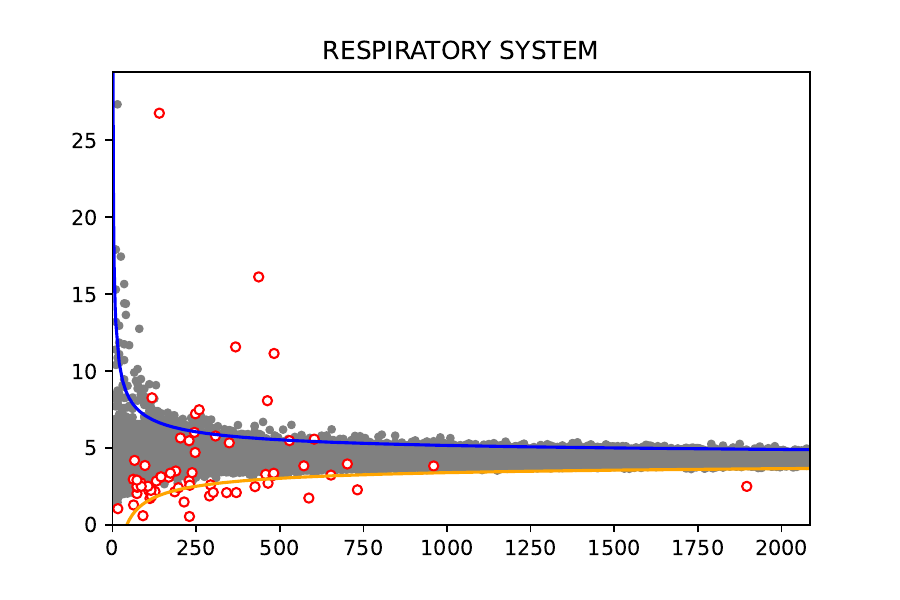}
\includegraphics[width=0.3\linewidth]{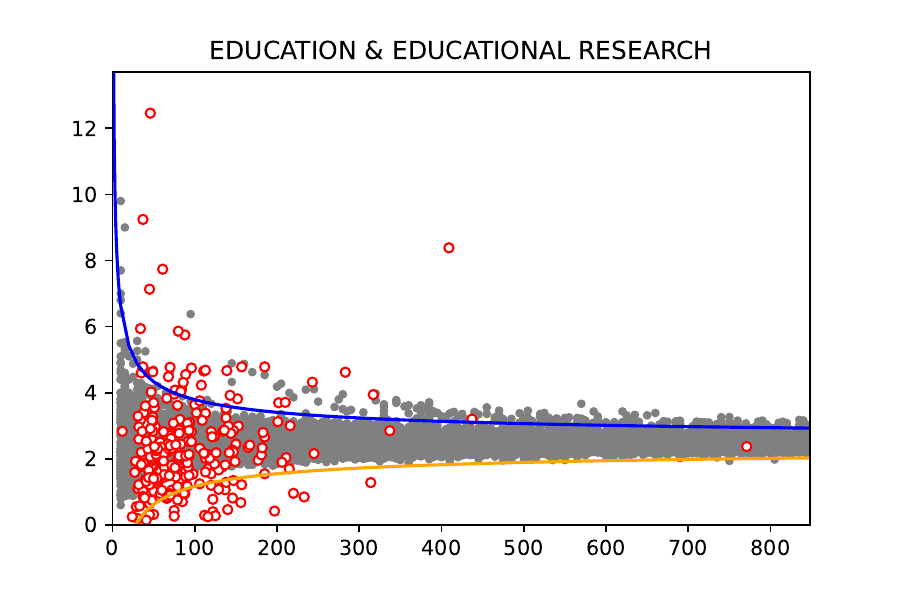}
\includegraphics[width=0.3\linewidth]{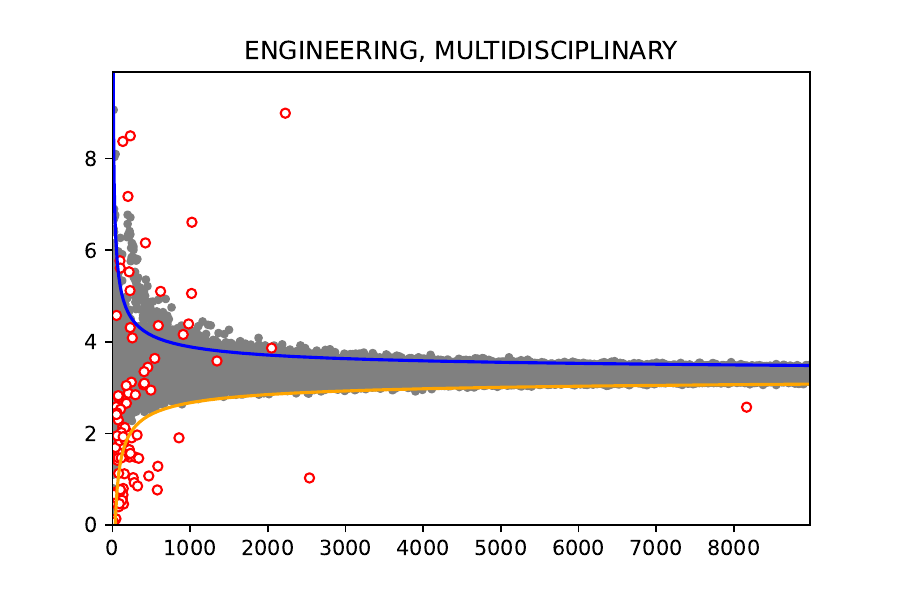}
\includegraphics[width=0.3\linewidth]{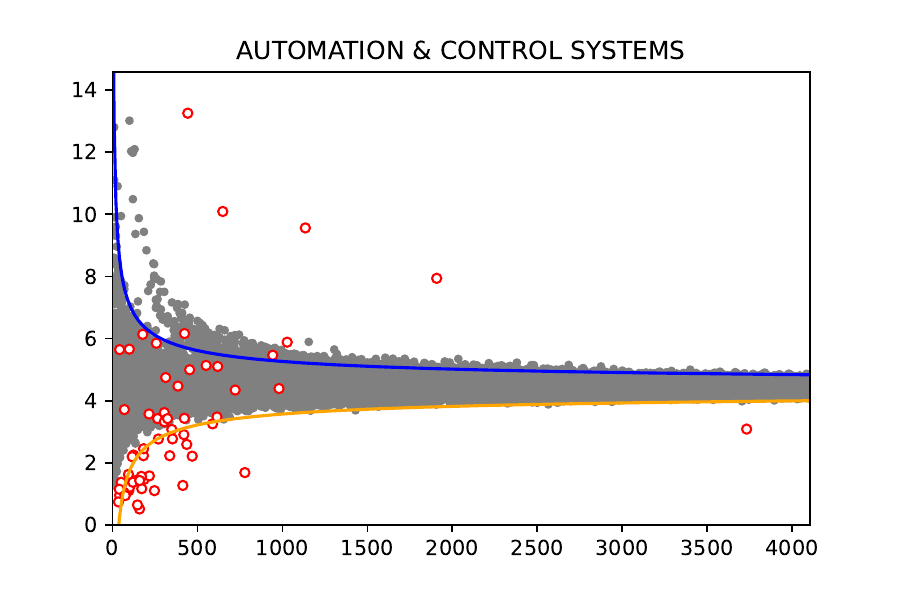}
\includegraphics[width=0.3\linewidth]{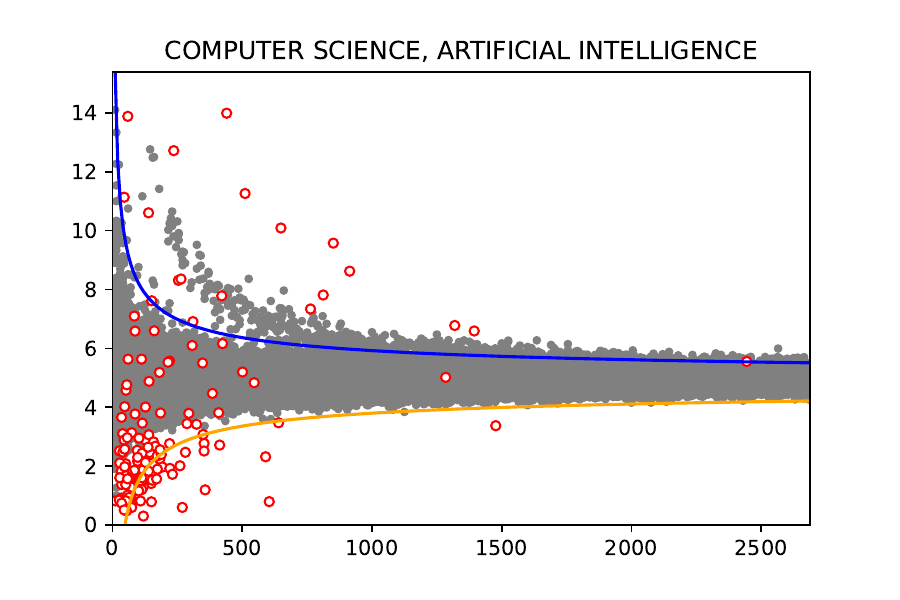}
\includegraphics[width=0.3\linewidth]{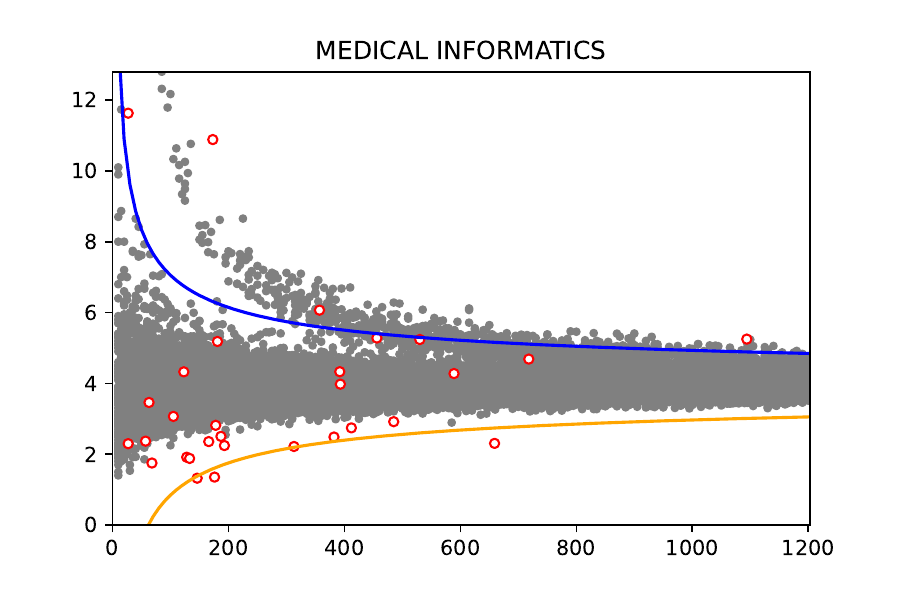}
\includegraphics[width=0.3\linewidth]{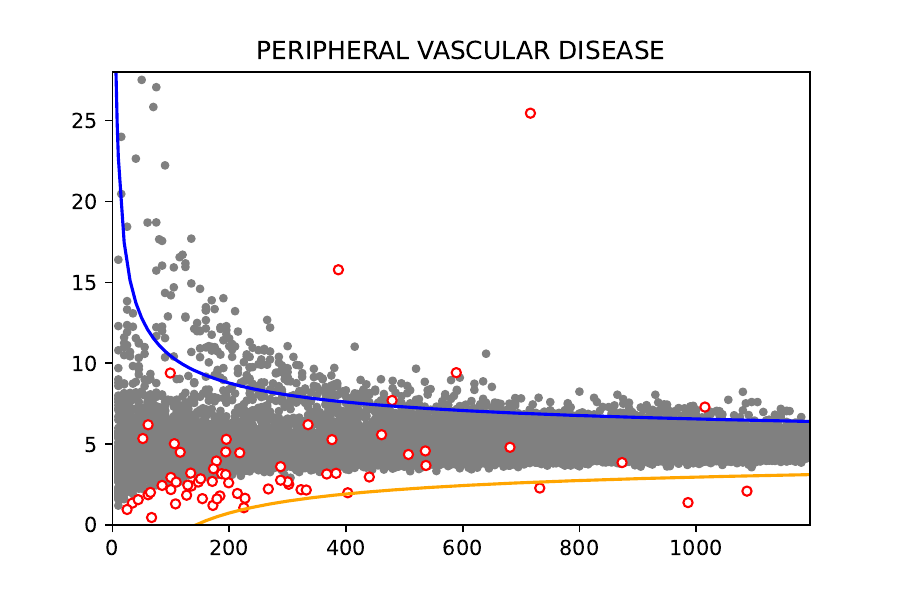}
\includegraphics[width=0.3\linewidth]{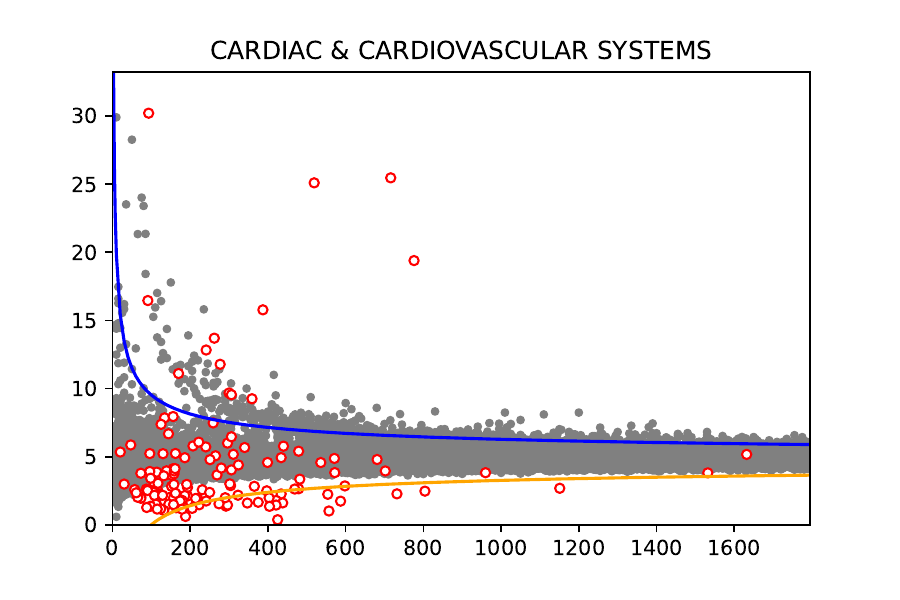}
\includegraphics[width=0.3\linewidth]{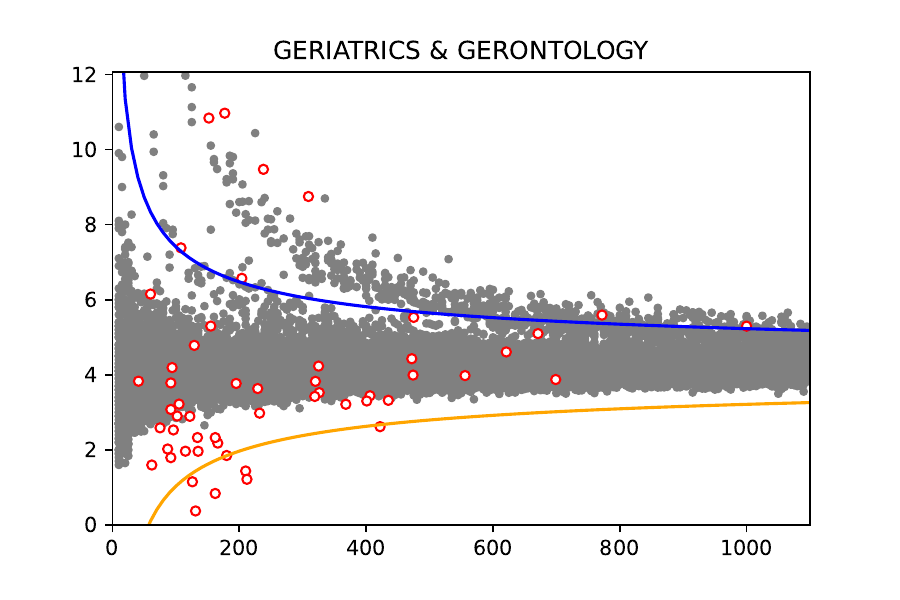}
\includegraphics[width=0.3\linewidth]{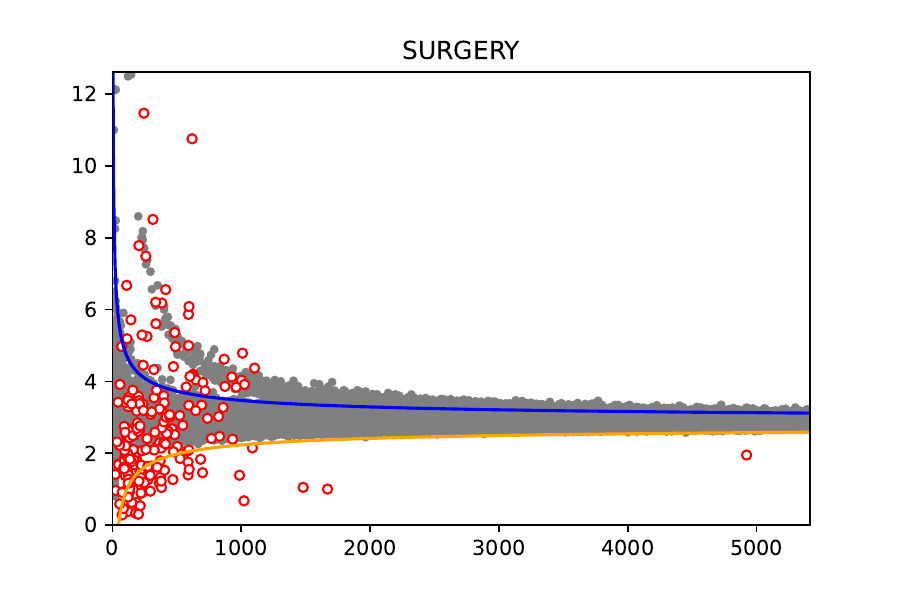}
\includegraphics[width=0.3\linewidth]{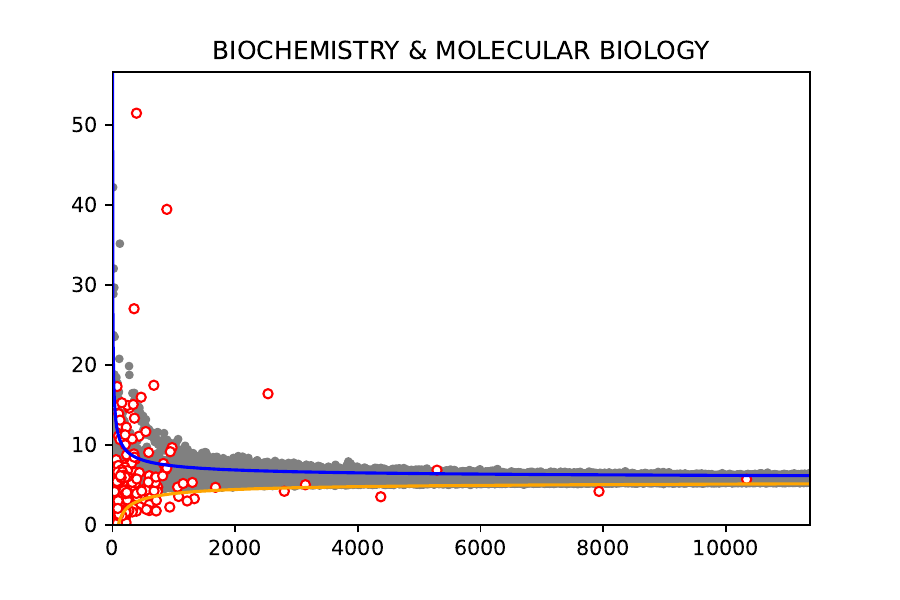}
\includegraphics[width=0.3\linewidth]{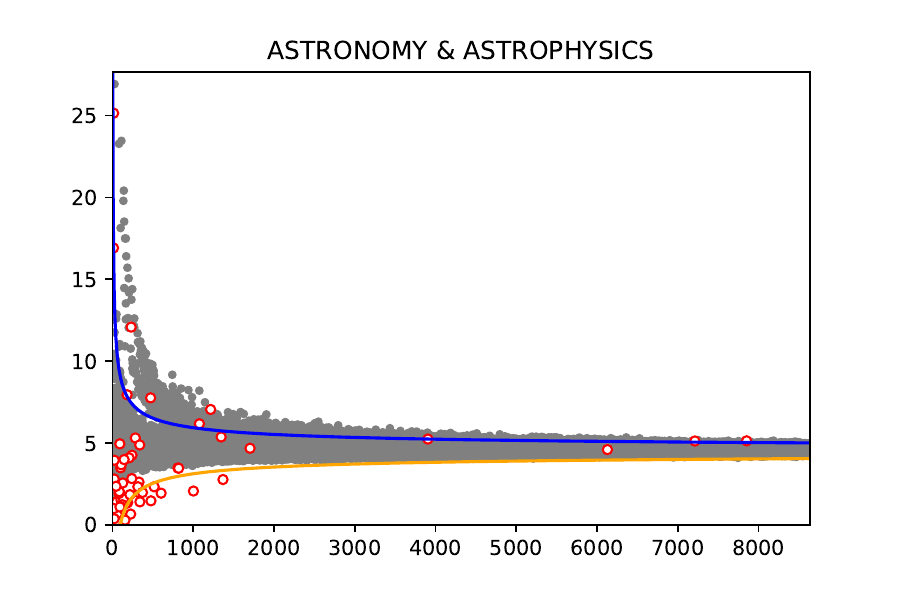}
\includegraphics[width=0.3\linewidth]{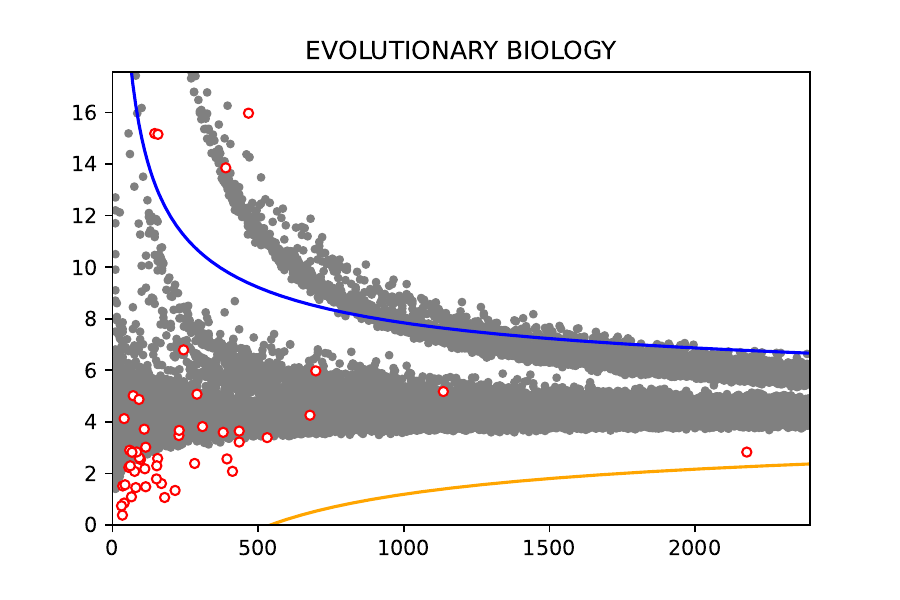}
\includegraphics[width=0.3\linewidth]{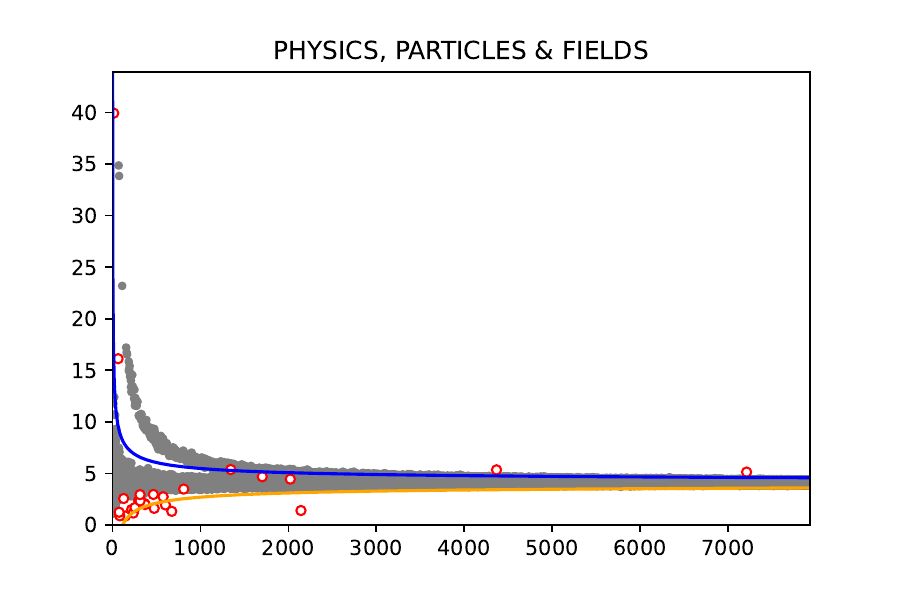}
\includegraphics[width=0.3\linewidth]{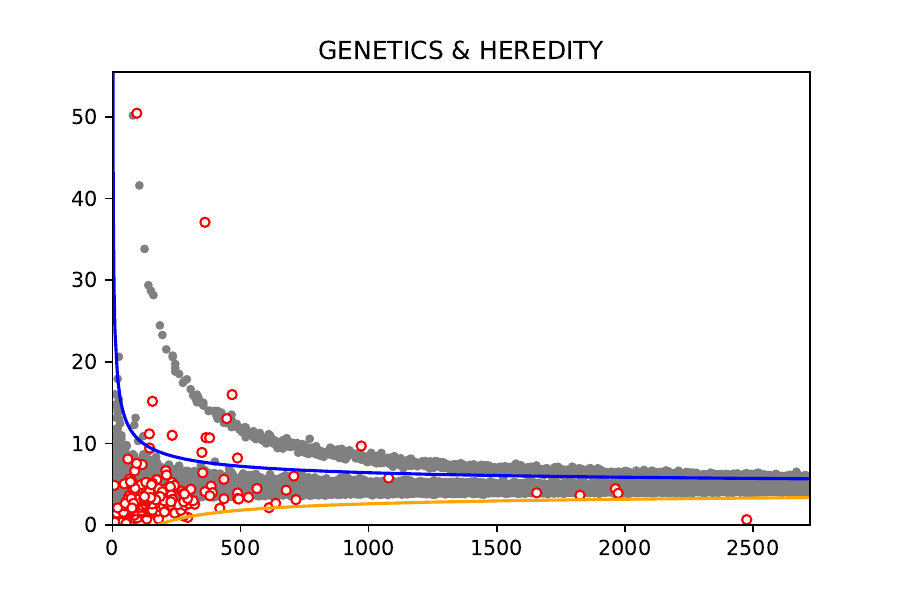}
\includegraphics[width=0.3\linewidth]{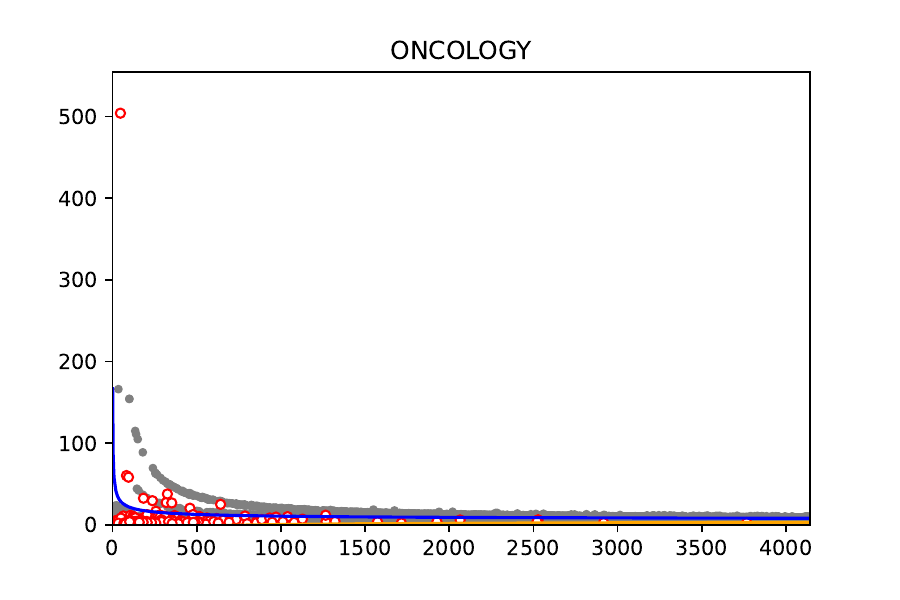}
\caption{Dependence of citation average $f_n$ (gray dots) on journal size $n$ for randomly formed journals in 18 field categories in the 2020 JCR. The fields are selected for skewness values ranging from high to the highest among the 229 fields. 
Shown in blue and orange lines are the upper and lower limits, respectively, predicted by the Central Limit Theorem for random journals. Also shown (in red circles) are the citation averages, $f$, for actual journals in each category. The $f_n$ values tend to fall within the theorem's predicted limits, at least qualitatively, for skewness values up to $\approx 25$ (EDUCATION \& EDUCATIONAL RESEARCH), with gradually increasing departures from the theoretical limits at lower journal sizes due to outlier papers that result in very heavy-tailed distributions. Beyond the skewness value of 30 (ENGINEERING, MULTIDISCIPLINARY), the departure from the theoretical bounds at small journal sizes is more evident. It becomes extreme as the skewness value exceeds 100 (PHYSICS, PARTICLES \& FIELDS, GENETICS \& HEREDITY, and ONCOLOGY). 
The skewness increases from left to right and from row to row, from PHYSICS, NUCLEAR (skewness = 10) to ONCOLOGY (skewness = 244).
}
\label{fig:CLT_bounds_2}
\end{figure}

Figures \ref{fig:CLT_bounds_1} and \ref{fig:CLT_bounds_2} describe a numerical experiment. For the 18 fields shown in each figure, we form journals of various sizes $n$ by randomly drawing from the actual citations of papers in the field, and depict their citation average $f_n$ as gray dots. The blue and orange lines denote the upper and lower limits, respectively, predicted by the Central Limit Theorem for random journals (Eq. (\ref{eq:3}) with $k=3$). Also shown (in red circles) are the citation averages, $f$, for actual journals in each category (2020 JCR). 
For each figure, the skewness of the citations of papers in the field increases from left to right and from row to row. Figure \ref{fig:CLT_bounds_1} contains 12 fields with the lowest skewness values, from EDUCATION, SPECIAL (skewness = 2.7) to FORESTRY (skewness = 3.5). For the next 6 fields, the skewness increases from  PRIMARY HEALTH CARE (skewness = 4) to HEMATOLOGY (skewness = 9) in roughly integer units. Evidently, the $f_n$ values tend to fall within the theorem's predicted limits for all fields shown in Figure \ref{fig:CLT_bounds_1}. 

Figure \ref{fig:CLT_bounds_2} depicts 18 fields on the upper end of the skewness spectrum. From PHYSICS, NUCLEAR (skewness = 10) to MEDICAL INFORMATICS (skewness = 45), the skewness increases by roughly 5 integer units. The following 10 fields shown have the highest skewness values observed among all the 229 fields in our dataset, ranging from PERIPHERAL VASCULAR DISEASE (skewness = 47) to EVOLUTIONARY BIOLOGY (skewness = 94), while the last three fields are PHYSICS, PARTICLES \& FIELDS  (skewness = 119), GENETICS \& HEREDITY (skewness = 146), and ONCOLOGY (skewness = 244).    




Figures \ref{fig:CLT_bounds_1} and \ref{fig:CLT_bounds_2} show that the Central Limit Theorem provides a good description of the behavior of randomly constructed journals in the vast majority of fields. Even for fields with extremely heavy-tailed citation distributions, such as EDUCATION \& EDUCATIONAL RESEARCH (skewness = 25), the agreement remains broadly satisfactory, although the precision decreases as more points fall outside the theoretical bounds. Although the Central Limit Theorem is often assumed to apply for sample sizes of about 30 or more, highly skewed distributions require larger samples for the approximation to hold. The figures, therefore, provide a graphical indication of the journal size above which the citation averages of randomly formed journals are well approximated by the Central Limit Theorem. In our experience, the approximation holds for very small journal sizes for about 85\% of the fields (and at sufficiently large sizes for all fields). 

\begin{figure}
\centering
\includegraphics[width=0.5\linewidth]{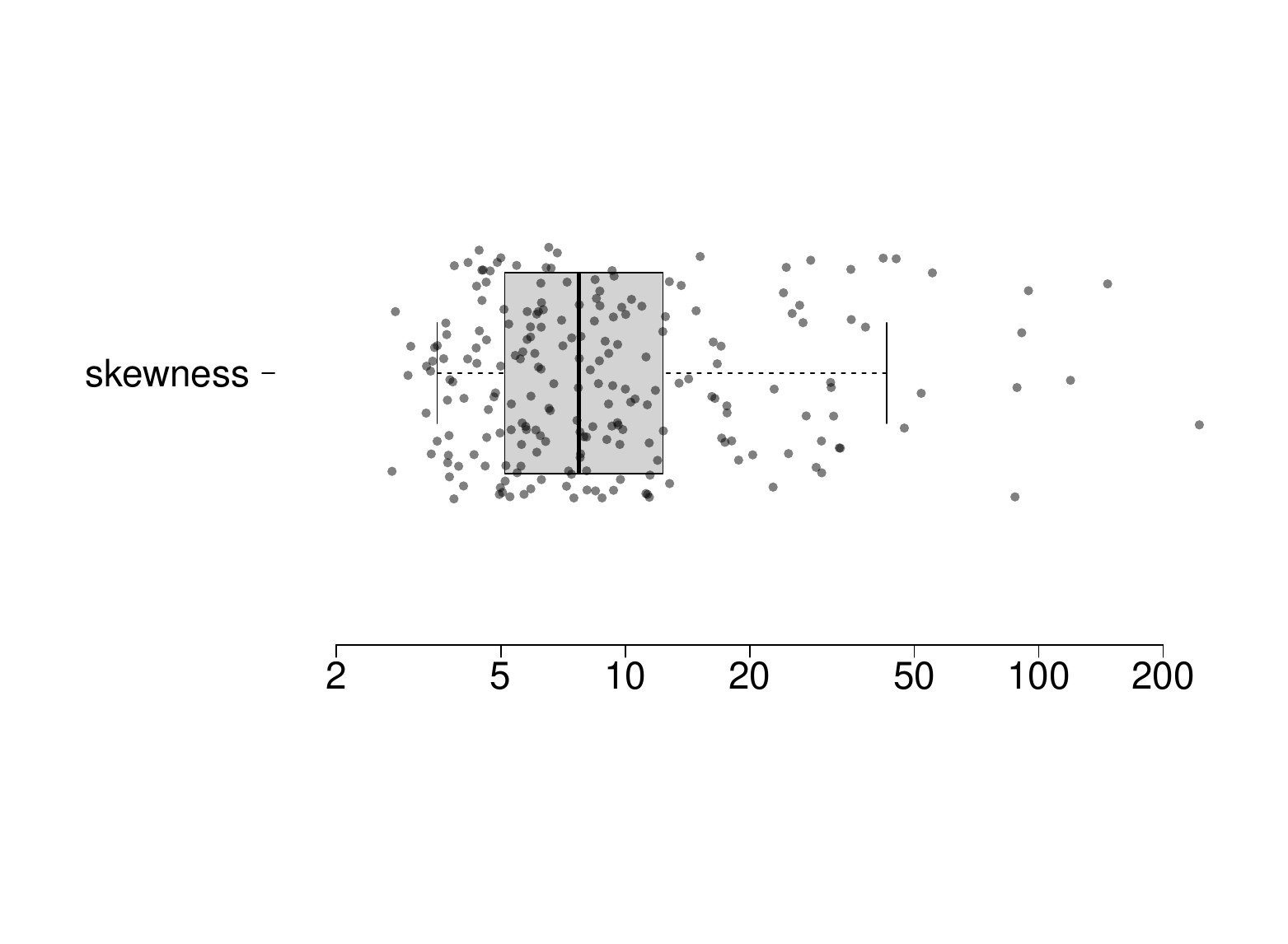}
\caption{Boxplot (in logarithmic scale) describing the distribution of skewness values across the 229 subject (field) categories in the Journal Citation Reports of Clarivate Analytics. The center line shows the median; box limits indicate the 25th and 75th percentiles as determined by R software; whiskers extend to 5th and 95th percentiles; data points are plotted as gray circles (Spitzer {\it et al.}, 2014).}
\label{fig:skewness_boxplot}
\end{figure}

In Figure \ref{fig:skewness_boxplot} we present a boxplot of the distribution of skewness values for the citations of all papers in each of the 229 fields in the 2020 JCR. 75\% of fields have skewness less than 12; while 5\% of fields have skewness greater than 40.

\subsection{Standardization of citation averages corrects their scale dependence }

The Central Limit Theorem allows us to standardize citation averages and remove their scale dependence. Here is how. Upon rewriting Eq. (\ref{eq:3}) as 
\begin{equation}
 -k \le    \frac{(f_n-\mu) \sqrt{n}}{\sigma} \le k \;, \qquad
    \begin{cases}
      k=1 & \text{for 68\% of random journals}\\
      k=2 & \text{for 95\% of random journals}\\
      k=3 & \text{for 99.7\% of random journals}
    \end{cases}       
\label{eq:6}
\end{equation}
we realize that the quantity
\begin{equation}
\Phi \equiv \frac{(f_n-\mu) \sqrt{n}}{\sigma},
\label{eq:Phi_def}
\end{equation}
which we hereby call the $\Phi$ {\it index}, is equal to the distance of the IF from the mean $(f_n-\mu)$,
{\it normalized} to the sample (i.e., journal) standard deviation $(\sigma/\sqrt{n})$ from Eq. (\ref{eq:2}). The expression in Eq. (\ref{eq:Phi_def}) may be familiar to some readers from the statistics of standardizing averages (De Veaux, Velleman, \& Bock, 2014, pp. 110--118). Indeed, the $\Phi$ index is a standardized average analogous to the $z$-score in statistics, which measures how many standard deviations a data point is away from the mean. Equation (\ref{eq:6}) shows that the $\Phi$ index is bounded by a constant (integer $k$), unlike the citation average $f_n$ that has size-dependent bounds, as shown in Eq. (\ref{eq:3}). This is a first indication that the $\Phi$ index is size-independent; we will shortly provide a numerical demonstration as well. 

We can calculate a $\Phi$ index for any sample drawn from a distribution of research papers with a well-defined ($\mu, \sigma$). Therefore, we can also use it for standardizing citation averages across different research fields. To compare $\Phi$ indexes of journals in different fields, all we need are the citation mean ($\mu$) and standard deviation ($\sigma$) of the papers in each field. 

Comparison of $\Phi$ indexes {\it within} a given field is considerably easier: All we need is the field's citation mean ($\mu$). The field's standard deviation ($\sigma$) becomes just a scaling factor in this case, and can be set to unity. 

Beyond the analogy with the $z$-score in statistics, the $\Phi$ index also has an elegant geometric interpretation, as we explain below. 

\subsection{A geometric interpretation of the $\Phi$ index}

Let us plot of the citation average $f_n$ as a function of journal size $n$ for various randomly formed journals that are drawn from a distribution of research papers described by ($\mu, \sigma$). We wish to compare two different-sized journals represented by the points $A$ and $B$ in the plot, with respective coordinates ($f_1, n_1$) and ($f_2, n_2$). 

Consider the journal with the maximum citation average at size $n_1$, namely, $f_{1,max} = \mu + k \sigma/\sqrt{n_1}$ from Eq. (\ref{eq:3}), which we denote with the point $A'$.
The journals $A$ and $A'$ differ in selectivity: Journal $A'$ is the most selective journal possible at size $n_1$, because it contains all the highest-cited papers one can find for a journal of this size. A journal with a different size $n_2$ will have a different maximum citation average as dictated by Eq. (\ref{eq:3}), namely, $f_{2,max} = \mu + k \sigma/\sqrt{n_2}$; we denote this journal with point $B'$ in the plot. The points $A'$ and $B'$ are equivalent: They represent the highest possible citation average for journals of different sizes, and correspond to a paper selection process that is maximally selective at each size. 

\begin{figure}
\centering
\includegraphics[width=0.45\linewidth]{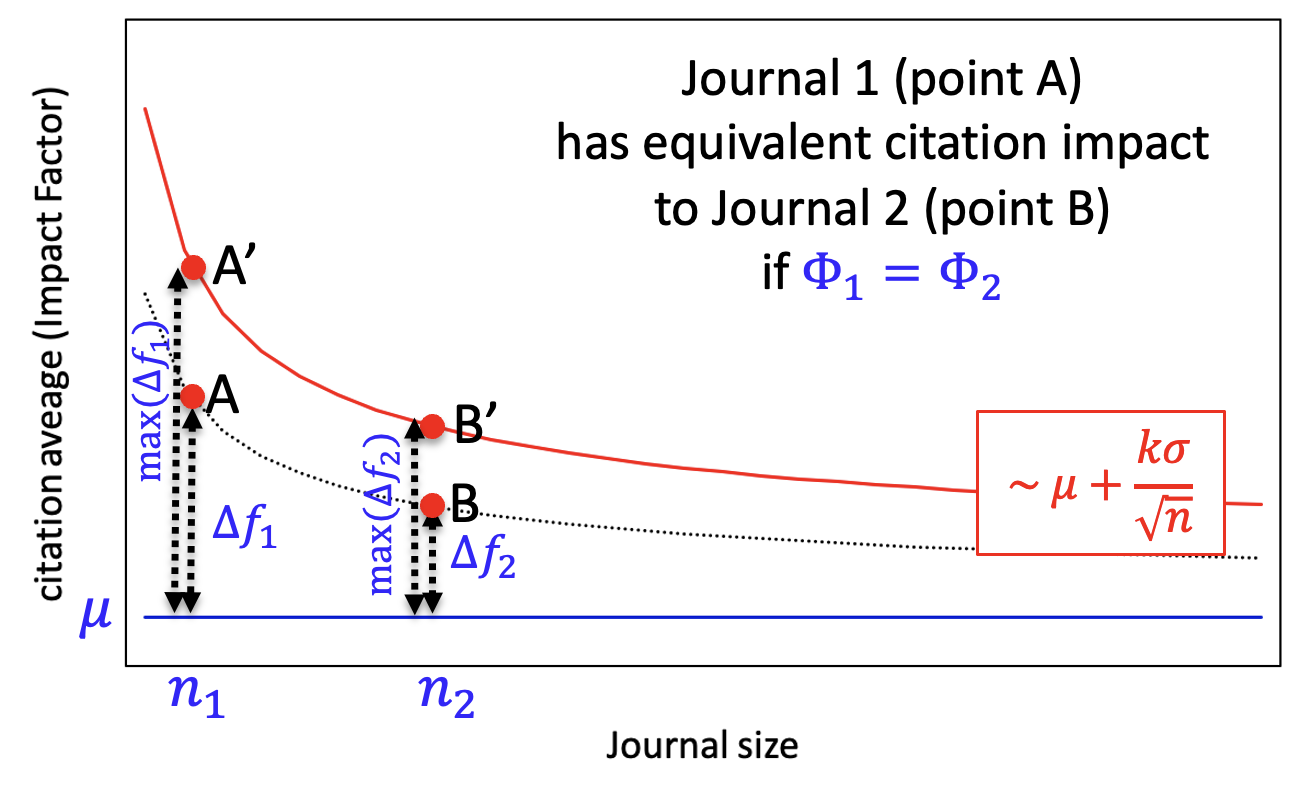}
\caption{Geometric interpretation of the $\Phi$ index.}
\label{fig:geom_interpretation}
\end{figure}

We posit that journals $A$ and $B$ are equivalent if they have the same score {\it in their respective scales}, which are size-dependent because their respective maximum scores attainable are different. As we can see from the geometrical construction in Figure \ref
{fig:geom_interpretation}, points $A$ and $B$ are equivalent citation-wise if they are {\it equidistant} from the global citation mean  $\mu$, where {\it distances} are normalized to the IF range at any given size. In geometrical terms,  $A$ and $B$ are equivalent if 
\begin{equation}
\frac{\Delta f_1}{\text{max}(\Delta f_1)} = 
\frac{\Delta f_2}{\text{max}(\Delta f_2)}
\label{eq:A_equiv_B}
\end{equation}
and therefore 
\begin{equation}
\frac{(f_1-\mu) \sqrt{n_1}}{\sigma} =
\frac{(f_2-\mu) \sqrt{n_2}}{\sigma}
\end{equation}
or
\begin{equation}
\Phi_1 = \Phi_2.
\end{equation}
In other words, journals $A$ and $B$ are equivalent if they are equally selective {\it at their size}. This geometric interpretation provides an elegant and intuitive understanding of the $\Phi$ index.

\subsection{The $\Phi$ index as a $z$ score.}

The geometric interpretation of the $\Phi$ index agrees with our intuitive expectation that journals A and B in Figure \ref{fig:geom_interpretation} ought to be equivalent in terms of their citation accrual. But it also has a more rigorous interpretation, as the $z$ score from statistics. It tells us how far from the average a value is, where the distance from the average is normalized to the standard deviation of the population of $n$-sized journals. Thus, a $z$ score of 1 tells us that our journal is 1 standard deviation away from the mean. Interpreted as a $z$ score, the $\Phi$ index is a standardized metric and thus allows also comparison of scores (citation averages) that are measured on different scales (different $\mu, \sigma$) in analogy with the use of standardized averages to compare athletes' performances across different sports or students' scores across different academic disciplines (De
Veaux, Velleman, \& Bock, 2014, pp. 111--112).

In the most general form, we have proposed (Antonoyiannakis, 2018) the $\Phi$ index as 
\begin{equation}
\Phi=\frac{f-\mu_s}{\sigma_s/\sqrt{N_{2Y}}}, \label{eq:Phi}
\end{equation}
where $\mu_s, \sigma_s$ are the global average and standard deviation of the citation distribution of all papers in the {\it field} of the journal in question. Here, we denoted with $f$ and $N_{2Y}$ the journal's citation average (IF) and biennial size, respectively. Also, we have used $k=1$ for simplicity and without loss of generality. 

Throughout this paper, we use the subscript ``s'' to denote field-standardized (or field-normalized) quantities, i.e., quantities defined relative to the distribution of citation indicators within a scientific field (subject). 

\section{A standardized citation indicator accounts for scale {\it and} field}
\label{sec:2}

To calculate a journal's $\Phi$ index from Eq. (\ref{eq:Phi}), we need to know the quantities $\mu_s$ and $\sigma_s$ for the journal's field. For example, if we treated all 3,537,118 papers published in all journals in 2018--2019 as if they form a single field, we would have $\mu=4.11, \sigma=12.5$. A comparison of the $\Phi$ index of all journals in this case would be free from the size-dependence bias of citation averages, and would thus be fairer than a comparison of Impact Factors. 

\begin{figure}
\centering
\includegraphics[width=0.45\linewidth]{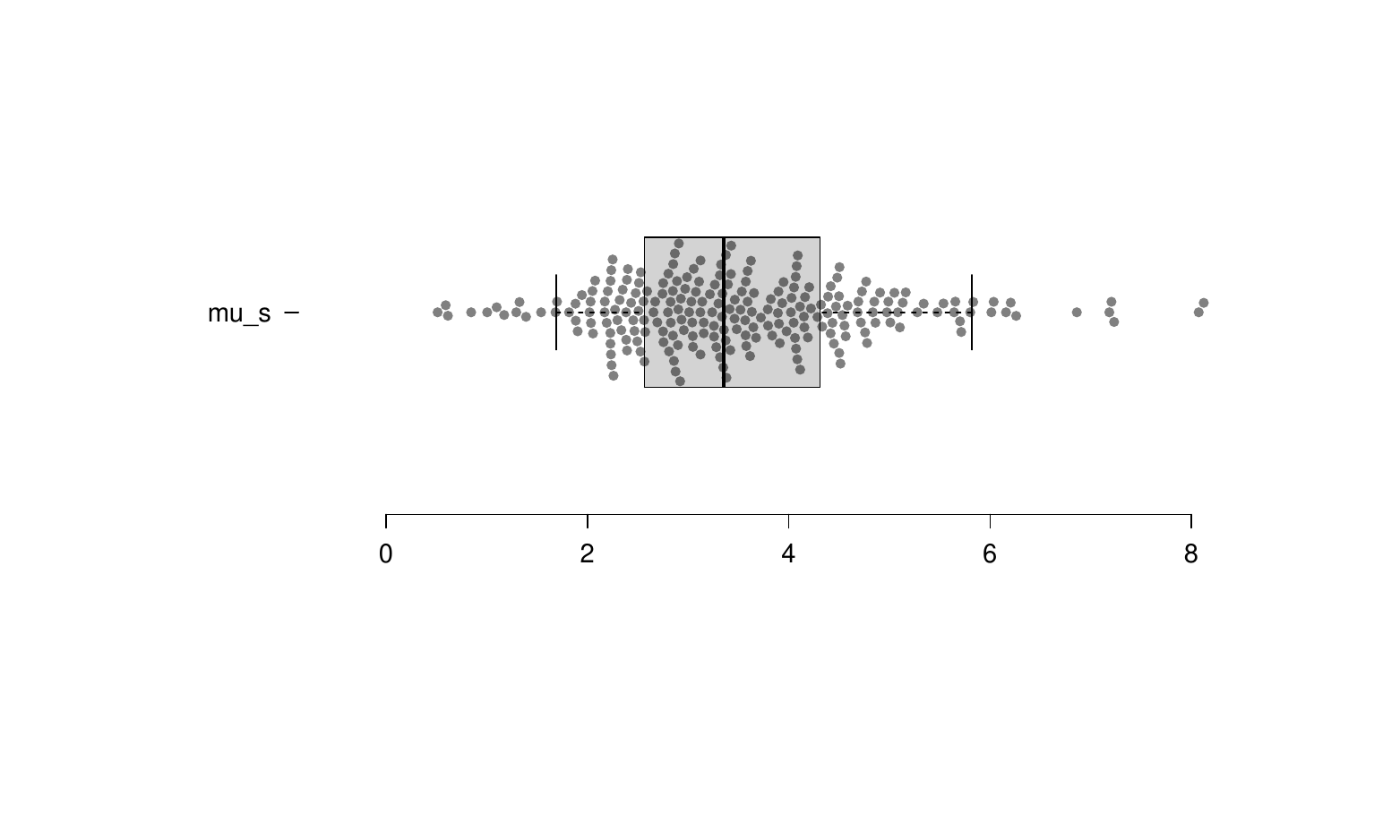}
\includegraphics[width=0.45\linewidth]{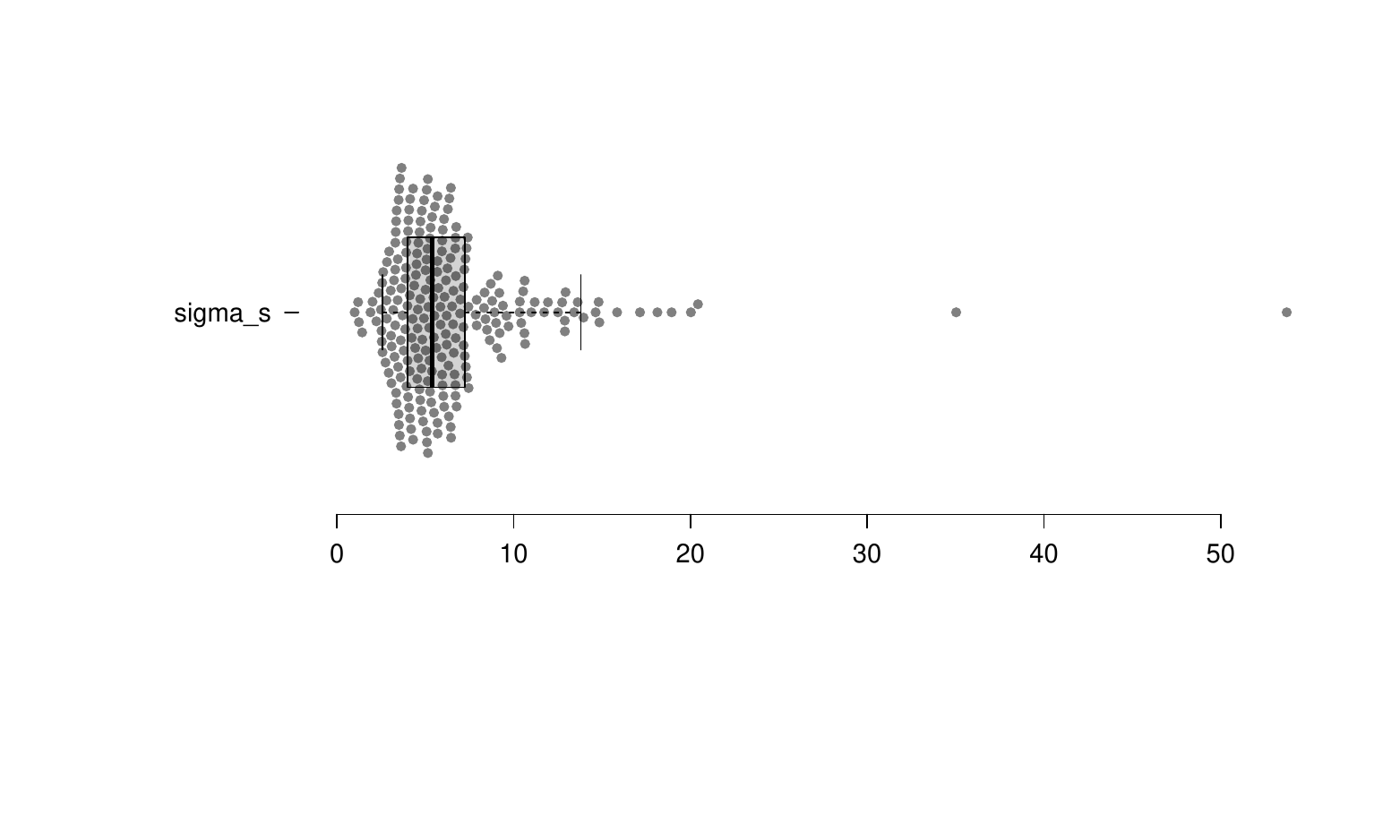}
\caption{Boxplots describing the distribution of $\mu_s$ (left) and $\sigma_s$ (right) values across the 229 subject (field) categories in the Journal Citation Reports of Clarivate Analytics, plus an ``all field" category that contains all journals. The center lines show the medians; box limits indicate the 25th and 75th percentiles as determined by R software; whiskers extend to 5th and 95th percentiles; data points are plotted as gray circles (Spitzer {\it et al.}, 2014).}
\label{fig:mu_sigma_boxplots}
\end{figure}

But we can do better. Citation indicator comparisons need to account for differences in citation practices across research fields. A glance at Eq. (\ref{eq:Phi}) shows how we can readily use the $\Phi$ index to standardize both for size {\it and} field. To compare journals in different research fields, all we need is to find the $\mu_s$ and $\sigma_s$ of each field, and use them to calculate the field-specific $\Phi$ index of each journal. In Table \ref{table:mu_sigma_various_subjects}  we show different ($\mu_s, \sigma_s$) for several subject (field) categories in the 2020 Journal Citation Reports (JCR) of Clarivate Analytics. As we can see, both quantities span more than one order of magnitude, with $\mu_s$ ranging from 0.51 to 8.1 and $\sigma_s$ ranging from 1 to 53.7. See Figure  \ref{fig:mu_sigma_boxplots}.

Equation (\ref{eq:Phi}) says that a journal's standardized citation average depends on its {\it community} of journals, all of which define the field and determine the {\it global} quantities $\mu_s$ and $\sigma_s$, which are computed for all papers in the field. The notion that a journal's indicator is not defined in absolute terms but in relation to its community is key to Eq. (\ref{eq:Phi}). Indeed, what is the significance of the statement that a journal’s citation indicator (here, Impact Factor) is equal to 1, 10, or 100? In itself and without context, the number does not mean much; it gains meaning in connection with its community of peer journals as a benchmark. This is the rationale behind standardized averages, and we now see why the $\Phi$ index allows naturally for field standardization, in addition to size standardization.  

But if a journal is measured in relation to its peers, then the choice of peers affects the journal's score by definition. So the selection of the journals that form a field is critical. Clearly, if we include in a field those journals with higher or lower citation averages, we push the field's global average $\mu_s$ to correspondingly higher or lower values. The value of $\mu_s$ affects in turn the sign of $(f-\mu_s)$, which is multiplied by a size-sensitive factor $\sqrt{N_{2Y}}$ in the numerator of the $\Phi$ index. Thus, whenever the addition or removal of journals in a field causes the term $(f-\mu_s)$ to flip sign, the corresponding journal's $\Phi$ index can jump abruptly in the rankings, particularly for large journals. This caveat is important to keep in mind, especially if we try to over-analyze fields by assigning them to ever-decreasing journal lists. At any rate, the decision of which journals to include in each field category has important ramifications in field-based citation rankings. But this issue is hardly unique to the $\Phi$ index rankings. 

In this paper, we rely on the 229 subject (field) categories in the Clarivate Analytics 2020 JCR.

\begin{table}[h]
\centering
\begin{tabular}{l l l l l}
\hline
Subject (field) & $\mu_s$ & $\sigma_s$ & no. papers & no. journals \\
\hline
ALL & 4.11 & 12.50 & 3,537,118 & 12173 \\
AREA STUDIES & 1.09 & 2.01 & 5301 & 80 \\
ASTRONOMY \& ASTROPHYSICS & 4.53 & 14.80 & 42725 & 60 \\
BIOCHEMISTRY \& MOLECULAR BIOLOGY & 5.65 & 18.12 & 118555 & 279 \\
CARDIAC \& CARDIOVASCULAR SYSTEMS & 4.77 & 15.85 & 40771 & 142 \\
\hline
CELL BIOLOGY & 7.20 & 14.84 & 61131 & 191 \\
CHEMISTRY, MULTIDISCIPLINARY & 6.86 & 12.93 & 156779 & 171 \\
COMPUTER SCIENCE, CYBERNETICS & 5.53 & 17.14 & 3210 & 22 \\
CULTURAL STUDIES & 1.00 & 2.231 & 3167 & 45 \\
ENERGY \& FUELS & 7.23 & 11.72 & 85899 & 108 \\
\hline
ENGINEERING, ENVIRONMENTAL & 8.07 & 10.64 & 40179 & 54 \\
EVOLUTIONARY BIOLOGY & 4.51 & 35.03 & 12603 & 48 \\
GENETICS \& HEREDITY & 4.50 & 20.02 & 42825 & 170 \\
HISTORY & 0.61 & 1.24 & 5517 & 100 \\
INFORMATION SCIENCE \& LIBRARY SCIENCE & 3.38 & 6.16 & 8823 & 85 \\
\hline
LAW & 1.29 & 2.36 & 9059 & 148 \\
LOGIC & 0.51 & 1.00 & 1599 & 19 \\
MATHEMATICS & 1.17 & 2.56 & 55051 & 316 \\
MEDICINE, GENERAL \& INTERNAL & 4.38 & 20.42 & 62300 & 167 \\
MULTIDISCIPLINARY SCIENCES & 6.03 & 13.95 & 120752 & 71 \\
\hline
NANOSCIENCE \& NANOTECHNOLOGY & 8.12 & 12.88 & 85204 & 106 \\
ONCOLOGY & 5.70 & 53.73 & 92083 & 240 \\
PHYSICS, MULTIDISCIPLINARY & 3.83 & 9.20 & 42287 & 78 \\
PSYCHOLOGY, PSYCHOANALYSIS & 0.59 & 1.19 & 833 & 12 \\
SOCIAL SCIENCES, INTERDISCIPLINARY & 2.25 & 3.55 & 12451 & 108 \\
\hline
\end{tabular}
\caption{Citation averages $\mu_s$ and standard deviations $\sigma_s$ for papers in various research fields, which are defined here according to the subject (field) categories in the 2020 Journal Citation Reports (JCR) of Clarivate Analytics. The third column shows the number of papers, and the fourth column shows the number of journals in each field. The first entry corresponds to treating all 12173 journals in the 2020 JCR as a single field.}
\label{table:mu_sigma_various_subjects}
\end{table}

\section{Using a `random sample test' to check for scale independence}

As we showed earlier, the citation average $f_n$ of a randomly formed journal takes a range of values that varies as $\sigma/\sqrt{n}$ measured from the population mean $\mu$. We demonstrated this effect with the numerical experiment of Figure \ref{fig:f-mu_1/sqrt(n)}, where we plotted $f_n$ vs. $n$ for a total of more than 118,000 randomly drawn `journals' in the field of 
biochemistry \& molecular biology.

\begin{figure}
\centering
\includegraphics[width=0.45\linewidth]{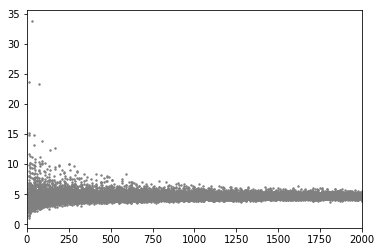}
\includegraphics[width=0.45\linewidth]{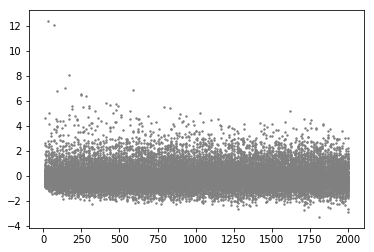}
\caption{Citation average (left) and $\Phi$ index (right) vs. size for a randomly sampled journal in multidisciplinary sciences. The citation average is scale dependent, the $\Phi$ index not.}
\label{fig:MULTIDISCIPLINARY}
\end{figure}

\begin{figure}
\centering
\includegraphics[width=0.45\linewidth]{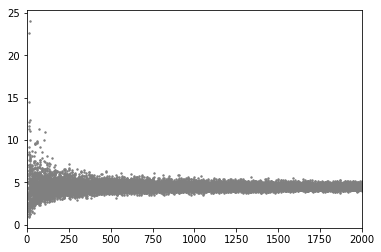}
\includegraphics[width=0.45\linewidth]{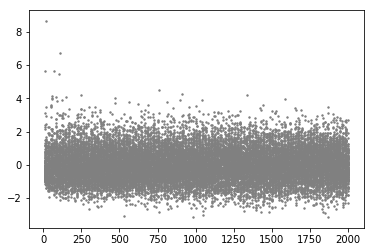}
\caption{Citation average (left) and $\Phi$ index (right) vs. size for a randomly sampled journal in condensed matter physics. The citation average is scale dependent, the $\Phi$ index not.}
\label{fig:PHYSICS_CONDENSED}
\end{figure}

\begin{figure}
\centering
\includegraphics[width=0.45\linewidth]{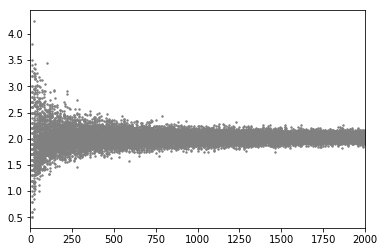}
\includegraphics[width=0.45\linewidth]{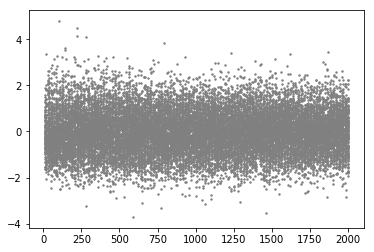}
\caption{Citation average (left) and $\Phi$ index (right) vs. size for a randomly sampled journal in psychology. The citation average is scale dependent, the $\Phi$ index not.}
\label{fig:PSYCHOLOGY}
\end{figure}

\begin{figure}
\centering
\includegraphics[width=0.45\linewidth]{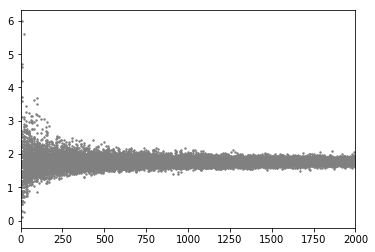}
\includegraphics[width=0.45\linewidth]{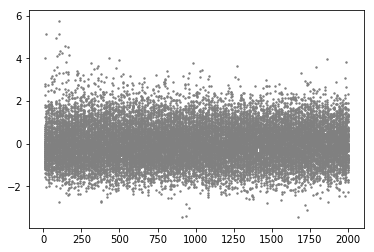}
\caption{Citation average (left) and $\Phi$ index (right) vs. size for a randomly sampled journal in information science and library science. The citation average is scale dependent, the $\Phi$ index not.}
\label{fig:INFORMATION}
\end{figure}

\begin{figure}
\centering
\includegraphics[width=0.45\linewidth]{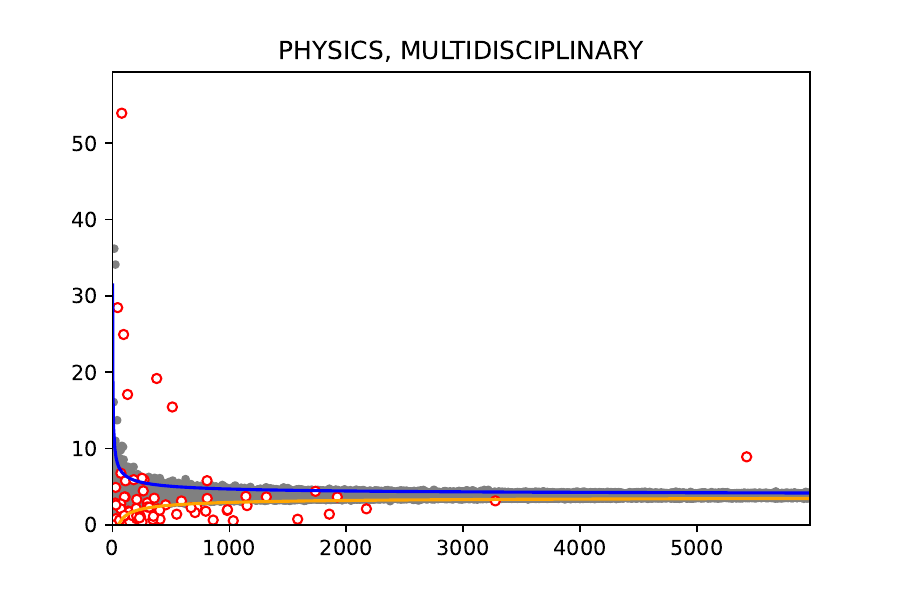}
\includegraphics[width=0.45\linewidth]{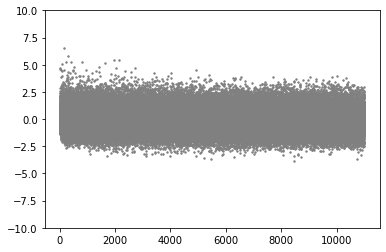}
\caption{Citation average (left) and $\Phi$ index (right) vs. size for randomly sampled journals in multidisciplinary physics (2020 JCR). The citation average is scale dependent, the $\Phi$ index not. Furthermore, the randomly formed journals have citation averages that range between the theoretically expected upper and lower values from Eq. (\ref{eq:3}), shown here (for $k=3$) as continuous blue and orange lines, respectively. The $\Phi$ index values are confined within the range $[-3,3]$, as expected from Eq. (\ref{eq:6}) for 99.7\% of journals. Shown in red are the citation averages of actual journals in physics.
}
\label{fig:PHYSICS}
\end{figure}

Does the $\Phi$ index pass this `random sample test'? If so, then plotting the $\Phi$ index vs. size would yield no size dependence. In Figures \ref{fig:MULTIDISCIPLINARY}, \ref{fig:PHYSICS_CONDENSED}, \ref{fig:PSYCHOLOGY}, \ref{fig:INFORMATION}, and \ref{fig:PHYSICS}, we plot the citation average $f_n$ (left panels) and $\Phi$ index (right panels) vs. journal size for 40,000 `journals' drawn randomly from papers in the fields of multidisciplinary sciences, condensed matter physics, psychology, information science \&
library science, and multidisciplinary physics, respectively. Evidently, the range of values of the $\Phi$ index in all these cases shows no dependence with journal size, unlike the citation average. Furthermore, the vast majority of $\Phi$ index values are bounded within the range $[-3,3]$. These results make sense and confirm our expectation from Eq. (\ref{eq:6}), where the inequality covers 99.7\% of random journals for $k=3$.

\section{$\Phi$ index for journals}

Having established the mathematical foundation and the statistical relevance of the $\Phi$ index, we now proceed to calculate it for journals in several fields.

\subsection{Method \& Data}

First, for each journal in the 2020 JCR, we obtained its individual Journal Citation Report, which contained the 2020 citations to each of its citable papers (i.e., articles and reviews) published in 2018--2019. Of the 12,316 journals in the JCR, 143 whose individual citation reports would not download due to technical glitches were excluded. These missing journals were mostly sparse across the 229 JCR subject categories---95\% of categories had 7 or fewer missing journals---but concentrated in four fields: BIOCHEMISTRY \& MOLECULAR BIOLOGY; MATHEMATICS; CHEMISTRY, PHYSICAL; and CHEMISTRY, ORGANIC, each with 10 or more missing titles (maximum 18). More than half of the missing titles in these four categories are Russian or Eastern European translated journals, and the pattern is especially pronounced in MATHEMATICS (12 out of 14 missing titles). This suggests a common technical characteristic of translated journals---such as a non-standard ISSN format or DOI structure---that caused Clarivate's individual report download to fail systematically for this class of titles.

We thus ended up with a master list of 12,173 journals with citation distributions. For each journal in the master list we calculated its citation average, $f$, which approximates the IF and becomes identical to it when there are no ``free'' or ``stray'' citations in the numerator (these are citations to front-matter items such as editorials, letters to the editor, commentaries, etc., or citations to the journal without specific reference of volume and page or article number). We will thus use the terms ``IF'' and ``citation average'' interchangeably, for simplicity. Collectively, the 12,173 journals in our master list published 3,537,118 papers in 2018--2019, which received 14,550,739 citations in 2020. This is our dataset.

A note on data vintage is warranted. Our results are based on the 2020 JCR, but the $\Phi$ index methodology is general: it applies to any publication and citation time windows, and to any dataset of citation averages, not just Impact-Factor-style metrics. Moreover, even at a quantitative level, the key population parameters are stable across time. For example, according to Dimensions, the global citation mean across all fields increased by approximately 14\% from 2020 to 2025 (articles and reviews; all fields combined)---a modest shift that would leave the qualitative structure of our rankings and the CLT bounds essentially unchanged.

\subsection{Results I: Scale-independent $\Phi$ index}

As a first step, we calculate the $\Phi$ index for all journals in the JCR as if they belong to the same field. 

Do the $\Phi$  rankings make sense? In Tables \ref{table:Top50_Phi} and \ref{table:Top50_Phi_A}  we list the top-50 journals ranked by the overall $\Phi$ index and the article-only $\Phi_A$ index, respectively; we also list the corresponding citation average rankings ($f$-ranks and $f_A$-ranks, respectively), for comparison. In Table \ref{table:Top50_Phi} we also list the values and rankings of two other indicators, the Source Normalized Impact per Paper (SNIP) and the Journal Citation Indicator (JCI). 

We can make the following observations. 

(a) The $\Phi$ rankings are sufficiently distinct from the $f$ rankings. For example, 20 of the top-50 journals ranked by $\Phi$ are new entries. Likewise, 123 of the top-500 $\Phi$-ranked journals are new entries. 

(b) Importantly, several entries in the top 50 $\Phi$ rankings improved considerably in their positions compared to $f$ rankings. Of particular note are the following transitions from rank($f$) to rank($\Phi$): 
\newline 
NATURE COMMUNICATIONS (\#196 $\to$ \#10), 
\newline 
JOURNAL OF THE AMERICAN CHEMICAL SOCIETY (\#173 $\to$ \#15), 
\newline 
ANGEWANDTE CHEMIE (\#193 $\to$ \#16), 
\newline 
ADVANCED FUNCTIONAL MATERIALS (\#128 $\to$ \#18), 
\newline 
APPLIED CATALYSIS B-ENVIRONMENTAL (\#118 $\to$ \#19), 
\newline 
NUCLEIC ACIDS RESEARCH (\#151 $\to$ \#24), 
\newline 
JOURNAL OF MATERIALS CHEMISTRY A (\#250 $\to$ \#26), 
\newline 
CHEMICAL ENGINEERING JOURNAL (\#226 $\to$ \#28), 
\newline 
ACS NANO (\#170 $\to$ \#29), 
\newline 
NANO ENERGY (\#131 $\to$ \#30), 
\newline 
PROC NATL ACAD SCI USA (\#334 $\to$ \#36), 
\newline 
ACS APPLIED MATERIALS \& INTERFACES (\#453 $\to$ \#41), and 
\newline 
RENEWABLE \& SUSTAINABLE ENERGY REVIEWS (\#189 $\to$ \#49).

Qualitatively, many of the revised rankings appear justified, as they correspond with the prevailing views of the expert community, to the best of our knowledge. For example, NATURE COMMUNICATIONS is a highly regarded multidisciplinary journal. The JOURNAL OF THE AMERICAN CHEMICAL SOCIETY is considered the flagship journal of the American Chemical Society, while ANGEWANDTE CHEMIE is one of the most respected journals in chemistry. ADVANCED MATERIALS, ADVANCED ENERGY MATERIALS, and ADVANCED FUNCTIONAL MATERIALS are among the most prestigious journals in materials science, while APPLIED CATALYSIS B-ENVIRONMENTAL is one of the leading journals in catalysis, environmental science, and chemical engineering. Another example is PROC NATL ACAD SCI USA, a highly respected journal that is the official publication of the U.S. National Academy of Sciences. 

(c) The journals NATURE, SCIENCE, and CELL improve their rankings and enter the top-10 slots. Again, these improved ranking results align with the reputation of these three titles as among the most selective journals worldwide. 

\begin{figure}
\centering
\includegraphics[width=0.45\linewidth]{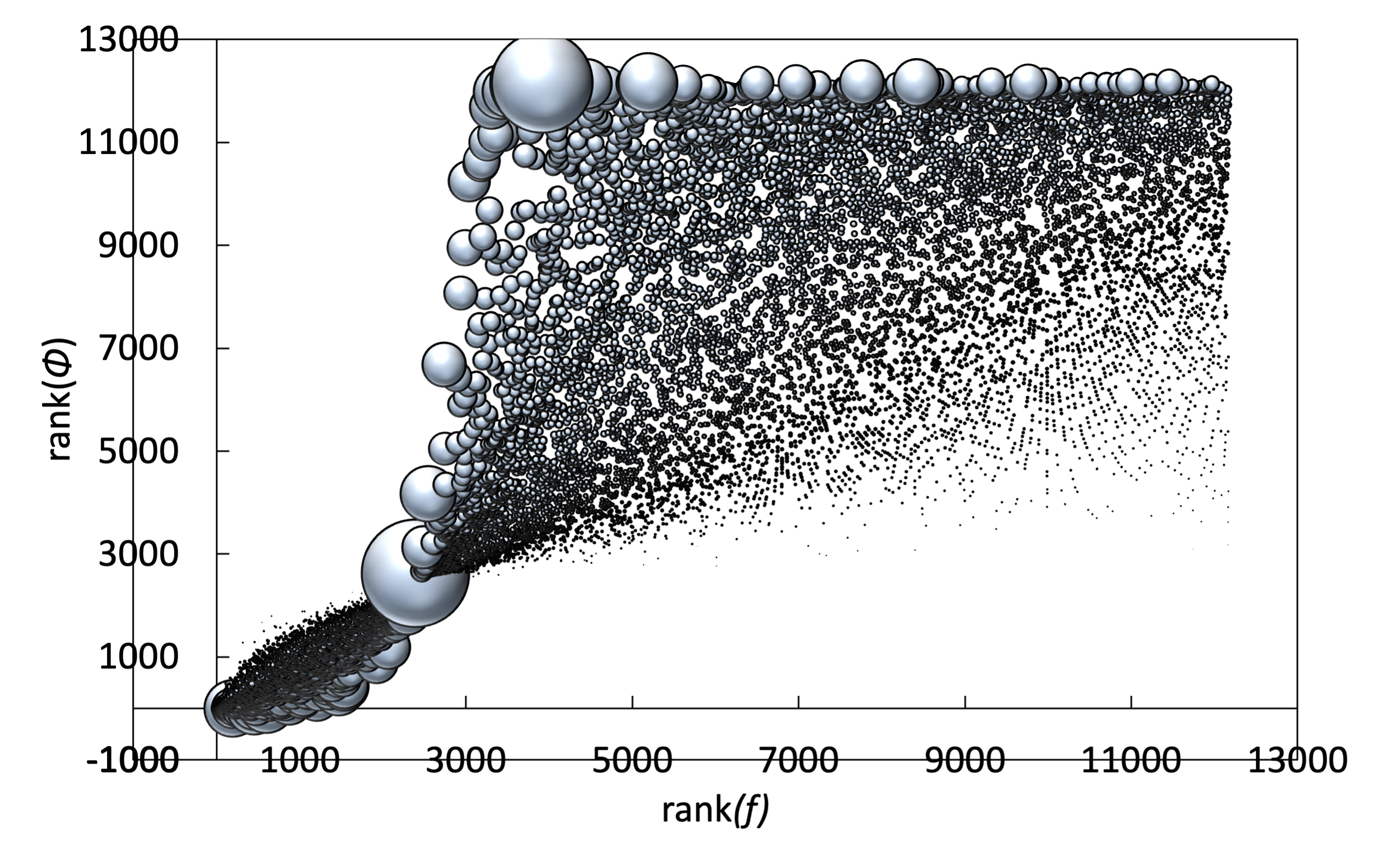}
\includegraphics[width=0.45\linewidth]{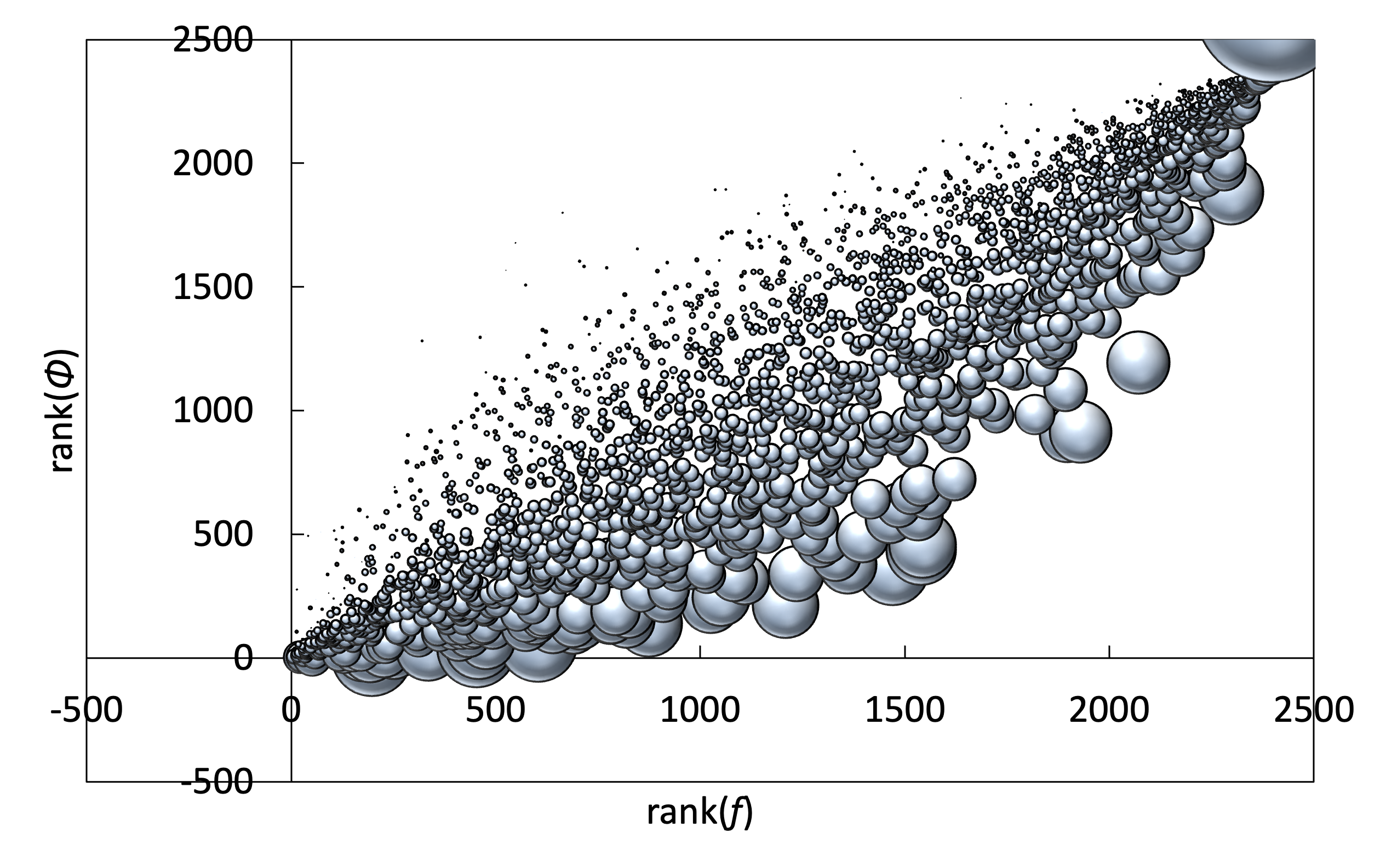}
\caption{Left panel: Comparison of $\Phi$ rankings vs. $f$ rankings for all 12173 journals. The area size of each bubble is proportional to the journal size. The convex-shaped ``lobe'' on the lower left contains the 2382 journals with $\Phi >0$ or $f > \mu$. The large concave-shaped ``lobe'' on the upper right contains the 9791 journals with $\Phi <0$ or $f < \mu$. A measure of rank correlation between the two rankings is given by the Kendall coefficient $\tau =  0.62$. Right panel: Detail of left panel for the 2382 ranks whereby $\Phi >0$ or $f > \mu$. For this set, $\tau =0.73$. Clearly, a higher journal size rewards above-average performing journals but penalizes below-average journals.}
\label{fig:rank_f_phi}
\end{figure}

(d) Figure \ref{fig:rank_f_phi} shows $\Phi$ rankings plotted against $f$ rankings for all 12173 journals (with articles and reviews counted together). The area size of each bubble is proportional to the journal size. We note that the two rankings converge at rank \#2382, the ``inflection'' point shown in Figure \ref{fig:rank_f_phi}. At the inflection point, the $\Phi$ index flips sign and the citation average $f$ crosses the global citation average $\mu$. So, to the {\it left} of the inflection point we have all 2382 journals with above-average performance, i.e., $f > \mu$ (or $\Phi > 0$). To the  {\it right} of the inflection point lie all the 9791 journals with below-average performance, i.e., $f < \mu$ (or $\Phi < 0$). 

As expected from Eq. \ref{eq:Phi_def}, a higher journal size is advantageous for above-average performing journals, but becomes disadvantageous for below-average performing journals. And this is what happens in Figure \ref{fig:rank_f_phi}: Larger journals are rewarded and get pulled to better ranks when $f > \mu$, but are penalized and pushed to worse ranks for $f < \mu$.

(e) How similar are the two rankings ($\Phi$ vs. $f$)? 
A measure of rank correlation---the similarity of journal orderings when ranked by $f$ vs. $\Phi$---is given by the Kendall $\tau$ coefficient, which takes the value $\tau = 0.62$ for the full set of 12173 journals, while $\tau = 0.73$ for the 2382 journals of positive $\Phi$. While the two rankings have some similarity, it is clear from the Kendall coefficient that substantial re-ordering occurs, as we also see from Figure \ref{fig:rank_f_phi}.  

\begin{table}[h]
\centering
\begin{tabular}{l l l l l l l l l l l}
\hline
\hline
Journal & $N_{2Y}$ & $f$ & JCI & SNIP & $\Phi$ & rank($f$) & rank(JCI) & rank(SNIP)  & rank($\Phi$) \\
\hline
CA-A CANCER JOURNAL FOR CLINICIANS & 47 & 503.9 & 77.6 & 142.3 & 274 & 1 & 1 & 1 & 1 \\
NEW ENGLAND JOURNAL OF MEDICINE & 649 & 79.8 & 26.1 & 14.78 & 154.1 & 3 & 2 & 8 & 2 \\
NATURE & 1807 & 45.4 & 8.7 & 9.27 & 140.5 & 21 & 20 & 32 & 3 \\
SCIENCE & 1556 & 42.1 & 7.6 & 7.93 & 119.9 & 23 & 26 & 44 & 4 \\
LANCET & 539 & 68.5 & 20.1 & 23.63 & 119.6 & 5 & 4 & 3 & 5 \\
\hline
ADVANCED MATERIALS & 2808 & 29.6 & 4.5 & 4 & 107.8 & 51 & 115 & 197 & 6 \\
CHEMICAL REVIEWS & 431 & 59.9 & 3.6 & 12.24 & 92.6 & 8 & 191 & 13 & 7 \\
CHEMICAL SOCIETY REVIEWS & 482 & 53.7 & 3.1 & 9.11 & 87.1 & 15 & 296 & 34 & 8 \\
CELL & 886 & 39.4 & 7.1 & 8.06 & 84.1 & 30 & 30 & 42 & 9 \\
NATURE COMMUNICATIONS & 10531 & 14.3 & 2.7 & 3.1 & 83.8 & 196 & 437 & 354 & 10 \\
\hline
NATURE MEDICINE & 391 & 51.5 & 10 & 7.83 & 74.9 & 16 & 13 & 45 & 11 \\
ADVANCED ENERGY MATERIALS & 1428 & 28.3 & 4.3 & 3.17 & 73 & 57 & 133 & 331 & 12 \\
NATURE REVIEWS MOLECULAR CELL BIOLOGY & 90 & 90.7 & 7 & 14.25 & 65.7 & 2 & 31 & 9 & 13 \\
ENERGY \& ENVIRONMENTAL SCIENCE & 564 & 38 & 6 & 4.95 & 64.3 & 33 & 53 & 130 & 14 \\
JOURNAL OF THE AMERICAN CHEMICAL SOCIETY & 4924 & 15.3 & 2.8 & 2.58 & 62.9 & 173 & 406 & 585 & 15 \\
\hline
ANGEWANDTE CHEMIE-INTERNATIONAL EDITION & 5567 & 14.6 & 2.4 & 2.23 & 62.4 & 193 & 587 & 884 & 16 \\
NATURE ENERGY & 204 & 58 & 8.2 & 7.34 & 61.5 & 11 & 21 & 52 & 17 \\
ADVANCED FUNCTIONAL MATERIALS & 2751 & 18.1 & 2.9 & 2.44 & 58.4 & 128 & 338 & 683 & 18 \\
APPLIED CATALYSIS B-ENVIRONMENTAL & 2136 & 19.2 & 3 & 2.73 & 55.8 & 118 & 308 & 515 & 19 \\
JAMA-J AMER MED ASSOC & 412 & 36.5 & 11.5 & 10.87 & 52.6 & 39 & 9 & 17 & 20 \\
\hline
NATURE MATERIALS & 308 & 41.1 & 7.1 & 6.95 & 51.9 & 24 & 29 & 61 & 21 \\
NATURE BIOTECHNOLOGY & 228 & 45.7 & 10 & 6.97 & 50.2 & 20 & 12 & 59 & 22 \\
NATURE GENETICS & 361 & 37.1 & 8.8 & 6.54 & 50.1 & 37 & 18 & 66 & 23 \\
NUCLEIC ACIDS RESEARCH & 2535 & 16.4 & 3.2 & 4.46 & 49.5 & 151 & 253 & 154 & 24 \\
JOULE & 339 & 37.4 & 5 & 4.81 & 49 & 35 & 84 & 140 & 25 \\
\hline
JOURNAL OF MATERIALS CHEMISTRY A & 5182 & 12.6 & 1.9 & 1.66 & 48.9 & 250 & 1035 & 1963 & 26 \\
LANCET ONCOLOGY & 326 & 38 & 8 & 9.9 & 48.9 & 32 & 22 & 28 & 27 \\
CHEMICAL ENGINEERING JOURNAL & 4657 & 13.1 & 2 & 2.18 & 48.9 & 226 & 882 & 949 & 28 \\
ACS NANO & 2670 & 15.6 & 2.5 & 2.39 & 47.3 & 170 & 535 & 722 & 29 \\
NANO ENERGY & 1835 & 17.7 & 3 & 2.25 & 46.6 & 131 & 313 & 862 & 30 \\
\hline
NATURE NANOTECHNOLOGY & 300 & 37.5 & 5.5 & 6.31 & 46.2 & 34 & 68 & 73 & 31 \\
CIRCULATION & 716 & 25.5 & 6.2 & 5.39 & 45.7 & 67 & 48 & 108 & 32 \\
NATURE REVIEWS MATERIALS & 91 & 63.5 & 4.1 & 13.09 & 45.3 & 6 & 147 & 10 & 33 \\
NATURE REVIEWS MICROBIOLOGY & 109 & 58.4 & 5 & 10.49 & 45.3 & 10 & 82 & 22 & 34 \\
NATURE REVIEWS DRUG DISCOVERY & 72 & 69 & 10.9 & 12.93 & 44 & 4 & 11 & 11 & 35 \\
\hline
PROC NATL ACAD SCI USA & 6620 & 10.8 & 2 & 2.88 & 43.3 & 334 & 884 & 435 & 36 \\
JOURNAL OF CLINICAL ONCOLOGY & 642 & 25.3 & 5.3 & 6.23 & 42.8 & 70 & 75 & 74 & 37 \\
NATURE REVIEWS CANCER & 95 & 58.5 & 7.6 & 10.23 & 42.4 & 9 & 25 & 24 & 38 \\
NATURE CATALYSIS & 214 & 39.9 & 4.7 & 5.34 & 41.9 & 28 & 101 & 109 & 39 \\
NATURE REVIEWS CLINICAL ONCOLOGY & 83 & 60.5 & 7.7 & 7.65 & 41.1 & 7 & 24 & 48 & 40 \\
\hline
ACS APPLIED MATERIALS \& INTERFACES & 10072 & 9.2 & 1.5 & 1.45 & 40.7 & 453 & 1873 & 2720 & 41 \\
NATURE PHOTONICS & 218 & 36.9 & 8.8 & 9.95 & 38.7 & 38 & 19 & 27 & 42 \\
EUROPEAN HEART JOURNAL & 519 & 25.1 & 5.8 & 5.9 & 38.2 & 72 & 57 & 86 & 43 \\
IMMUNITY & 330 & 30.3 & 5.6 & 5.31 & 38 & 47 & 66 & 110 & 44 \\
ACS ENERGY LETTERS & 723 & 21.7 & 3.2 & 2.41 & 37.9 & 93 & 263 & 704 & 45 \\
\hline
NATURE REVIEWS IMMUNOLOGY & 104 & 49.8 & 5.7 & 8.77 & 37.3 & 19 & 60 & 37 & 46 \\
ACCOUNTS OF CHEMICAL RESEARCH & 645 & 22.1 & 1.3 & 3.74 & 36.6 & 89 & 2710 & 227 & 47 \\
LANCET NEUROLOGY & 159 & 40.3 & 9.7 & 12.45 & 36.5 & 25 & 14 & 12 & 48 \\
RENEWABLE \& SUSTAINABLE ENERGY REVIEWS & 1801 & 14.8 & 1.1 & 4.67 & 36.2 & 189 & 3987 & 144 & 49 \\
NATURE REVIEWS GENETICS & 95 & 50.4 & 5.6 & 8.8 & 36.1 & 18 & 65 & 36 & 50 \\
\hline
\hline
\end{tabular}
\caption{Top 50 journals ranked by their overall $\Phi$ index. We also listed the citation average ($f$), the Journal Citation Indicator (JCI), the SNIP indicator, and their respective rankings.}
 \label{table:Top50_Phi}
\end{table}

(f) But $\Phi$ rankings are also quite different from the SNIP and JCI rankings. Looking at Table \ref{table:Top50_Phi}, we see several qualitative similarities between the rankings by $f$, SNIP, and JCI. For example, only $\Phi$ rankings place NATURE and SCIENCE in the top-5 positions, while the other rankings place them in positions ranging from 20 to 44. 

(g) Also, only $\Phi$ rankings elevate large journals (publishing more than 10,000 papers in two years) and mid-sized journals  (publishing from 2,000--10,000 papers in two years) to the top 50 positions. Specifically, we see two large journals (NATURE COMMUNICATIONS, ACS APPLIED MATERIALS \& INTERFACES) and 10 mid-sized journals (JOURNAL OF THE AMERICAN CHEMICAL SOCIETY, ANGEWANDTE CHEMIE-INTERNATIONAL EDITION
ADVANCED FUNCTIONAL MATERIALS, APPLIED CATALYSIS B-ENVIRONMENTAL, NUCLEIC ACIDS RESEARCH, JOURNAL OF MATERIALS CHEMISTRY A, CHEMICAL ENGINEERING JOURNAL, ACS NANO, 
NANO ENERGY, PROC NATL ACAD SCI USA). 

(h) The rankings by $f$, SNIP, and JCI are more correlated with each other than with $\Phi$ rankings. Looking only at the journals in Table \ref{table:Top50_Phi}, the Kendall $\tau$ coefficient between the $\Phi$ ranks and any of the $f$, SNIP, or JCI ranks ranges from 0.15--0.17. In contrast, the Kendall $\tau$ coefficient between any two pairs of the $f$, SNIP, or JCI ranks in this Table ranges from 0.57--0.73.

(i) What about different document types?
We know that articles and reviews are cited differently (Garfield, 1996; MacRoberts \& MacRoberts, 1996), which calls for separate journal rankings for articles and reviews. This concern can be easily accommodated. One can obtain $\Phi$ index rankings for journals that publish only articles or reviews, or a mixture of both. See Table \ref{table:Top50_Phi_A}, where we list the top 50 ranked journals in terms of their article-only $\Phi_A$ index. Compared to Table \ref{table:Top50_Phi}, we note that many journals retain their positions in the top-50 slots, which means that the $\Phi$ and  $\Phi_A$ indexes are largely correlated---indeed, Kendall $\tau( \text{rank}(\Phi), \text{rank}(\Phi_A)) =0.91$. But upon closer inspection, we note that review journals are over-represented in the top $\Phi$ ranks of Table  \ref{table:Top50_Phi}, which is a clear consequence of the well-known fact that review articles skew the rankings (Miranda \& Garcia-Carpintero, 2018). For example, 11 journals of the top-50 titles in Table  \ref{table:Top50_Phi} publish only reviews. Some notable transitions in the top-50 list as we move from rank($f_A$) to rank($\Phi_A$) are the following: 
\newline 
JOURNAL OF MATERIALS CHEMISTRY A  (\#168 $\to$ \#27), 
\newline 
SCIENCE ADVANCES (\#137 $\to$ \#34), 
\newline 
JOURNAL OF CLEANER PRODUCTION (\#339 $\to$ \#39), 
\newline 
PHYSICAL REVIEW LETTERS (\#313 $\to$ \#43), and
\newline 
SCIENCE OF THE TOTAL ENVIRONMENT (\#504 $\to$ \#48).

(j) As a cautionary tale about data quality and data cleaning, we note in Table \ref{table:Top50_Phi_A} the titles CA-A CANCER JOURNAL FOR CLINICIANS, NATURE REVIEWS DISEASE PRIMERS, and REVIEWS OF MODERN PHYSICS. All these entries reflect a cataloging error in the \texttt{Document Type} by Clarivate Analytics, since these journals publish only {\it reviews} and should thus be absent from {\it article-only} $\Phi_A$ rankings of journals. 

\begin{table}[h]
\centering
\begin{tabular}{l l l l l l}
\hline
\hline
Journal	&	$N_{2Y,A}$ & $f_{A}$	&	$\Phi_{A}$			&	rank($f_{A}$)	&	rank($\Phi_{A}$)	\\
\hline
CA-A CANCER JOURNAL FOR CLINICIANS & 35 & 659.1 & 321.1 & 1 & 1 \\
NEW ENGLAND JOURNAL OF MEDICINE & 553 & 84.7 & 157.6 & 3 & 2 \\
NATURE & 1755 & 44.7 & 141.9 & 15 & 3 \\
SCIENCE & 1434 & 40.2 & 114.3 & 18 & 4 \\
LANCET & 383 & 65.8 & 100.5 & 4 & 5 \\
\hline
ADVANCED MATERIALS & 2450 & 27.2 & 96.1 & 42 & 6 \\
NATURE COMMUNICATIONS & 10471 & 14.2 & 88.2 & 130 & 7 \\
CELL & 815 & 37.5 & 79.6 & 20 & 8 \\
NATURE MEDICINE & 373 & 48.6 & 71.8 & 11 & 9 \\
ADVANCED ENERGY MATERIALS & 1296 & 25.9 & 65.8 & 46 & 10 \\
\hline
JOURNAL OF THE AMERICAN CHEMICAL SOCIETY & 4848 & 15 & 64.9 & 116 & 11 \\
ANGEWANDTE CHEMIE-INTERNATIONAL EDITION & 5260 & 13.4 & 57.9 & 143 & 12 \\
APPLIED CATALYSIS B-ENVIRONMENTAL & 2094 & 18.8 & 56.9 & 79 & 13 \\
ADVANCED FUNCTIONAL MATERIALS & 2591 & 16.9 & 55.4 & 89 & 14 \\
ENERGY \& ENVIRONMENTAL SCIENCE & 494 & 32.9 & 53.6 & 29 & 15 \\
\hline
NUCLEIC ACIDS RESEARCH & 2534 & 16.4 & 52.6 & 92 & 16 \\
NATURE GENETICS & 361 & 37.1 & 52.4 & 22 & 17 \\
LANCET ONCOLOGY & 255 & 43.3 & 52.2 & 16 & 18 \\
NATURE BIOTECHNOLOGY & 218 & 45.4 & 50.9 & 13 & 19 \\
NATURE ENERGY & 174 & 50.1 & 50.6 & 10 & 20 \\
\hline
ACS NANO & 2632 & 15.4 & 49.3 & 113 & 21 \\
CHEMICAL ENGINEERING JOURNAL & 4518 & 12.4 & 48.2 & 159 & 22 \\
NATURE MATERIALS & 294 & 37.1 & 47.3 & 21 & 23 \\
PROC NATL ACAD SCI USA & 6619 & 10.8 & 47.1 & 204 & 24 \\
CIRCULATION & 690 & 25.3 & 46.7 & 49 & 25 \\
\hline
NANO ENERGY & 1771 & 17.2 & 46.7 & 88 & 26 \\
JOURNAL OF MATERIALS CHEMISTRY A & 4896 & 11.8 & 46.5 & 168 & 27 \\
JOURNAL OF CLINICAL ONCOLOGY & 581 & 26.8 & 45.9 & 44 & 28 \\
NATURE NANOTECHNOLOGY & 286 & 36.4 & 45.7 & 23 & 29 \\
ACS APPLIED MATERIALS \& INTERFACES & 10032 & 9.1 & 44.1 & 299 & 30 \\
\hline
JOULE & 269 & 35.5 & 43.1 & 24 & 31 \\
JAMA-J AMER MED ASSOC & 304 & 31.9 & 40.6 & 33 & 32 \\
ACS ENERGY LETTERS & 701 & 21 & 37.8 & 68 & 33 \\
SCIENCE ADVANCES & 2103 & 13.5 & 37 & 137 & 34 \\
NATURE CATALYSIS & 193 & 35.5 & 36.5 & 25 & 35 \\
\hline
NATURE REVIEWS DISEASE PRIMERS & 82 & 51.2 & 35.6 & 9 & 36 \\
EUROPEAN HEART JOURNAL & 449 & 23.8 & 35.2 & 54 & 37 \\
LANCET NEUROLOGY & 101 & 45.2 & 34.4 & 14 & 38 \\
JOURNAL OF CLEANER PRODUCTION & 7409 & 8.6 & 34.4 & 339 & 39 \\
ACS CATALYSIS & 2317 & 12.3 & 34.1 & 161 & 40 \\
\hline
NATURE PHOTONICS & 202 & 31.9 & 33.1 & 32 & 41 \\
JAMA ONCOLOGY & 280 & 27.2 & 32.4 & 43 & 42 \\
PHYSICAL REVIEW LETTERS & 5425 & 8.9 & 31.4 & 313 & 43 \\
REVIEWS OF MODERN PHYSICS & 43 & 60.1 & 30.6 & 6 & 44 \\
IMMUNITY & 281 & 25.6 & 30.3 & 47 & 45 \\
\hline
CANCER CELL & 216 & 28.2 & 29.8 & 40 & 46 \\
J AMER COLL CARDIOL & 610 & 18.2 & 29.5 & 82 & 47 \\
SCIENCE OF THE TOTAL ENVIRONMENT & 10054 & 7.3 & 29.3 & 504 & 48 \\
NATURE METHODS & 289 & 24.3 & 28.9 & 52 & 49 \\
NATURE NEUROSCIENCE & 339 & 22.7 & 28.9 & 61 & 50 \\
\hline
\hline
\end{tabular}
\caption{Top 50 journals ranked by their article-only $\Phi_{A}$ index. Note that three of these entries---CA-A CANCER JOURNAL FOR CLINICIANS, NATURE REVIEWS DISEASE PRIMERS, and REVIEWS OF MODERN PHYSICS---are listed as article-publishing journals due to a cataloging error in the \texttt{Document Type} by Clarivate Analytics.}
\label{table:Top50_Phi_A}
\end{table}

(k) As an aside, we expect the above conclusions to hold also for 5-year citation averages. Previous studies (Leydesdorff {\it et al.}, 2019; Rousseau, 2009) and our own experience indicate that there are no significant changes in journal rankings when we shift from 2-year to 5-year Journal Impact Factors.  

(l) What about confidence intervals on the $\Phi$ index? Since we calculate $\mu$ and $\sigma$ from the population of all papers, we consider them as fixed, non-random constants. In this case, the uncertainty in estimating $\Phi$ arises solely from the sampling variability of the citation average $f$. Then, under standard assumptions of independent data drawn from the same distribution, it follows that the standard error on the $\Phi$ index is 
\begin{equation}
\text{SE}(\Phi) = \text{SE}\left(\frac{(f_n-\mu) \sqrt{n}}{\sigma}\right) 
= \text{SE}\left(\frac{f_n \sqrt{n}}{\sigma}\right) 
=  \frac{\sqrt{n}}{\sigma} \text{SE}(f_n) 
=  \frac{\sqrt{n}}{\sigma}  \frac{\sigma}{\sqrt{n}} 
= 1
\label{eq:Phi_SE}
\end{equation}
and therefore the confidence interval on the $\Phi$ index at the 95\% level is 
\begin{equation}
\text{CI}_{95\%}(\Phi) = \Phi \pm 1.96.
\label{eq:Phi_CI}
\end{equation}
Similarly,  the confidence interval at the 99\% level is $\text{CI}_{99\%}(\Phi) = \Phi \pm 2.576$, while at the 99.7\% level we have $\text{CI}_{99.7\%}(\Phi) = \Phi \pm 3$, which agrees with Eq. \ref{eq:6}.

We note that $\text{SE}(\Phi) = 1$ holds strictly when papers within a journal are treated as independent and identically distributed (i.i.d.) draws from the field distribution. Here, $\mu$ and $\sigma$ are not estimates but exact population parameters, calculated from the full census of papers indexed in the 2020 JCR; so, no estimation uncertainty attaches to them. For real journals, however, the i.i.d. condition does not hold exactly, since papers are editorially selected rather than randomly drawn. The confidence intervals derived above therefore apply strictly to randomly formed journals. Nevertheless, the fact that 78.6\% of real journals in the 2020 JCR have citation averages within the random-journal range (at $k=3$) suggests that the random-journal model is not merely a theoretical construct but a meaningful empirical benchmark.

Let us put these confidence intervals into perspective. Among the 12173 journals studied here, 46\% have $\Phi$ index outside the range expected for randomly drawn journals at the 95\% confidence interval (i.e., $|\Phi | \ge 1.96$). Furthermore, 28.5\% journals have $\Phi$ index outside the random range at the 99\% confidence interval ($|\Phi | \ge 2.576$), while 21.4\% journals have $\Phi$ index outside the random range at the 99.7\% confidence interval ($|\Phi | \ge 3$). (The latter statistic is in agreement with our earlier observation that 78.6\% journals have citation average $f$ within the values expected from the Central Limit Theorem for $k=3$.)

The fact that so many journals have $\Phi$ values (and thus also citation averages, $f$ values) within the range expected for a randomly formed journal gives us pause. It means that we should avoid the temptation to overinterpret rankings, especially for journals whose $\Phi$ values lie within the random journal range (for instance, $-1.96 < \Phi < 1.96$ at the 95\% confidence interval, or $-2.576 < \Phi < 2.576$ at the 99\% confidence interval). For those journals---and, to a lesser extent, for the other journals that are outside the random range---it would make more sense to think in terms of {\it ratings}, not rankings, classifying journals in categories according to their citation performance levels. For example, we could classify journals in 3 tiers, as shown in Table \ref{table:Tiered_journals}. A more nuanced rating system with 6 tiers (or bands) is shown in Table \ref{table:Banded_journals}. The idea of a ratings system for journals goes beyond the purposes of this paper and will be explored in future work.

\begin{table}[h]
\centering
\begin{tabular}{l l l l l}
\hline
\hline
Tier & Criterion & Journals & fraction \\
\hline
high impact & $\Phi > 1.96$ & 1231 & 10.1\% \\
average impact & $-1.96  \le \Phi \le 1.96$ & 6542 &53.7\% \\
low impact & $\Phi < -1.96$ & 4400 & 36.1\% \\
\hline
\hline
\end{tabular}
\caption{Suggested 3-tier rating of journals based on their $\Phi$ index.}
\label{table:Tiered_journals}
\end{table}


\begin{table}[h]
\centering
\begin{tabular}{l l l l l}
\hline
\hline
Tier & Criterion & Journals & fraction \\
\hline
significant & $\Phi > 3$ & 933 & 7.7\% \\
strong  & $2  < \Phi \le 3$ & 279 & 2.3\% \\
high average & $0  < \Phi \le 2$ & 1170 & 9.6\% \\
low average & $-2  < \Phi \le 0$ & 5546 & 45.6\%  \\
weak & $-3  < \Phi \le -2$ & 2576 & 21.2\% \\
marginal & $\Phi \le -3$ & 1669 & 13.7\% \\
\hline
\hline
\end{tabular}
\caption{Suggested 6-tier rating of journals based on their $\Phi$ index.}
\label{table:Banded_journals}
\end{table}


\subsection{Results II: Scale-independent {\it and} field-normalized $\Phi$ index for monodisciplinary journals}


\begin{table}[h]
\centering
\begin{tabular}{l l l l l l l l l}
\hline
\hline
Subject (field) category & $\mu_s^{art}$ & $\mu_s^{rev}$ & $\sigma_s^{art}$ & $\sigma_s^{rev}$ & $N_s^{art}$ & $N_s^{rev}$ & Journals \\
\hline
AGRICULTURE, DAIRY \& ANIMAL SCIENCE & 2.08 & 5.84 & 2.64 & 8.69 & 17089 & 759 & 62 \\
PSYCHIATRY & 3.45 & 7.17 & 5.37 & 9.89 & 38282 & 5186 & 217 \\
CLINICAL NEUROLOGY & 3.5 & 6.22 & 7.91 & 13.57 & 51637 & 9751 & 208 \\
MEDICINE, GENERAL \& INTERNAL & 4.01 & 5.97 & 20.68 & 19.21 & 50457 & 11843 & 167 \\
PARASITOLOGY & 3.31 & 6.44 & 5.15 & 11.23 & 11195 & 1028 & 37 \\
\hline
AGRICULTURAL ENGINEERING & 5.59 & 17.32 & 6.16 & 17.21 & 7917 & 295 & 14 \\
GEOCHEMISTRY \& GEOPHYSICS & 3.13 & 8.59 & 4.32 & 13.57 & 23175 & 389 & 79 \\
MATERIALS SCIENCE, CHAR. \& TESTING & 2.79 & 6.2 & 4.48 & 10.45 & 6621 & 117 & 30 \\
ALLERGY & 4.72 & 6.42 & 7.57 & 8.85 & 4225 & 1234 & 28 \\
CRITICAL CARE MEDICINE & 4.63 & 6.4 & 10.74 & 9.22 & 8584 & 1568 & 36 \\
\hline
SOCIAL SCIENCES, BIOMEDICAL & 2.71 & 6.08 & 3.89 & 6.84 & 6509 & 313 & 44 \\
ENVIRONMENTAL SCIENCES & 4.64 & 12.27 & 6.77 & 19.15 & 145587 & 6773 & 273 \\
ENGINEERING, PETROLEUM & 2.58 & 5.41 & 3.51 & 6.91 & 4804 & 170 & 17 \\
PHYSICS, ATOMIC, MOLECULAR \& CHEMICAL & 3.5 & 10.13 & 5.05 & 14.29 & 33278 & 325 & 34 \\
ENERGY \& FUELS & 6.56 & 18.45 & 9.89 & 25.91 & 81064 & 4835 & 108 \\
\hline
GENETICS \& HEREDITY & 4.07 & 8.5 & 20.54 & 13.89 & 38614 & 4211 & 170 \\
LIMNOLOGY & 2.99 & 5.81 & 3.79 & 9.56 & 4084 & 79 & 21 \\
CELL BIOLOGY & 6.41 & 11.95 & 12.49 & 24.06 & 52345 & 8786 & 191 \\
PSYCHOLOGY, DEVELOPMENTAL & 3.17 & 7.16 & 3.97 & 9.51 & 11390 & 609 & 74 \\
COMPUTER SCIENCE, INTERDISC APPL & 3.93 & 9.56 & 8.3 & 16.65 & 31545 & 799 & 111 \\
\hline
LINGUISTICS & 1.5 & 3.4 & 2.69 & 5.48 & 11878 & 283 & 192 \\
FISHERIES & 2.6 & 6.78 & 3.47 & 9.17 & 12564 & 473 & 54 \\
DEVELOPMENT STUDIES & 3.04 & 5.87 & 4.9 & 7.67 & 4529 & 135 & 42 \\
HOSPITALITY, LEISURE, SPORT \& TOURISM & 4.84 & 11.59 & 7.56 & 17.46 & 6773 & 307 & 58 \\
SOCIAL ISSUES & 2.11 & 4.36 & 3.02 & 7.52 & 3736 & 133 & 44 \\
\hline
PSYCHOLOGY, MATHEMATICAL & 3.73 & 5.77 & 13.76 & 8.62 & 1338 & 122 & 13 \\
FAMILY STUDIES & 2.43 & 5.34 & 3.13 & 8.79 & 6199 & 290 & 46 \\
CHEMISTRY, PHYSICAL & 6.75 & 20.22 & 10.74 & 28.11 & 144909 & 4799 & 150 \\
MATERIALS SCIENCE, BIOMATERIALS & 5.31 & 13.53 & 6.14 & 15.03 & 16647 & 1140 & 41 \\
MATERIALS SCIENCE, MULTIDISCIPLINARY & 5.37 & 18.13 & 9.17 & 27.66 & 240166 & 8437 & 327 \\
\hline
\hline
\end{tabular}
\caption{Global citation averages, standard deviations, and document numbers for articles and reviews in 30 subject (field) categories. The number of journals in each category is listed in the last column.}
\label{table:stats_30subjects}
\end{table}

So far, we have ranked $\Phi$ indexes of all journals as if they belong to a unified field, with a global citation average $\mu$ and a global citation standard deviation $\sigma$. Of course, different fields have different citation practices, which one needs to account for to meaningfully compare citation averages across fields. As we mentioned in section \ref{sec:2}, we can readily apply the $\Phi$ index methodology to this task. The different citation practices of each field category $s$ are simply encoded in the numbers ($\mu_s$, $\sigma_s$), which are used to calculate a journal's $\Phi_s$ index, as shown in Eq. (\ref{eq:Phi}). 

For monodisciplinary journals that publish in a single academic discipline, this is a straightforward step to take. We can calculate and compare $\Phi_s$ indexes across fields, because the $\Phi$ index is uniquely suited to this task, in the same way that we use $z$ scores to compare athletes' performances across different sports. 

Likewise, we can readily calculate the composite $\Phi_s$ index of a journal whose {\it entirety} is concurrently allocated to two or more fields. In this case, all papers published by the journal belong to each field. For example, take NATURE PHOTONICS. All of its papers are concurrently allocated by Clarivate Analytics in two field categories, OPTICS and PHYSICS, APPLIED. So, the composite $\Phi_s$ index of the journal is simply the average of its $\Phi_s$ indexes in each field. And since NATURE PHOTONICS has $\Phi_s = 68.3$ in OPTICS and $\Phi_s = 42.5$ in PHYSICS, APPLIED, the journal has a composite $\Phi_s = 55.4$.

\begin{table}[h]
\centering
\begin{tabular}{l l l l l l l l l}
\hline
\hline
Journal & $n_A$ & $f_A$ & $\Phi_A$ & $\Phi_{s,A}$ & rank($f_A$) & rank($\Phi_A$) & rank($\Phi_{s,A}$) \\
\hline
NATURE & 1755 & 44.7 & 141.9 & 121 & 15 & 3 & 1 \\
ADVANCED MATERIALS & 2450 & 27.2 & 96.1 & 103.3 & 42 & 6 & 2 \\
SCIENCE & 1434 & 40.2 & 114.3 & 96.8 & 18 & 4 & 3 \\
NEW ENGLAND JOURNAL OF MEDICINE & 553 & 84.7 & 157.6 & 91.7 & 3 & 2 & 4 \\
NATURE MEDICINE & 373 & 48.6 & 71.8 & 73.2 & 11 & 9 & 5 \\
\hline
ADVANCED ENERGY MATERIALS & 1296 & 25.9 & 65.8 & 72.5 & 46 & 10 & 6 \\
ENERGY \& ENVIRONMENTAL SCIENCE & 494 & 32.9 & 53.6 & 72 & 29 & 15 & 7 \\
CA-A CANCER JOURNAL FOR CLINICIANS & 35 & 659.1 & 321.1 & 67.1 & 1 & 1 & 8 \\
NATURE BIOTECHNOLOGY & 218 & 45.4 & 50.9 & 65.4 & 13 & 19 & 9 \\
NATURE NEUROSCIENCE & 339 & 22.7 & 28.9 & 63.2 & 61 & 50 & 10 \\
\hline
NATURE COMMUNICATIONS & 10471 & 14.2 & 88.2 & 63.1 & 128 & 7 & 11 \\
JOURNAL OF THE AMER CHEM SOC & 4848 & 15 & 64.9 & 63.1 & 114 & 11 & 12 \\
NATURE ENERGY & 174 & 50.1 & 50.6 & 61.2 & 10 & 20 & 13 \\
CELL & 815 & 37.5 & 79.6 & 60.8 & 20 & 8 & 14 \\
APPLIED CATALYSIS B-ENVIRONMENTAL & 2094 & 18.8 & 56.9 & 60.6 & 78 & 13 & 15 \\
\hline
LANCET & 383 & 65.8 & 100.5 & 58.5 & 4 & 5 & 16 \\
NATURE PHOTONICS & 202 & 31.9 & 33.1 & 55.4 & 32 & 41 & 17 \\
ADVANCED FUNCTIONAL MATERIALS & 2591 & 16.9 & 55.4 & 54.8 & 87 & 14 & 18 \\
ANGEWANDTE CHEMIE-INTERN EDIT & 5260 & 13.4 & 57.9 & 54.1 & 141 & 12 & 19 \\
NATURE MATERIALS & 294 & 37.1 & 47.3 & 54.1 & 21 & 23 & 20 \\
\hline
LANCET NEUROLOGY & 101 & 45.2 & 34.4 & 52.9 & 14 & 38 & 21 \\
NATURE NANOTECHNOLOGY & 286 & 36.4 & 45.7 & 50.3 & 23 & 29 & 22 \\
IMMUNITY & 281 & 25.6 & 30.3 & 49.5 & 47 & 45 & 23 \\
IEEE COMMUN SURV AND TUTOR & 236 & 24.6 & 26.5 & 48.7 & 51 & 58 & 24 \\
JOULE & 269 & 35.5 & 43.1 & 48.6 & 24 & 31 & 25 \\
\hline
NANO ENERGY & 1771 & 17.2 & 46.7 & 47.2 & 86 & 26 & 26 \\
NEURON & 582 & 14.6 & 21.7 & 47.1 & 123 & 85 & 27 \\
CHEMICAL ENGINEERING JOURNAL & 4518 & 12.4 & 48.2 & 46.6 & 157 & 22 & 28 \\
PHYSICAL REVIEW LETTERS & 5425 & 8.9 & 31.4 & 46.2 & 307 & 43 & 29 \\
ACS NANO & 2632 & 15.4 & 49.3 & 45.3 & 111 & 21 & 30 \\
\hline
MMWR SURVEILLANCE SUMMARIES & 26 & 57.8 & 22.8 & 44.5 & 7 & 75 & 31 \\
REVIEWS OF MODERN PHYSICS & 43 & 60.1 & 30.6 & 44 & 6 & 44 & 32 \\
NATURE CLIMATE CHANGE & 268 & 20.9 & 23.2 & 43.4 & 70 & 70 & 33 \\
CARBOHYDRATE POLYMERS & 2396 & 8.7 & 19.8 & 42 & 326 & 98 & 34 \\
ACS ENERGY LETTERS & 701 & 21 & 37.8 & 41.7 & 68 & 33 & 35 \\
\hline
MMWR-MORBID MORTAL WEEKLY REP & 456 & 14.9 & 19.7 & 41.1 & 118 & 102 & 36 \\
NATURE PLANTS & 215 & 14.6 & 13.1 & 40.7 & 124 & 163 & 37 \\
LANCET INFECTIOUS DISEASES & 198 & 22.8 & 22.2 & 40.6 & 59 & 80 & 38 \\
JOURNAL OF MATERIALS CHEMISTRY A & 4896 & 11.8 & 46.5 & 39.7 & 166 & 27 & 39 \\
SCIENCE OF THE TOTAL ENVIRONMENT & 10054 & 7.3 & 29.3 & 39.7 & 495 & 48 & 40 \\
\hline
EUROPEAN UROLOGY & 278 & 15.6 & 16.3 & 37.4 & 106 & 123 & 41 \\
NATURE GEOSCIENCE & 283 & 15 & 15.6 & 37.3 & 115 & 131 & 42 \\
NATURE CATALYSIS & 193 & 35.5 & 36.5 & 37.2 & 25 & 35 & 43 \\
FOOD CHEMISTRY & 3881 & 7.2 & 17.8 & 36.3 & 511 & 113 & 44 \\
NATURE METHODS & 289 & 24.3 & 28.9 & 36.1 & 52 & 49 & 45 \\
\hline
SENSORS AND ACTUATORS B-CHEMICAL & 4307 & 7.3 & 18.9 & 36 & 504 & 105 & 46 \\
BIOMATERIALS & 1046 & 11.8 & 21.6 & 35.7 & 165 & 86 & 47 \\
LANCET GLOBAL HEALTH & 187 & 19.2 & 17.4 & 35.5 & 76 & 116 & 48 \\
NATURE IMMUNOLOGY & 217 & 21.7 & 21.8 & 35.5 & 63 & 83 & 49 \\
CELL HOST \& MICROBE & 233 & 18.4 & 18.4 & 35.4 & 79 & 109 & 50 \\
\hline
\hline
\end{tabular}
\caption{Top 50 journals ranked by field-specific $\Phi$ index (articles only), $\Phi_{s,A}$. Here, $n_A$ is each journal's size. Shown for comparison are the rankings by citation average $f_{A}$ and $\Phi_{A}$. Note that two of these entries---CA-A CANCER JOURNAL FOR CLINICIANS and REVIEWS OF MODERN PHYSICS---are listed as article-publishing journals due to a cataloging error in the \texttt{Document Type} by Clarivate Analytics.}
\label{table:top50_Phi_sA}
\end{table}

\begin{table}[h]
\centering
\begin{tabular}{l l l l l l l l l}
\hline
\hline
Journal & $n_A$ & $f_A$ & $\Phi_A$ & $\Phi_{s,A}$ & rank($f_A$) & rank($\Phi_A$) & rank($\Phi_{s,A}$) \\
\hline
BLOOD & 752 & 14.1 & 23.4 & 34.6 & 129 & 67 & 51 \\
OPHTHALMOLOGY & 338 & 10.1 & 9.6 & 34.5 & 236 & 244 & 52 \\
COMPOSITES PART B-ENGINEERING & 2167 & 8.3 & 17.3 & 33.6 & 364 & 118 & 53 \\
NEW PHYTOLOGIST & 878 & 7.8 & 9.9 & 33.5 & 425 & 237 & 54 \\
PLANT PHYSIOLOGY & 826 & 7.9 & 9.9 & 33.5 & 401 & 234 & 55 \\
\hline
LIGHT-SCIENCE \& APPLICATIONS & 241 & 15.9 & 15.6 & 33.2 & 100 & 132 & 56 \\
NATURE PHYSICS & 369 & 18.1 & 22.8 & 33.1 & 81 & 74 & 57 \\
NATURE MICROBIOLOGY & 357 & 16.2 & 19.4 & 32.8 & 92 & 103 & 58 \\
NATURE BIOMEDICAL ENGINEERING & 132 & 23.2 & 18.5 & 32.7 & 57 & 108 & 59 \\
NATURE ELECTRONICS & 100 & 25.5 & 18 & 32.6 & 48 & 112 & 60 \\
\hline
JOURNAL OF CLEANER PRODUCTION & 7409 & 8.6 & 34.4 & 32.5 & 332 & 39 & 61 \\
FOOD HYDROCOLLOIDS & 1244 & 9 & 15.1 & 32.5 & 306 & 136 & 62 \\
JAMA PEDIATRICS & 189 & 11.3 & 8.5 & 32.2 & 190 & 277 & 63 \\
PHYSICAL REVIEW X & 511 & 15.5 & 21.9 & 31.8 & 108 & 81 & 64 \\
JAMA PSYCHIATRY & 194 & 15.7 & 13.8 & 31.8 & 103 & 154 & 65 \\
\hline
CIRCULATION & 690 & 25.3 & 46.7 & 31.7 & 49 & 25 & 66 \\
PEDIATRICS & 981 & 6.1 & 6.1 & 31.4 & 735 & 399 & 67 \\
ENERGY CONVERSION AND MANAGEMENT & 2310 & 9 & 20.9 & 31.4 & 297 & 89 & 68 \\
NUCLEIC ACIDS RESEARCH & 2534 & 16.4 & 52.6 & 31.3 & 90 & 16 & 69 \\
ANALYTICAL CHEMISTRY & 3862 & 6.6 & 14.6 & 31.3 & 605 & 144 & 70 \\
\hline
GLOBAL CHANGE BIOLOGY & 774 & 9.8 & 13.9 & 31.2 & 251 & 149 & 71 \\
SCIENCE ROBOTICS & 112 & 21 & 15.1 & 31.2 & 69 & 138 & 72 \\
JOURNAL OF MEDICINAL CHEMISTRY & 1411 & 7.3 & 11 & 31.1 & 490 & 205 & 73 \\
WATER RESEARCH & 1679 & 9.8 & 20.5 & 31 & 253 & 93 & 74 \\
AEROSPACE SCIENCE AND TECHNOLOGY & 1228 & 5 & 3.5 & 30.8 & 1202 & 665 & 75 \\
\hline
NATURE GENETICS & 361 & 37.1 & 52.4 & 30.5 & 22 & 17 & 76 \\
LANCET PSYCHIATRY & 82 & 21.1 & 13 & 29.8 & 67 & 165 & 77 \\
IEEE TRANSACTIONS ON INDUSTRIAL ELECTRONICS & 1907 & 7.9 & 15 & 29.5 & 403 & 139 & 78 \\
PROC NATL ACAD SCI USA & 6619 & 10.8 & 47.1 & 29.5 & 202 & 24 & 79 \\
ORGANIC LETTERS & 3942 & 6 & 11.4 & 29.4 & 788 & 197 & 80 \\
\hline
NATURE CHEMISTRY & 277 & 23.6 & 27.3 & 29.2 & 56 & 55 & 81 \\
WORLD PSYCHIATRY & 34 & 29.9 & 12.6 & 28.7 & 36 & 171 & 82 \\
IEEE INTERNET OF THINGS JOURNAL & 1321 & 8.7 & 14.7 & 28.7 & 325 & 142 & 83 \\
ELIFE & 2781 & 7.8 & 17.7 & 28.6 & 418 & 115 & 84 \\
ANNALS OF THE RHEUMATIC DISEASES & 389 & 12.2 & 13.7 & 28.5 & 161 & 156 & 85 \\
\hline
JOURNAL OF CLINICAL INVESTIGATION & 694 & 12.8 & 19.7 & 28.5 & 152 & 101 & 86 \\
AMERICAN JOURNAL OF PSYCHIATRY & 142 & 16.1 & 12.1 & 28 & 93 & 176 & 87 \\
ANNALS OF SURGERY & 583 & 10.3 & 13.1 & 27.9 & 228 & 164 & 88 \\
RADIOLOGY & 614 & 9.5 & 11.7 & 27.4 & 268 & 185 & 89 \\
ACS APPLIED MATERIALS \& INTERFACES & 10032 & 9.1 & 44.1 & 27.4 & 293 & 30 & 90 \\
\hline
IEEE TRANSACTIONS ON SMART GRID & 1261 & 8.6 & 14.1 & 27.3 & 333 & 148 & 91 \\
BIOSENSORS \& BIOELECTRONICS & 1599 & 9.6 & 19.3 & 27.1 & 264 & 104 & 92 \\
INORGANIC CHEMISTRY & 3466 & 5.1 & 6.5 & 27 & 1127 & 376 & 93 \\
CELL METABOLISM & 286 & 23.1 & 27.1 & 27 & 58 & 56 & 94 \\
EUROPEAN HEART JOURNAL & 449 & 23.8 & 35.2 & 26.9 & 54 & 37 & 95 \\
\hline
BRAIN & 471 & 11.7 & 14.3 & 26.8 & 167 & 146 & 96 \\
SCIENCE ADVANCES & 2103 & 13.5 & 37 & 26 & 135 & 34 & 97 \\
SCIENCE TRANSLATIONAL MEDICINE & 484 & 16.5 & 23.2 & 26 & 89 & 69 & 98 \\
OPTICA & 448 & 10.4 & 11.6 & 25.7 & 222 & 188 & 99 \\
REMOTE SENSING OF ENVIRONMENT & 979 & 10 & 16 & 25.7 & 245 & 126 & 100 \\
\hline
\hline
\end{tabular}
\caption{Top 51-100 journals ranked by field-specific $\Phi$ index (articles only), $\Phi_{s,A}$. Here, $n_A$ is each journal's size. Shown for comparison are the rankings by citation average $f_{A}$ and $\Phi_{A}$.}
\label{table:top51_100_Phi_sA}
\end{table}

Once we calculate field-specific $\Phi$ indexes (denoted $\Phi_s$), we can compare journals in different fields. In Tables \ref{table:top50_Phi_sA}, \ref{table:top51_100_Phi_sA}, we list the top 100 journals ranked by their article-only $\Phi_s$ index, i.e., $\Phi_{s,A}$. This ranking corrects for {\it field} disparities, in addition to scale disparities that are already corrected for by use of the general $\Phi$ index. Because journals in less-cited fields are compared to their in-field journals, their citation averages are field-standardized, which allows them to rise in rankings, if they are sufficiently cited {\it within} their field. 

As our standardization process removes citation disparity across different fields, we expect journals from a wider range of fields to reach the higher $\Phi_s$ ranks. And indeed this is what happens. For example, the top 1000 slots of $\Phi_{s,A}$ rankings include journals from nearly all field (220 subject (field) categories out of 229 JCR categories, or 96\% coverage). Compare this with 180 categories in the top-1000 $\Phi_{A}$ rankings (79\% coverage), or 178 categories in the top-1000 $f_{A}$ rankings (78\% coverage). 

In particular, the top 1000 ranks of $\Phi_{s,A}$ 
include more journals from mathematics, sociology, entomology, family studies, agronomy, history, linguistics, and law, which are some of the underrepresented categories when we rank by citation average, $f_A$. At the same time, the top 1000 ranks of $\Phi_{s, A}$ list fewer journals from materials science, management, oncology, business, cell biology, biochemistry, and chemistry. An indicative list of some new entries in the top 1000 slots (compared to when we ranked by citation average, $f_A$) is shown in Table \ref{table:new_entries_top1000}. 

\begin{table}[h]
\centering
\begin{tabular}{l l }
\hline
\hline
Subject (field) & Title \\
\hline
MATHEMATICS & ANNALS OF MATHEMATICS \\ 
 & INVENTIONES MATHEMATICAE \\
 & JOURNAL OF THE AMERICAN MATHEMATICAL SOCIETY \\
\hline
SOCIOLOGY & AMERICAN JOURNAL OF SOCIOLOGY \\
& JOURNAL OF MARRIAGE AND FAMILY \\
& WORK EMPLOYMENT AND SOCIETY \\
\hline
ENTOMOLOGY & INSECT SYSTEMATICS AND DIVERSITY \\ 
& INSECT BIOCHEMISTRY AND MOLECULAR BIOLOGY \\
& PEST MANAGEMENT SCIENCE \\
\hline
FAMILY STUDIES &  JOURNAL OF MARRIAGE AND FAMILY \\ 
& CHILD ABUSE \& NEGLECT \\
\hline
AGRONOMY &  FIELD CROPS RESEARCH \\
& EUROPEAN JOURNAL OF AGRONOMY \\
& INDUSTRIAL CROPS AND PRODUCTS \\
\hline
HISTORY &  PAST \& PRESENT \\
\hline
HISTORY \& PHILOSOPHY OF SCIENCE &  BRITISH JOURNAL FOR THE PHILOSOPHY OF SCIENCE \\ 
& SOCIAL STUDIES OF SCIENCE \\
\hline
LINGUISTICS &  APPLIED LINGUISTICS \\ 
& STUDIES IN SECOND LANGUAGE ACQUISITION \\
& MODERN LANGUAGE JOURNAL \\
\hline
LAW &  HARVARD LAW REVIEW \\
& STANFORD LAW REVIEW \\
& YALE LAW JOURNAL \\
\hline
\hline
\end{tabular}
\caption{An indicative list of some {\it new} entries in the top 1000 ranks when we rank by $\Phi_{s,A}$ (field-specific $\Phi$ index) as opposed to ranking by $f_A$ (citation average).}
\label{table:new_entries_top1000}
\end{table}

\section{Discussion}

\noindent
\subsection{Cautionary note: Over-analyzing field categories}

A cautionary note is warranted here. The assignment of journals into increasingly more specialized field categories can be problematic in two ways. First, it could give a flawed sense of accuracy in how a field is defined by including only the core journals and excluding interdisciplinary or more peripheral journals that still publish papers in the field. Second, if the list of journals that define a field becomes too small, then the field citation average $\mu_s$ could become volatile at the omission or inclusion of a single journal in the list. And this in turn could introduce some volatility in the $\Phi_s$ index of those journals whose citation average, $f$, lies close to $\mu_s$, since the term $(f - \mu_s)$ could then flip sign. 
As we mentioned earlier in section \ref{sec:2}, the process of allocating journals in each field is critical because it defines the benchmark against which each journal is measured. 

For example, to calculate field-specific $\Phi$ indexes of physics journals, we can use the Clarivate Analytics classification. Clarivate lists physics journals in 8 subcategories. In Table \ref{table:mu_sigma_physics}, we have calculated the field-specific citation mean $\mu_s$ and standard deviation $\sigma_s$ for each physics subcategory, as well as for all subcategories together that we combined into the category ``Physics.''  Looking at that table, we cannot fail to notice a large citation disparity across physics subcategories, as $\mu_s$ and $\sigma_s$ each vary by almost a factor of 3 ($\mu_s$ ranges from 2.3--6.3 and $\sigma_s$ ranges from 3.5--12.7). Is this a genuine disparity in citation practices across the physics subcategories, or could it indicate a less-than-ideal classification of some journals into these physics subcategories? In our experience, the latter scenario is more likely. Also concerning is the fact that multidisciplinary physics has a considerably lower citation average $\mu_s$ than the two subcategories of applied physics and condensed matter physics, even though journals in multidisciplinary physics are generally understood to be the most prestigious and should therefore be the most cited in their field. These observations serve as another reminder that the assignment of journals to various subcategories should be done carefully. Problematic classification of journals into field categories could adversely affect $\Phi$ index rankings, as it already happens with quartile rankings of Impact Factors (V\^{\i}iu, 2021). 

Although a detailed discussion of field classification lies beyond the scope of this study, we briefly illustrate how journal rankings change as the level of field aggregation becomes more refined. Specifically, we consider the 8 physics subfields---from mathematical physics to condensed matter physics--and compare journal orderings obtained using the global $\Phi$ index (all fields), the field-specific $\Phi_s$ index (physics), and the subfield-specific $\Phi_s$ index.

For the full set of 413 physics journals, the rankings remain essentially identical whether they are computed using the global or the physics-specific $\Phi$ index, yielding a Kendall correlation of $\tau =0.99$. This reflects the fact that the mean citation rate in physics, $\mu_s = 4.10$, is nearly identical to the global mean across all papers,  $\mu_s = 4.11$.

Moving one level further down the classification hierarchy, we compare rankings using the subfield-specific $\Phi_{s, A}$ index and the all-field $\Phi_{A}$ index (articles only). Table \ref{table:mu_sigma_physics} shows the Kendall $\tau$ coefficient between the two rankings for all 8 subfields, ranging from moderate ($\tau = 0.54$) to excellent ($\tau = 0.98$). 

Taken together, these comparisons indicate that $\Phi$-based rankings are relatively robust to reasonable levels of field aggregation. However, substantial disagreement in rankings is observed with increasing disparity between the field and subfield citation means (as expected). Isolated cases of substantial rank shifts (e.g., 24 $\rightarrow$ 64, 48  $\rightarrow$ 136, 63  $\rightarrow$ 104) merit closer examination.

\begin{table}[h]
\centering
\begin{tabular}{l l l l l l l l l}
\hline
\hline
Subject (field)	&	$\mu_s$	&	$\sigma_s$	&	no.  papers 	& 	no. journals 	& 	Kendall $\tau$	\\
\hline
PHYSICS & 4.1 & 9.5 & 302089 & 413 & 0.99  \\
PHYSICS, MATHEMATICAL & 2.3 & 4.8 & 20953 & 53 & 0.54 \\
PHYSICS, NUCLEAR & 2.5 & 4.7 & 11049 & 17 & 0.76 \\
\hline
PHYSICS, FLUIDS \& PLASMAS & 2.6 & 3.5 & 18892 & 30 & 0.58 \\
PHYSICS, ATOMIC, MOLECULAR \& CHEMICAL & 3.6 & 5.3 & 33603 & 34 & 0.91 \\
PHYSICS, MULTIDISCIPLINARY & 3.8 & 9.2 & 42287 & 78 & 0.98 \\
\hline
PHYSICS, PARTICLES \& FIELDS & 4.1 & 14.6 & 25240 & 26 & 0.98 \\
PHYSICS, APPLIED & 5 & 10.4 & 148133 & 152 & 0.86 \\
PHYSICS, CONDENSED MATTER & 6.3 & 12.7 & 69916 & 66 & 0.81 \\
\hline
\hline
\end{tabular}
\caption{Citation average, $\mu_s$, and standard deviation, $\sigma_s$, for physics papers (articles \& reviews) in various subject (field) categories. Also shown (last column) is the Kendall $\tau$ coefficient between the all-field  and subfield-specific $\Phi_A$ rankings. Note that some journals are listed in more than one field category.}
\label{table:mu_sigma_physics}
\end{table}

\noindent
\subsection{Scale-independent {\it and} field-specific $\Phi$ index for multidisciplinary journals}

How do we calculate field-specific $\Phi$ indexes for cross-disciplinary journals that publish in various fields, such as {\it Nature}, {\it Science}, {\it PLOS ONE}, {\it Scientific Reports}, etc? Here, while every paper published in the journal may belong to a single field, the journal itself publishes papers from various fields. So, the calculation of a journal-specific ($\mu_s, \sigma_s$) is not straightforward, and we need to exercise caution. 

There are two approaches to this problem: (a) 
Treat a multidisciplinary journal as a {\it composite} of monodisciplinary `sub-journals' and calculate a $\Phi_s$ index for each sub-journal. This way, we would end up with as many $\Phi_s$ indexes for a journal as the number of fields it covers. (b) Calculate a composite $\Phi$ index for all the fields that comprise the journal, and use that single metric to describe the multidisciplinary journal. 

Both approaches have pros and cons. In the former approach, a journal such as {\it Science} will have several $\Phi_s$ indexes. This situation is not unlike recognizing that, say, a university with different departments has a different ranking score in each area of activity, so it can rank higher in some areas than others. 

In the latter approach, the task is to integrate various field-specific $\Phi_s$ indexes to produce a single, composite $\Phi$ index for the whole journal. While this approach may seem simpler at a technical level---all we need to do is to build a composite that makes statistical sense, using, e.g., weighted averages---its conceptual challenge is the use of a single metric to describe the citation performance of {\it disparate} fields in the same journal. On the other hand, we can envision practical situations where users may find a composite $\Phi$ index advantageous for a quick and rough impression, in the same vein that a prospective student may be interested in a university's overall position in academic rankings (not in specific fields), a journalist in a country's overall health index score (not in specific categories such as air pollution or infant mortality), and an economist in a country's overall unemployment rate (not in age- or sector-specific rates). 

In future work, we will explore the $\Phi$ index of multidisciplinary journals further.

\noindent
\subsection{ $\Phi$ index for other citation averages}

We have seen how the $\Phi$ index arises naturally from the Central Limit Theorem, as the distance of a journal's citation average from the field mean, divided (rescaled) by the standard deviation of the expected citations for an equal-sized randomly formed journal. 
Although we have applied the $\Phi$ index to journals so far, there is nothing particular to journals in how the $\Phi$ index was developed. We can thus readily adapt the $\Phi$ index methodology to standardize for scale and field and compare the citation averages of individual researchers, research teams, universities, and countries. A preliminary investigation reveals considerable changes (and plausible corrections) in university rankings of citation averages that are standardized in the fashion of the $\Phi$ index (Antonoyiannakis, 2025). 

\noindent
\subsection{A quick way to obtain $\Phi$ index rankings within a field}

As we have discussed earlier, to calculate the $\Phi$ index of a set of $n$ papers, we need their citation average $f_n$ as well as the mean and standard deviation $\mu, \sigma$ of citations of all papers in the field, as shown in Eq. \ref{eq:Phi_def}. This way we can compare the $\Phi$ indexes of entities (journals, universities, etc.) across various fields of different $\mu, \sigma$. 

But if we are only interested in rankings {\it within} a given field, then all entities share the same $\mu, \sigma$, and the situation is simplified computationally. In this case, the standard deviation $\sigma$ serves as a scaling factor in the $\Phi$ index calculation, and we can set it to unity (or any other value) for simplicity. All we need to obtain the rankings then are the publication counts $n$ and the citation averages $f_n$ of all entities comprising the field, from which we can deduce $\mu$. This represents a significant improvement from a computational perspective, because many databases (e.g., Web of Science or Dimensions) readily provide these numbers as an outcome of a simple search, whereas to calculate $\sigma$, one needs the citation counts of each paper in the field. The price we pay for this simplicity is that the $\Phi$ indexes thus calculated are meaningful only within a field and cannot be used to compare across fields. But in many practical applications, this suffices.  

To illustrate this approach, we show in Table \ref{table:publishers} the $\Phi$ rankings of the top 20 publishers in the broad area of Physical Sciences, using data from Dimensions. For this demonstration, we treat all papers (research articles and reviews) in Physical Sciences as forming a single field, for which we find $\mu = 13.3$, while we set $\sigma=10$. 
Compared to their $f$ rankings, the $\Phi$ rankings of the following publishers show clear improvement: Springer Nature, the American Physical Society, the American Astronomical Society, Wiley, the American Chemical Society, Oxford University Press, EDP Sciences, the Royal Society of Chemistry, Optica Publishing Group, The Royal Society, Elsevier, and the American Geophysical Union. 

Overall, these improvements make sense. For example, Springer Nature publishes a large number (56945) of papers across 346 journals that vary widely in terms of prestige and editorial standards. In order of decreasing selectivity, these journals range from Nature to Nature Communications, the Nature journal series (Nature Physics, Nature Chemistry, etc.), the Communications journal series (Communications Physics, Communications Chemistry, etc.), the npj journal series, the European Physical Journal (EPJ) series, and Scientific Reports, among other titles. It is also reasonable that the 56945 papers published in Springer Nature journals are cited less, on average, than the 1753 papers published in the journals of the American Association for the Advancement of Science (AAAS)---90\% of these AAAS papers are published in the renowned Science journal and the highly respected Science Advances. 

In contrast, rankings by citation averages are problematic. For example, we find it hard to justify the 15-rank gap in $f$ rankings between AAAS and Springer Nature papers. It would make more sense here if Springer Nature is ranked lower (but not {\it far} lower) than AAAS, and certainly not lower than publishers such as Opto-Electronic Advances, Radiological Society of North America (RSNA), Iskender AKKURT, Institute of Physical Optics, Tsinghua University Press, Johnson Matthey, or Quantum OA Verein, which is what occurs in $f$ rankings. 

On a similar vein, the American Physical Society, the American Astronomical Society, Wiley, and the American Chemical Society are known for their rigorous peer review and high editorial standards. They publish a large number of papers--- ranging from 57266 to  15773 in 2020--2022--- so their improved positions in $\Phi$ rankings compared to $f$  rankings make sense. On the other hand, Annual Reviews is a publisher of high-quality, authoritative reviews that are well-cited, but given its small publication count (157 reviews in 2020--2022), it is not unreasonable that its position is pushed downward somewhat in $\Phi$ rankings.

\begin{table}[h]
\centering
\begin{tabular}{l l l l l l}
\hline
\hline
Name &$n$ & $f$ &$\Phi^{(*)}$ &rank ($f$) &rank($\Phi$) \\
\hline
American Association for the Advancement of Science (AAAS) &1753 &79.2 &275.7 &2 &1 \\
Springer Nature &56945 &21.2 &187.7 &17 &2 \\
American Physical Society (APS) &57266 &19.3 &142.8 &22 &3 \\
American Astronomical Society &15104 &24.8 &140.9 &12 &4 \\
Wiley &16815 &24.2 &140.9 &14 &5 \\
\hline
American Chemical Society (ACS) &15773 &23.7 &130.2 &15 &6 \\
Annual Reviews &157 &89.4 &95.3 &1 &7 \\
Proceedings of the National Academy of Sciences of the USA &1032 &42.2 &92.7 &4 &8 \\
Oxford University Press (OUP) &13702 &19.7 &74.5 &21 &9 \\
EDP Sciences &11086 &17.7 &46 &29 &10 \\
\hline
Tsinghua University Press &550 &29.9 &38.8 &8 &11 \\
Quantum OA Verein &619 &28.7 &38.2 &10 &12 \\
Royal Society of Chemistry (RSC) &6674 &16.7 &27.5 &33 &13 \\
Optica Publishing Group &16645 &15 &21.5 &38 &14 \\
The Royal Society &868 &18.2 &14.3 &27 &15 \\
\hline
Radiological Society of North America (RSNA) &34 &37 &13.8 &5 &16 \\
Elsevier &78545 &13.8 &13.1 &42 &17 \\
Opto-Electronic Advances &11 &45.2 &10.5 &3 &18 \\
Chinese Chemical Society &41 &28.4 &9.6 &11 &19 \\
American Geophysical Union (AGU) &3733 &14.9 &9.5 &39 &20 \\
\hline
\hline
\end{tabular}
\caption{Top 20 publishers ranked by $\Phi$ index in Physical Sciences, using data from Dimensions. Publication years 2020--2022. Citations were counted from the publication date up until May 27, 2025. Document types are articles and reviews. $^{(*)}$ For this dataset, $\mu_s=13.3$. The value of $\sigma_s$ is a scaling factor that does not affect the rankings within a given field; here, we assumed $\sigma_s = 10$.}
\label{table:publishers}
\end{table}

We can apply this approach to journal rankings, too. In Tables \ref{table:journals_condmat} and \ref{table:journals_organic_chemistry} we list the top 30 journals ranked by $\Phi_A$ index (articles only) in Condensed Matter Physics and Organic Chemistry, respectively. Again, the Dimensions database is used, which has paper-level identification of the field and can thus list the field-specific papers from multidisciplinary journals such as Nature, Science, Nature Communications, etc. The earlier-mentioned caveat on what constitutes a field applies.

\begin{table}[h]
\centering
\begin{tabular}{l l l l l l}
\hline
\hline
Name &$n$ & $f_A$ &$\Phi_A^{(*)}$ &rank ($f_A$) &rank($\Phi_A$) \\
\hline
Nature &193 &208.7 &269.8 &1 &1 \\
Science &124 &165.8 &168.5 &2 &2 \\
Nature Communications &979 &57.7 &135.3 &10 &3 \\
Physical Review Letters &1371 &40.3 &95.7 &20 &4 \\
Nature Materials &122 &95.8 &89.8 &5 &5 \\
\hline
Science Advances &249 &57 &67.1 &11 &6 \\
Nature Nanotechnology &64 &95.4 &64.7 &6 &7 \\
Physical Review X &198 &58.3 &61.7 &8 &8 \\
Advanced Materials &383 &44.6 &59 &18 &9 \\
Nature Physics &113 &69.9 &58.9 &7 &10 \\
\hline
ACS Nano &657 &35.1 &52.9 &32 &11 \\
Nano Letters &847 &29.1 &42.6 &46 &12 \\
npj Computational Materials &252 &39.8 &40.2 &22 &13 \\
Proceedings of the National Academy of Sciences of the USA &197 &39 &34.4 &25 &14 \\
Advanced Functional Materials &336 &32.8 &33.6 &35 &15 \\
\hline
Energy \& Environmental Science &14 &100.7 &32.2 &4 &16 \\
Acta Materialia &435 &28.9 &30.1 &47 &17 \\
Carbon &493 &27 &27.8 &50 &18 \\
Light: Science \& Applications &42 &48.2 &21.8 &16 &19 \\
ACS Applied Materials \& Interfaces &796 &22 &21.3 &73 &20 \\
\hline
Nature Photonics &21 &58.3 &20 &9 &21 \\
Science Bulletin &54 &39.6 &18.4 &23 &22 \\
npj Quantum Information &63 &36 &17.1 &29 &23 \\
npj Quantum Materials &240 &25.1 &16.5 &60 &24 \\
Journal of Alloys and Compounds &2571 &17.7 &16.5 &117 &25 \\
\hline
Applied Surface Science &1139 &19.3 &16.4 &105 &26 \\
Physical Review Research &1122 &19.1 &15.6 &107 &27 \\
Materials Today Physics &237 &24.4 &15.3 &64 &28 \\
National Science Review &47 &35.7 &14.5 &30 &29 \\
ACS Photonics &132 &27.1 &14.5 &49 &30 \\
\hline
\hline
\end{tabular}
\caption{Top 30 journals ranked by $\Phi_A$ index in Condensed Matter Physics, using data from Dimensions. Publication years 2020--2022. Citations were counted from the publication date up until May 27, 2025. Document types are articles. $^{(*)}$ For this dataset, $\mu_s=14.4$. The value of $\sigma_s$ is a scaling factor that does not affect the rankings within a given field; here, we assumed $\sigma_s = 10$. The following journals were in the original search results by Dimensions but have been removed by hand since they only publish review articles: Reviews of Modern Physics, Nature Reviews Physics,
Nature Reviews Methods Primers, 
Liquid Crystals Reviews, 
Reviews on Advanced Materials and Technologies, 
Annual Review of Condensed Matter Physics. Dimensions also identified articles in the following preprint repositories in this search, but we have removed them since preprint repositories are not journals: 
arXiv, 
bioRxiv, 
ChemRxiv, 
TechRxiv,
medRxiv, 
ScienceOpen Preprints,
OSF Preprints, 
Preprints.org.
}
\label{table:journals_condmat}
\end{table}

\begin{table}[h]
\centering
\begin{tabular}{l l l l l l}
\hline
\hline
Name &$n$ & $f_A$ &$\Phi_A^{(*)}$ &rank ($f_A$) &rank($\Phi_A$) \\
\hline
Journal of the American Chemical Society &2084 &54 &181.6 &7 &1 \\
Angewandte Chemie International Edition &2737 &40.8 &139.1 &13 &2 \\
Nature &59 &141.6 &97.8 &1 &3 \\
ACS Catalysis &1581 &37.9 &94.2 &15 &4 \\
Science &76 &119.7 &91.9 &2 &5 \\
\hline
Nature Communications &526 &48.3 &78.2 &9 &6 \\
Nature Chemistry &149 &68.1 &65.7 &3 &7 \\
Nature Catalysis &92 &66.5 &50.1 &4 &8 \\
Chemical Science &1195 &26.2 &41.4 &48 &9 \\
Organic Letters &4819 &20.1 &40.9 &77 &10 \\
\hline
Green Chemistry &757 &28.1 &38.2 &41 &11 \\
Chem &187 &36.6 &30.6 &16 &12 \\
Applied Catalysis B Environment and Energy &51 &49.2 &24.9 &8 &13 \\
CCS Chemistry &191 &29 &20.4 &37 &14 \\
Journal of Medicinal Chemistry &323 &25.6 &20.4 &50 &15 \\
\hline
Advanced Materials &19 &55.9 &18.1 &6 &16 \\
ACS Sustainable Chemistry \& Engineering &321 &23.7 &17 &58 &17 \\
ACS Applied Materials \& Interfaces &131 &26.7 &14.3 &46 &18 \\
Chinese Chemical Letters &397 &21.3 &14.1 &70 &19 \\
ACS Nano &45 &34.6 &13.6 &22 &20 \\
\hline
ACS Central Science &48 &32.3 &12.5 &27 &21 \\
Science Advances &33 &35.9 &12.4 &20 &22 \\
ChemSusChem &263 &21.9 &12.4 &65 &23 \\
Environmental Science and Technology &30 &36.3 &12.1 &17 &24 \\
Proceedings of the National Academy of Sciences of the USA &37 &33.5 &11.7 &25 &25 \\
\hline
Analytical Chemistry &86 &26.7 &11.5 &47 &26 \\
National Science Review &11 &47.7 &11.1 &10 &27 \\
Science China Chemistry &118 &23.9 &10.5 &56 &28 \\
Nano Letters &14 &41.9 &10.3 &12 &29 \\
Materials Horizons &10 &46.5 &10.2 &11 &30 \\
\hline
\hline
\end{tabular}
\caption{Top 30 journals ranked by $\Phi_A$ index in Organic Chemistry, using data from Dimensions. Publication years 2020--2022. Citations were counted from the publication date up until May 27, 2025. Document types are articles. $^{(*)}$ For this dataset, $\mu_s=14.2$. The value of $\sigma_s$ is a scaling factor that does not affect the rankings within a given field; here, we assumed $\sigma_s = 10$. The following journals were in the original search results by Dimensions but have been removed by hand since they only publish review articles: Mini-Reviews in Organic Chemistry, 
Nature Reviews Chemistry, 
Mini-Reviews in Medicinal Chemistry, 
Chemical Society Reviews, 
Russian Chemical Reviews, 
Phytochemistry Reviews, 
Wiley Interdisciplinary Reviews Computational Molecular Science, 
Accounts of Chemical Research. Dimensions also identified articles in the following preprint repositories in this search, but we have removed them since preprint repositories are not journals:  
ChemRxiv, 
bioRxiv, 
arXiv, 
Preprints.org.
}
\label{table:journals_organic_chemistry}
\end{table}

\section{Conclusions and outlook}

In this work, we introduced the $\Phi$ index, a scale- and field-standardized citation indicator derived from the Central Limit Theorem. By standardizing citation averages in a manner analogous to the $z$-score in statistics, the $\Phi$ index corrects for the well-documented but 
persistently unaddressed size dependence of the Impact Factor, while simultaneously accounting for differences in citation practices across fields. It has an elegant geometric interpretation: it measures how far a journal's citation average lies from the field mean, in units of the 
standard deviation expected for a randomly formed journal of the same size.

We validated the $\Phi$ index using a Monte Carlo random sample test---a principled, easy-to-apply diagnostic that we propose as a 
standard check for any citation indicator. Applying the $\Phi$ index to 12,173 journals in the 2020 Journal Citation Reports, we showed that it corrects for size bias and broadens the representation of high-performing journals from traditionally underrepresented fields such as mathematics, sociology, law, history, and linguistics in the top rankings. We also derived a closed-form standard error, $\text{SE}(\Phi) = 1$, which yields a ready-made uncertainty framework: the 95\% confidence interval is simply $\Phi \pm 1.96$. This allows practitioners to move beyond strict numerical rankings toward statistically principled ratings, an approach we illustrated with a proposed tiered classification of journals. A preliminary study (Antonoyiannakis, 2023) further showed that the $\Phi$ index outperforms the IF as a predictor of a journal's capacity to publish papers that reach the 99$^\text{th}$ percentile of citations in their field---a promising direction for future validation.

Because the $\Phi$ index requires only a publication count and a citation average for within-field comparisons---quantities readily available in most bibliometric databases---it is straightforward to implement and equally applicable to departments, universities, and countries. Future work will explore these extensions, examine the $\Phi$ index's robustness to gaming, and develop a more detailed comparison with other field-normalized indicators such as SNIP and JCI. We also intend to apply the random sample test more broadly as a community diagnostic tool.

We hope that the $\Phi$ index, and the random sample test that accompanies it, will encourage the bibliometrics community to adopt more statistically rigorous standards for evaluating and comparing citation indicators. We further hope that funding agencies, universities, and publishers will consider scale- and field-standardized metrics as a fairer basis for research assessment.

\vskip 0.5cm
\noindent
{\bf Acknowledgements}

\noindent 
I am grateful to Jerry I. Dadap and Hugues Chat{\'e} for stimulating discussions and encouragement, and to the late Richard Osgood Jr. and Irving Herman for hospitality at Columbia University. This work uses data from the Web of Science and the Journal Citation Reports (2020), accessed through Columbia University. The $\Phi$ index and related analyses are part of a broader bibliostatistics research program (https://bibliostatistics.org).

\vskip 0.5cm
\noindent
{\bf Funding}

\noindent 
This research did not receive any grant from funding agencies in the public, commercial, or not-for-profit sectors.

\vskip 0.5cm
\noindent
{\bf Author Contributions}

\noindent 
Manolis Antonoyiannakis: Conceptualization, data curation, formal analysis, methodology, writing.

\vskip 0.5cm
\noindent
{\bf Competing Interests}

\noindent
The author is a Senior Associate Editor and a Bibliostatistics Analyst at the American Physical Society. 
The manuscript expresses the views of the author and not of any journals, societies, or institutions where he may serve.
The methodology was developed independently of the author's editorial role.

\vskip 0.5cm



\vskip 0.5cm
\noindent
{\bf References}
%
%

\noindent 
Adams, J., McVeigh, M., Pendlebury, D., \&  Szomszor, M. (2019). Profiles, not metrics. \par 
Available from: 
\url{https://clarivate.com/webofsciencegroup/campaigns/profiles-not-metrics/}

\noindent 
Amin, M., \& Mabe, M. (2004). 
Impact factors: Use and abuse. 
{\it International Journal of Environmental Science  \par and Technology, 1,} 1--6.

\noindent 
Antonoyiannakis, M., \& Mitra, S. (2009). Editorial: Is PRL too large to have an `impact'? \par 
{\it Physical Review Letters}, {\it 102}, 060001. \url{https://doi.org/10.1103/PhysRevLett.102.060001}




\noindent 
Antonoyiannakis, M. (2018). Impact Factors and the Central Limit Theorem: Why citation averages are scale \par dependent, {\it Journal of Informetrics}, {\it 12}, 1072--1088.
\url{https://doi.org/10.1016/j.joi.2018.08.011}

\noindent 
Antonoyiannakis, M. (2019).
How a Single Paper Affects the Impact Factor: Implications for Scholarly \par Publishing,
{\it 
Proceedings of the 17th Conference of the International Society of Scientometrics and Informetrics, \par vol. II,} 2306--2313. 
Available from: 
\url{http://tinyurl.com/535s8sdn}

\noindent 
Antonoyiannakis, M. (2020). Impact Factor volatility due to a single paper: A comprehensive analysis, \par {\it  Quantitative Science Studies}, {\it 1}, 639--663.
\url{https://direct.mit.edu/qss/article/1/2/639/96141}

\noindent
Antonoyiannakis, M. (2023). 
The journal $\Phi$ index and highly cited papers. APS March Meeting. \par Available from: 
\url{https://ui.adsabs.harvard.edu/abs/2023APS..MARQ02004A/abstract}

\noindent
Antonoyiannakis, M. (2025). 
The $\Phi$ index and world university rankings. APS March Meeting. \par Available from: \url{https://schedule.aps.org/smt/2025/events/MAR-C69/3}





\noindent 
Campbell, P. (2008). 
Escape from the impact factor. 
{\it Ethics in Science and Environmental
Politics, 8}, 5--7. \par 
 \url{https://doi.org/10.3354/esep00078}

\noindent 
Clarivate. (2017).
A Closer Look at the Eigenfactor$^{\rm TM}$ Metrics.
Available from:  \par  
\url{https://clarivate.com/academia-government/blog/closer-look-eigenfactor-metrics/}

\noindent 
Clarivate. (2021).
Introducing the Journal Citation Indicator. 
Available from:  \url{http://tinyurl.com/ywzjudyb}





 
\noindent 
De Veaux, R. D., Velleman, P. D., \& Bock, D. E. (2014). Stats: Data and models (3rd Edition). \par Pearson Education Limited.







\noindent
Garfield, E. (1996). How can impact factors be improved?
{\it BMJ, 313}, 411--413.  \url{https://doi.org/10.1038/520429a}

\noindent
Gaind, N. (2018). Few UK universities have adopted rules against impact-factor abuse. {\it Nature News}. \par Available from:  \url{https://www.nature.com/articles/d41586-018-01874-w}

\noindent
Gingras, Y. (2016). Bibliometrics and Research Evaluation: Uses and Abuses (MIT Press).

\noindent
Hicks, D., Wouters, P., Waltman, L., de Rijcke, S.,  \& Rafols, I. (2015). Bibliometrics: The Leiden Manifesto for \par research metrics. \textit{\it Nature, 520}, 429--431.  \url{https://doi.org/10.1038/520429a}

\noindent 
Journal Citation Reports, Clarivate Analytics. Available from: 
 \url{https://jcr.incites.thomsonreuters.com/}









\noindent
Larivi{\`e}re, V., \& Sugimoto, C. R. (2019). The Journal Impact Factor: A Brief History, Critique, and  Discussion \par of Adverse Effects. In Gl{\"a}nzel, W.,  {\it et al}. (Eds.), \textit{\it Springer Handbook of} {\it Science and Technology Indicators}, \par Springer Handbooks.  \url{https://doi.org/10.1007/978-3-030-02511-3\_1}

\noindent 
Leydesdorff, L., Bornmann, L., and Adams, J. (2019). The integrated impact indicator revisited (I3*):
A non‑parametric \par alternative to the journal impact factor. {\it Scientometrics, 119}, 1669--1694.
\par  \url{https://doi.org/10.1007/s11192-019-03099-8}

\noindent 
MacRoberts, M.H., MacRoberts, B.R. Problems of citation analysis. (1996) {\it Scientometrics 36}, 435--444. \par 
 \url{https://doi.org/10.1007/BF02129604}

\noindent
Miranda, R., \& Garcia-Carpintero, E. (2018).
Overcitation and overrepresentation of review papers in the most cited \par papers,
{\it Journal of Informetrics, 12,} 1015--1030.
\url{https://doi.org/10.1016/j.joi.2018.08.006}

\noindent 
Moed, H. F. (2005). Citation analysis of scientific journals and journal impact measures. {\it Current Science}, \textit{89} (12), \par1990--1996.  \url{http://www.jstor.org/stable/24111059}

\noindent 
Moed, H. F. (2010). Measuring contextual citation impact of scientific journals,
{\it Journal of Informetrics, 4}, 265--277,
\par  \url{https://doi.org/10.1016/j.joi.2010.01.002}

\noindent 
Moed, H. F., Colledge, L., Reedijk, J., Moya-Anegon, F., Guerrero-Bote, V., Plume, A., \&Amin, M.  (2012).
\par Citation-based metrics are appropriate tools in journal assessment provided that they are accurate and used in an \par informed way.
{\it Scientometrics}, {\it 92}, 367--376.




\noindent 
Pulverer, B. (2013). Impact fact-or fiction?
{\it The EMBO Journal  32,} 1651--1652. 
\par  \url{https://doi.org/10.1038/emboj.2013.126}

\noindent 
Rousseau, R. (2009). What does the Web of Science five-year synchronous impact factor have to offer?. {\it Chinese \par Journal of Library and Information Science, 2} 1--7.


\noindent 
(2012). San Francisco Declaration on Research Assessment. Available from: 
\url{https://sfdora.org/read/}

\noindent 
Siler, K. \& Larivi{\` e}re, V. (2022).
Who games metrics and rankings? Institutional niches and journal impact factor \par inflation. {\it Research Policy 51}, 104608.
\url{https://doi.org/10.1016/j.respol.2022.104608}

\noindent 
Szomszor, M. (2021). 
Introducing the Journal Citation Indicator: A new, field-normalized measurement of journal \par citation impact. 
Available from: 
\url{http://tinyurl.com/fzb4ut76}

\noindent 
Spitzer, M., Wildenhain, J., Rappsilber, J., \& Tyers, M. (2014). BoxPlotR: a web tool for generation of  box plots. \par 
{\it Nature Methods, 11}, 121--122.
\url{https://doi.org/10.1038/nmeth.2811}


\noindent 
V\^{\i}iu, GA., P\u{a}unescu, M. The lack of meaningful boundary differences between journal impact factor quartiles 
 \par undermines their independent use in research evaluation. {\it Scientometrics} 126, 1495--1525 (2021). 
  \par \url{https://doi.org/10.1007/s11192-020-03801-1}








\noindent 
Waltman, L., van Eck, N. J., van Leeuwen, T. N., Visser, M. S. (2013). Some modifications to the SNIP journal \par impact indicator, {\it Journal of Informetrics, 7}, 272--285.  \url{https://doi.org/10.1016/j.joi.2012.11.011}

\noindent
Wouters, P., Sugimoto, C.R., Larivi{\`e}re, V., McVeigh, M.E., Pulverer, B., de Rijcke, S.,  Waltman, L. (2019). \par Rethinking impact factors: better ways to judge a journal.
{\it Nature 569}, 621-623.





\end{document}